\documentclass[aps,pra,superscriptaddress]{revtex4-1}

\usepackage{fullpage}
\usepackage{mathtools}
\usepackage{amsmath}
\usepackage{xcolor}
\usepackage{amssymb}
\usepackage{graphicx}
\usepackage{nameref}
\usepackage{braket}
\usepackage{tocloft}
\usepackage{bbold}
\usepackage{siunitx}
\usepackage{float}
\usepackage[english]{babel}
\usepackage{accents}
\usepackage[textwidth=2\marginparwidth]{todonotes}
\usepackage{adjustbox}
\usepackage{xparse,xcoffins}
\usepackage{soul}
\usepackage[authormarkup=none,final]{changes} 
\definechangesauthor[name={Joshua T Cantin}, color=red]{JTC}
\definechangesauthor[name={Joshua T Cantin, for Proof}, color=blue]{JTC_2}
\setdeletedmarkup{}

\DeclareMathOperator{\Tr}{Tr}

\newcommand\bigzero{\makebox(0,0){\text{\huge0}}}
\newcommand\oH{\emph{o}H$_{2}$}
\newcommand\oHmath{\mathrm{\emph{o}H_{2}}}

\newcommand{\RamHam}{\hat H^{\text{R}}}
\newcommand{\RamHamEl}{H^\text{R}}

\ExplSyntaxOn
\NewCoffin\imagecoffin
\NewCoffin\labelcoffin

\keys_define:nn { miguel/label }
{
	label   .tl_set:N = \l_miguel_label_tl,
	labelbox .bool_set:N = \l_miguel_label_box_bool,
	labelbox .default:n = true,
	fontsize .tl_set:N = \l_miguel_label_size_tl,
	fontsize .initial:n = \footnotesize,
	pos .choice:,
	pos/nw .code:n = \tl_set:Nn \l_miguel_label_pos_tl { left,up },
	pos/ne .code:n = \tl_set:Nn \l_miguel_label_pos_tl { right,up },
	pos/sw .code:n = \tl_set:Nn \l_miguel_label_pos_tl { left,down },
	pos/se .code:n = \tl_set:Nn \l_miguel_label_pos_tl { right,down },
	pos/n .code:n = \tl_set:Nn \l_miguel_label_pos_tl { hc,up },
	pos/w .code:n = \tl_set:Nn \l_miguel_label_pos_tl { left,vc },
	pos/s .code:n = \tl_set:Nn \l_miguel_label_pos_tl { hc,down },
	pos/e .code:n = \tl_set:Nn \l_miguel_label_pos_tl { right,vc },
	pos .initial:n = nw,
	unknown .code:n   = \clist_put_right:Nx \l_miguel_label_clist
	{ \l_keys_key_tl = \exp_not:n { #1 } }
}
\clist_new:N \l_miguel_label_clist
\box_new:N \l_miguel_label_box
\box_new:N \l_miguel_label_image_box

\NewDocumentCommand{\xincludegraphics}{O{}m}
{
	\group_begin:
	\tl_clear:N \l_miguel_label_tl
	\clist_clear:N \l_miguel_label_clist
	\keys_set:nn { miguel/label } { #1 }
	\tl_if_empty:NTF \l_miguel_label_tl
	{
		\miguel_includegraphics:Vn \l_miguel_label_clist { #2 }
	}
	{
		\SetHorizontalCoffin\imagecoffin
		{
			\miguel_includegraphics:Vn \l_miguel_label_clist { #2 }
		}
		\SetHorizontalCoffin\labelcoffin
		{
			\raisebox{\depth}
			{
				\bool_if:NTF \l_miguel_label_box_bool
				{ \fcolorbox{white}{white}{\l_miguel_label_size_tl\l_miguel_label_tl} }
				{ \l_miguel_label_size_tl\l_miguel_label_tl }
			}
		}
		\SetVerticalPole\imagecoffin{left}{3pt+\CoffinWidth\labelcoffin/2}
		\SetVerticalPole\imagecoffin{right}{\Width-7pt-\CoffinWidth\labelcoffin/2}
		\SetHorizontalPole\imagecoffin{up}{\Height-10pt-\CoffinHeight\labelcoffin/2}
		\SetHorizontalPole\imagecoffin{down}{3pt+\CoffinHeight\labelcoffin/2}
		\use:x{\JoinCoffins\imagecoffin[\l_miguel_label_pos_tl]\labelcoffin[vc,hc]} 
		\TypesetCoffin\imagecoffin
	}
	\group_end:
}
\NewDocumentCommand{\setlabel}{m}
{
	\keys_set:nn { miguel/label } { #1 }
}

\cs_new_protected:Nn \miguel_includegraphics:nn
{
	\includegraphics[#1]{#2}
}
\cs_generate_variant:Nn \miguel_includegraphics:nn { V }

\NewDocumentCommand{\xincludegraphicsB}{O{}m}
{
	\group_begin:
	\tl_clear:N \l_miguel_label_tl
	\clist_clear:N \l_miguel_label_clist
	\keys_set:nn { miguel/label } { #1 }
	\tl_if_empty:NTF \l_miguel_label_tl
	{
		\miguel_includegraphics:Vn \l_miguel_label_clist { #2 }
	}
	{
		\SetHorizontalCoffin\imagecoffin
		{
			\miguel_includegraphics:Vn \l_miguel_label_clist { #2 }
		}
		\SetHorizontalCoffin\labelcoffin
		{
			\raisebox{\depth}
			{
				\bool_if:NTF \l_miguel_label_box_bool
				{ \fcolorbox{white}{white}{\l_miguel_label_size_tl\l_miguel_label_tl} }
				{ \l_miguel_label_size_tl\l_miguel_label_tl }
			}
		}
		\SetVerticalPole\imagecoffin{left}{3pt+\CoffinWidth\labelcoffin/2}
		\SetVerticalPole\imagecoffin{right}{\Width-9pt-\CoffinWidth\labelcoffin/2}
		\SetHorizontalPole\imagecoffin{up}{\Height-10pt-\CoffinHeight\labelcoffin/2}
		\SetHorizontalPole\imagecoffin{down}{3pt+\CoffinHeight\labelcoffin/2}
		\use:x{\JoinCoffins\imagecoffin[\l_miguel_label_pos_tl]\labelcoffin[vc,hc]} 
		\TypesetCoffin\imagecoffin
	}
	\group_end:
}

\NewDocumentCommand{\xincludegraphicsC}{O{}m}
{
	\group_begin:
	\tl_clear:N \l_miguel_label_tl
	\clist_clear:N \l_miguel_label_clist
	\keys_set:nn { miguel/label } { #1 }
	\tl_if_empty:NTF \l_miguel_label_tl
	{
		\miguel_includegraphics:Vn \l_miguel_label_clist { #2 }
	}
	{
		\SetHorizontalCoffin\imagecoffin
		{
			\miguel_includegraphics:Vn \l_miguel_label_clist { #2 }
		}
		\SetHorizontalCoffin\labelcoffin
		{
			\raisebox{\depth}
			{
				\bool_if:NTF \l_miguel_label_box_bool
				{ \fcolorbox{white}{white}{\l_miguel_label_size_tl\l_miguel_label_tl} }
				{ \l_miguel_label_size_tl\l_miguel_label_tl }
			}
		}
		\SetVerticalPole\imagecoffin{left}{31pt+\CoffinWidth\labelcoffin/2}
		\SetVerticalPole\imagecoffin{right}{\Width-7pt-\CoffinWidth\labelcoffin/2}
		\SetHorizontalPole\imagecoffin{up}{\Height-10.5pt-\CoffinHeight\labelcoffin/2}
		\SetHorizontalPole\imagecoffin{down}{24pt+\CoffinHeight\labelcoffin/2}
		\use:x{\JoinCoffins\imagecoffin[\l_miguel_label_pos_tl]\labelcoffin[vc,hc]} 
		\TypesetCoffin\imagecoffin
	}
	\group_end:
}

\ExplSyntaxOff

\DeclareSIUnit\gauss{gauss}

\begin{document}
\raggedbottom
\title{
Transfer matrix theory of surface spin echo experiments with molecules
}
\author{J.~T.~Cantin}
\affiliation{Department of Chemistry, University of British Columbia, Vancouver, B.C., V6T 1Z1, Canada}
\author{G.~Alexandrowicz}
\affiliation{Schulich Faculty of Chemistry, Technion--Israel Institute of Technology, Technion City, Haifa 32000, Israel.}
\affiliation{Department of Chemistry, Swansea University, Singleton Park, Swansea SA2 8PP, UK.}
\author{R.~V.~Krems}
\affiliation{Department of Chemistry, University of British Columbia, Vancouver, B.C., V6T 1Z1, Canada}

\pacs{}
\date{\today}

\begin{abstract}
${}^3$He  beam  spin-echo experiments have been used to study surface morphology, molecular and atomic surface diffusion, phonon dispersions, phason dispersions and phase transitions of ionic liquids. However, the interactions between ${}^3$He atoms and surfaces or their adsorbates are typically isotropic and weak. To overcome these limitations, one can use  molecules instead of ${}^3$He in surface spin-echo experiments. The molecular degrees of freedom, such as rotation, may be exploited to provide additional insight into surfaces and the behaviour of their adsorbates. 
Indeed, a recent experiment has shown that \textit{ortho}-hydrogen can be used as a probe that is sensitive to the orientation of a Cu(115) surface [Godsi et al., Nat. Comm. \textbf{8}, 15357 (2017)]. However, the additional degrees of freedom offered by molecules also pose a theoretical challenge: a large manifold of molecular states and magnetic field-induced couplings between internal states. Here, we present a fully quantum mechanical approach to model molecular surface spin-echo experiments and connect the experimental signal to the elements of the time-independent molecule-surface scattering matrix.  We present a one-dimensional transfer matrix method that includes the molecular hyperfine degrees of freedom and accounts for the spatial separation of the molecular wavepackets due to the magnetic control fields. We apply the method to the case of \textit{ortho}-hydrogen, show that the calculated experimental signal is sensitive to the scattering matrix elements, and perform a preliminary comparison to experiment. This work sets the stage for Bayesian optimization to determine the scattering matrix elements from experimental measurements and for a framework that describes molecular surface spin-echo experiments to study dynamic surfaces.

\end{abstract}

\maketitle

%

\section{Introduction}

A major thrust of recent experimental work has been to achieve control over the longitudinal motion of atomic and molecular beams  \cite{Krems-book,lemeshko2013,friedrich2009,vanDeMeerakker2008,vanDeMeerakker2012,bethlem2003,dulieu2009,hogan2011}. Controlled beams can be used for a variety of applications, ranging from loading molecules into traps \cite{hutzler2012,bethlem2000,liu2015}, to measuring cross sections for molecular scattering with extremely high energy resolution \cite{onvlee2016,jankunas2015,stuhl2014,brouard2014},   
to precision spectroscopy \cite{tarbutt2013,cheng2016,acme_collaboration2018}, to controlled chemistry \cite{segev2019}. The development of methods for the initial state selection and control over both the longitudinal and transverse motion of molecular beams has also paved the way for matter-wave interferometry \cite{cronin2009,juffmann2013,hornberger2012}, nano-lithography \cite{gordon2003,rohwedder2007,dey2000}  and  precision studies of molecule-surface scattering. Although molecule-surface collisions have been a subject of numerous studies \cite{benedek2018,benedek1982,hulpke1992,tesaSerrate2016}, combining the latest advances in molecular beam control with surface scattering experiments opens opportunities for probing new regimes of molecule-surface energy exchange and obtaining detailed information about surface properties.  This is well exemplified by  
$^3$He spin echo (HeSE) experiments \cite{DeKieviet1995,DeKieviet1997,Jardine2009a,Jardine2009b} aiming to probe the structure of surfaces, as well as quantum matter adsorbed on surfaces, by scattering a beam of $^3$He in superpositions of nuclear spin states off a surface and observing the perturbation of the resulting interferometry signal. 
Analogous  to neutron spin echo experiments \cite{Mezei1980,Mezei2003}, HeSE experiments have been shown to detect the impact of gravity (on the energy scale of $\approx$ 10 neV) on the kinetic energy of atoms in the beam \cite{DeKieviet1995}. When used to study surfaces, HeSE experiments can be classified as a subset of quasi-elastic helium atom scattering experiments \cite{benedek2018}. Surface-sensitive HeSE experiments \cite{DeKieviet1997,Jardine2009a,Jardine2009b} have been used to study surface morphology \cite{Corem2013}, molecular and atomic surface diffusion \cite{Jardine2009a,Jardine2009b,Hedgeland2016,Godsi2015,hedgeland2011,lechner2013b,rotter2016}, inter-adsorbate forces \cite{Godsi2015,Kole2012}, phonon dispersions \cite{Kole2010,Jardine2009a,Jardine2009b}, phason dispersions \cite{Mcintosh2013}, structures and phase transitions of ionic liquids \cite{Mcintosh2014} and friction between adsorbates and surfaces \cite{Hedgeland2009a,Hedgeland2009b,lechner2013}. HeSE experiments have provided information about potential energy surfaces \cite{Jardine2004,Jardine2009a,Jardine2009b} and surface-adsorbate interactions \cite{Jardine2009a,Jardine2009b,Jardine2010} and are frequently combined with microscopic calculations to both test theory and gain insight into surface-adsorbate interactions \cite{Tamtoegl2018,Sacchi2017,hedgeland2011}.

The use of $^3$He as probe particles in HeSE experiments can sometimes be limited by the weak interaction strength between 
 $^3$He and surfaces or their adsorbates. In addition, $^3$He offers no internal degrees of freedom to absorb energy or induce anisotropic interactions. 
 Therefore, an important recent goal has been to extend surface spin-echo experiments to molecular beams \cite{Godsi2017}. 
   Molecules offer rotational degrees of freedom and anisotropic, state-dependent interactions, which could be exploited to gain new insights into surface dynamics. For example, it was recently shown that \emph{ortho}-hydrogen (\oH) can be used as a sensitive probe of surface morphology \cite{Godsi2017}: the experiment was able to discern how the interaction between an \oH~molecule and a Cu(115) surface depends on the orientation of the rotational plane of the hydrogen molecule relative to the surface. 
In addition, one could exploit the transfer of rotational energy from the probe molecules to surface adsorbates (or vice versa) in order to study the relative effects of the rotational and translational motion on the dynamics of the adsorbates. However, the increased complexity of molecules (compared to $^3$He atoms) makes the analysis of the spin-echo experiments complicated and requires one to account for the interplay of the translational, nuclear spin and rotational degrees of freedom in strong magnetic fields of differing orientations, in addition to the molecule-surface scattering event. 

Surface spin-echo experiments with molecules involve passing a molecular beam through a series of magnetic fields to control molecular wavepackets before and after the scattering event. 
A proper analysis of the resulting experimental signal must be based on (i) the solutions of the time-dependent Schr\"{o}dinger equation accounting for the development of entanglement between the translational motion and the internal states of molecules in the beam, as the beam transverses the magnetic fields of the spin-echo apparatus; (ii) the description of the molecule-surface  scattering events in the relevant frame of reference by the scattering matrix involving all relevant molecular states. This is a challenging task because the potential energy surfaces for molecule-surface interactions are difficult to compute with sufficient accuracy \cite{golibrzuch2014,krueger2015,park2019,yin2019,jiang2019,tchakoua2019,delcueto2019}, the calculations of the cross sections for molecule-surface scattering are extremely time consuming \cite{kroes2009,kroes2016} and because the orientation and strength of magnetic fields necessarily change throughout the spin-echo apparatus. An alternative formulation can be developed to treat the molecule-surface 
scattering matrix elements
as varying parameters to be determined from the experimental interferometry signal by one of the algorithms used in optimal control theory \cite{gordon1997,balintKurti2008,rabitz2000,brif2010,hofer2017,sola2018} or reinforcement machine learning designed to solve the inverse problem \cite{Vargas2017,rizzi2012}. In order for such a formulation to be practical, it is necessary to develop a rigorous method for the description of molecular dynamics inside the spin-echo apparatus, before and after the molecular wavepackets interact with the surface. This method must be efficient to allow for multiple feedback control loops, be accurate to ensure the proper description of interferometry dynamics and integrate rigorously the surface scattering matrix amplitudes into the resulting output signal.

In this paper, we exploit the transfer matrix method \cite{Walker1992,sanchez2012} to develop such
 a theoretical framework. The transfer matrix method \cite{Walker1992,sanchez2012} has been applied in various fields, such as for solving the 2D Ising model in statisical mechanics \cite{Baxter2007}, calculating reflection and transmission coefficients in optics \cite{born1980} and mesoscopic quantum transport \cite{mello2004}, determining photonic bandstructures \cite{Pendry1992}, and examining the tunnelling of a molecule through potential barriers \cite{saito1994,jarvis1998}.  The general and efficient framework we present can be used to analyze the coherent propagation of closed shell molecules through a series of static magnetic fields with different magnitudes and orientations, as well as through one or more scattering events.
 
 We apply this framework to surface-sensitive interferometry experiments that use closed shell molecules to study static surfaces. 
Specifically, we develop a fully quantum mechanical model of surface-sensitive molecular hyperfine interferometry experiments
by deriving a one-dimensional transfer matrix method that includes the internal hyperfine degrees of freedom of the probe molecules and that accounts for the eigenbasis changes between local regions of the magnetic field. We account for the experimental geometry with rotation matrices and describe the molecule-surface interaction with a scattering transfer matrix (a transformed version of the standard scattering matrix).

The method is applied to an \oH~hyperfine interferometry experiment.  By comparing the theoretical results with experimental measurements, we illustrate the importance of integrating over the velocity distribution of molecules in the beam. We further show that information about the scattering matrix elements is encoded in the experimental signal. 
In particular, we demonstrate that the experimental signal is sensitive to the magnitude and phase of the diagonal elements of the scattering transfer matrix. We also show that the signal is sensitive to scattering events that change the  projection quantum numbers of the molecular hyperfine states. Such dynamical processes are 
described by scattering transfer matrices with non-zero diagonal and off-diagonal matrix elements.
This sets the stage for determining, in part or in whole, the scattering transfer matrix elements of a particular molecule-surface interaction by comparing the computed and experimentally-measured signals. 

Finally, we compare our method with a semi-classical method, which is described briefly in the supplementary material of Ref.~\cite{Godsi2017} for \oH~and in more detail  in Ref.~\cite{Litvin2019} for spin 1/2 particles. Within this semi-classical method, the internal molecular degrees of freedom are treated quantum mechanically, while the centre of mass degree of freedom is treated classically. Through this comparison, we demonstrate that the present method can be extended to  study \textit{dynamic}, instead of static, surfaces by surface spin-echo experiments with molecules.


The remainder of this manuscript is organized as follows. In Section \ref{Sec_exptDescription}, we describe a generic molecular hyperfine interferometry experiment.
We then discuss, in Section \ref{Sec_magLensImpact}, the molecular state after  the state-selecting magnetic lens. In Section \ref{Sec_wavePacketPropagation},  we time-evolve the molecular state and integrate the result over the length of the detection window to obtain the relationship between the system eigenstates 
and the detector current. To obtain the system eigenstates,
we derive and apply, in Section \ref{Sec_transferMatrixFormalism}, a transfer matrix formalism that includes internal degrees of freedom. We also discuss the rotation and scattering transfer matrices used to account for the apparatus geometry and the molecule-surface interaction, respectively. In Section \ref{Sec_orthoHydrogen}, we demonstrate the application of this theoretical framework to the case of \oH, illustrate the need to integrate over the velocity distribution, illustrate the sensitivity of the calculated signal to various features of the scattering transfer matrix and perform a preliminary comparison with experiment. We compare the method of the present manuscript to the semi-classical method discussed by Godsi et al. \cite{Godsi2017} in Section \ref{Sec_method_comparison}.   Section \ref{Sec_Conclusion} concludes the work.

\section{Description of a Molecular Hyperfine Interferometry Experiment}
\label{Sec_exptDescription}

\begin{figure}[H]
	\centering
	\includegraphics[width=0.5\textwidth]{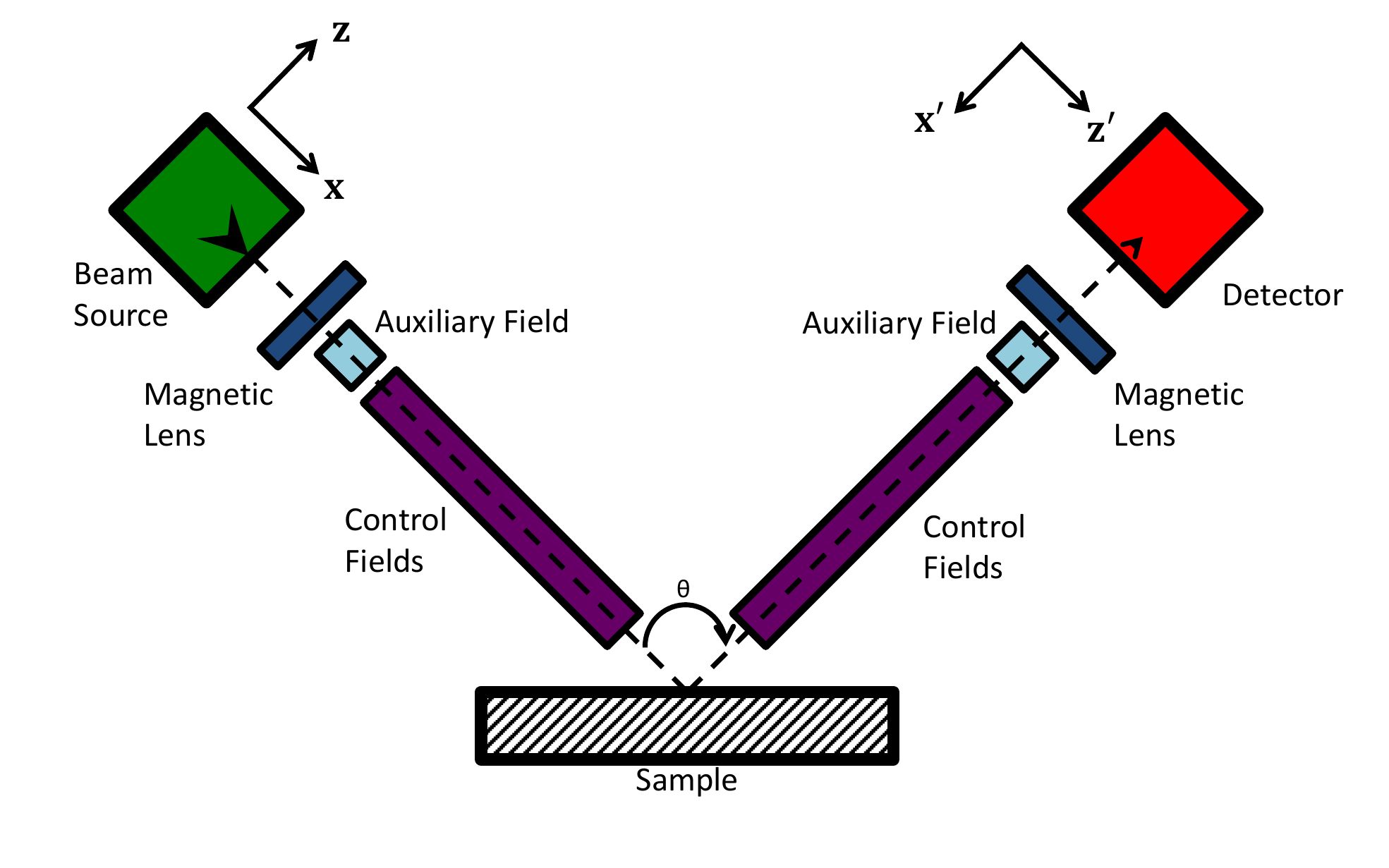}
	
	\caption{A generic molecular hyperfine interferometer consists of a beam source (green), magnetic lenses (dark blue), auxiliary fields (light blue), control fields (purple), the sample (hatched rectangle) in an ultra-high vacuum chamber, and the detector (red). See Section \ref{Sec_exptDescription} for more details on each component. The arrows and dashed line indicate the direction and path of the molecular beam, which is initially along the $+x$ direction and then along the $-x'$ direction after scattering. The two branches of the apparatus are separated by an angle $\theta$. $z$ and $z'$ denote the direction of the quantization axes before and after scattering, respectively. This definition of the quantization axes has been chosen to match the experiment by Godsi et al.~\cite{Godsi2017} and to simplify rotating the quantization axes in the transfer matrix method. The $y$ and $y'$ axes are identical and point into the page.}
	\label{Fig_experimentSchematic}
\end{figure}

A surface-sensitive molecular hyperfine interferometer uses a beam of molecules to probe various surface properties. To do this, a set of magnetic fields are used to simultaneously manipulate the internal hyperfine states of the probe molecules and  create a spatial superposition of molecular wavepackets. These wavepackets sequentially impact the sample surface and scatter in all directions. A second set of magnetic fields collects the molecules scattered in a narrow solid angle. This second set of magnetic fields further manipulates the molecular wavepackets, partially recombining them and allowing for molecular self-interference. Wavepackets with particular hyperfine states are then passed into a detector. A schematic of the experiment is depicted in Figure \ref{Fig_experimentSchematic}. We now discuss the different stages of the experiment in more detail.


The beam source must produce a continuous (or pulsed) beam of molecules with a sufficiently narrow velocity profile, mean velocity suitable for a particular experiment, sufficiently high flux,  and a density low enough to ensure that the molecules are non-interacting. One current apparatus \cite{Godsi2017} uses a supersonic expansion to produce such a beam. One can also envision experiments with slow molecular beams produced by extraction (sometimes with hydrodynamic enhancement) from a buffer-gas cooled cell \cite{hutzler2012} or with molecular beams controlled by electric-field \cite{BethlemPRL02} or magnetic-field deceleration \cite{WiederkehrMolPhys12}. Deceleration provides control over the mean velocity and narrows the velocity spread \cite{Krems-book}, which could be exploited for novel interferometry-based applications.  


The experiment selects molecules in particular hyperfine states  by employing a
magnetic lens whose magnetic field has a gradient in the radial direction. A cylindrically-symmetric field gradient is used to ensure sufficient molecular flux. The lens focusses molecules with low-field seeking states and defocusses molecules with high-field seeking states, allowing for purification of the molecular beam. After the lens, a unique quantization axis for the internal states is developed by using an auxiliary field that adiabatically rotates all magnetic moments until they lie along a single direction perpendicular to the beam path. The end of this auxiliary field is a strong dipolar field aligned along the $z$ direction that defines the quantization axis. Hexapole magnets can   be used as a  magnetic lens as their magnetic field gradients are sufficiently cylindrically symmetric \cite{Jardine2001,Dworski2004,Godsi2017}. More details about the internal states of the molecules immediately after the magnetic lens can be found in Section \ref{Sec_magLensImpact}.

Solenoids  whose magnetic fields are parallel to the beam propagation path are used to manipulate the molecular hyperfine states. These solenoids are helically-wrapped wire coils whose corresponding magnetic fields are generated by an electric current passing through each coil. These   solenoids are labelled as the control fields in Figure \ref{Fig_experimentSchematic}. Arbitrary magnetic field profiles can be obtained by changing the solenoid winding patterns and/or using multiple successive solenoids. 

The hyperfine states of a molecule change energy as the molecule enters a magnetic field. These changes to the hyperfine energy levels cause simultaneous changes in the molecular momenta, as the total energy is conserved. That is, when molecules enter a solenoid, molecules in low-field seeking states slow down and those in high-field seeking states speed up. Furthermore, because the direction of the magnetic field in a control field is not along the $z$ axis, the molecules are in a superposition of hyperfine states, with respect to the quantization axis defined by the magnetic field.  Thus, the differences in momenta cause the different components of each molecular wavepacket
to spatially separate as the wavepacket traverses the solenoid. 
Upon exiting the solenoid, the components of each wavepacket return to their original momenta, but remain spatially separated. That is, each wavepacket is now in an extended spatial superposition.

Each of these spatially separated wavepacket components comprise a superposition of the field-free hyperfine states. 
The exact superpositions of each wavepacket component, as well as the spatial separations between the components, depend on the magnetic field profile of the first branch. Each of the wavepacket components sequentially impacts the sample surface and 
scatters in all directions. However, the experiment only captures those molecules that pass through a particular solid angle.
While a current experiment \cite{Godsi2017} fixes the angle between the two branches, one can in principle explore many different scattering geometries by varying both the angle between the two branches of the apparatus and the orientation of the sample. 

After scattering, the collected molecules enter another set of control fields in the second branch of the apparatus. The hyperfine states again change in energy and momenta. In a helium-3 spin echo experiment, if the second magnetic field profile is identical but opposite in direction to the magnetic field profile of the first branch, the spatially separated wavepacket components realign (to first order) as they traverse the magnetic field(s), producing a spin echo. This allows the wavepacket components to interfere with each other. 
Interestingly, it has recently been shown \cite{Litvin2019} that echoes are also produced when the device operates with the fields in the same direction. 
With an arbitrary hyperfine Hamiltonian, such a realignment is only partial, though still useful. Experiments can be performed that explore either this spin-echo region or different relationships between the two magnetic field profiles, which  may allow for a variety of insights about the sample surface. For example, the two field profiles can be different or the field magnitudes can be varied simultaneously, keeping $B_1 = -B_2$. 
These different regimes of operation may produce different echoes, which can be collectively analyzed to provide more insight into molecule-surface interactions. 


Additionally, as the spatially separated wavepacket components hit the surface sequentially, rather than simultaneously, any temporal changes in the surface that are on the time scale of the impact-time separation can differentially impact the phases of each wavepacket component. This may result in different interference patterns or even loss of coherence. This loss of coherence is the basis for the sensitivity of HeSE measurements to surface motion \cite{Alexandrowicz2007}. Here, as in the recent experiment by Godsi et al. \cite{Godsi2017}, we focus on surfaces whose dynamics are either much faster or much slower then the molecule-surface or wavepacket-surface interaction times. Note, however, that the current framework is suitable for extension to interaction regimes where the surface dynamics are comparable to these time scales.

After leaving the last solenoid of the second branch, the wavepackets pass through another auxiliary field that begins with a strong dipolar field in the $z'$ direction. The auxiliary field then adiabatically connects magnetic moments aligned along the quantization axis to the radial direction of the final hexapole lens. This hexapole lens then focusses wavepackets with low-field seeking hyperfine states into the ionization detector and defocusses the rest. Finally, the ionization detector produces a current that is proportional to the molecular flux into the detector port. We describe how to calculate the molecular flux that enters the detector port in Section \ref{Sec_wavePacketPropagation} and the related transfer matrix formalism in Section \ref{Sec_transferMatrixFormalism}.


Analyzing the detector current as a function of the magnetic field profiles, the apparatus geometry, and the sample orientation can provide information about the interaction of the molecules with the sample surface. We discuss one possible analysis scheme in Section \ref{Sec_orthoHydrogen}. 



\subsection{Molecular Hyperfine Hamiltonian}
\label{Sec_molecularHami}

In principle, the only requirement for a molecular species to be suitable for molecular hyperfine interferometry is that the molecule have internal degrees of freedom whose energies are magnetic-field dependent. Such a requirement could be fulfilled by the presence of a nuclear spin, a rotational magnetic moment, or even an electronic spin. In practice, however, if the energy dependence on the magnetic field is too weak relative to the kinetic energy, state selection and state manipulation is difficult. On the other hand, if the dependence is too strong, the molecules may be difficult to control. Given these restrictions, we deem molecules that have a closed shell and are in an electronic state with zero orbital angular momentum to be most suitable for molecular hyperfine interferometry. In this case, the dominant interactions induced by magnetic fields are due to the nuclear magnetic spins of the molecules. 


The hyperfine states of such a closed shell molecule with zero orbital angular momentum arise from coupling between the nuclear spin and the rotational degrees of freedom. Interactions of these hyperfine states with a magnetic field arise from the response of the nuclear and rotational magnetic moments to the external magnetic field. We assume that the hyperfine Hamiltonian, also referred to here as the Ramsey Hamiltonian \cite{Ramsey1952}, is of the following form:
\begin{align}
\RamHam(\vec{B}) =  U\left(\hat{I}^2,\hat{J}^2,\hat{I}\cdot\hat{J},I,J\right) + V\left( \hat{I}^2, \hat{I}\cdot\vec{B},I,\vec{B}^2\right) + Q\left( \hat{J}^2, \hat{J}\cdot\vec{B},J,\vec{B}^2\right), \label{Eqn_RamHam}
\end{align}
where $\vec{B}$ is the vector of the external magnetic field, assumed to be uniform across the molecule; $\hat{I}$ and $\hat{J}$ are the nuclear spin and rotational angular momentum operators, respectively; $I$ and $J$ are the nuclear spin and rotational angular momentum quantum numbers, respectively; $U$ contains all spin-rotational couplings (such as $\hat{I}\cdot\hat{J}$ or $\hat I^2 \hat J^2$); $V$ contains all interactions of the nuclear spins with the magnetic field (such as $\hat{I}\cdot\vec{B}$); and $Q$ contains all interactions of the rotational angular momentum with the magnetic field (such as $\hat{J}\cdot\vec{B}$). Both $V$ and $Q$ are assumed to be proportional to positive powers of $|\vec{B}|$. 

At large magnetic fields, $V$ and $Q$ dominate, making the eigenbasis $\ket{Im_IJm_J}$, where \replaced[id=JTC_2]{$m_I$ and $m_J$ are the projections of the angular momenta $\vec{I}$ and $\vec{J}$ onto the external magnetic field direction, respectively}{ $m_{I/J}$ is the projection of the angular momentum $\vec{I}/\vec{J}$ onto the external magnetic field direction}.
At zero field, $\RamHam$ is diagonalized by $\ket{IJFM}$, where $\hat{F}=\hat{I}+\hat{J}$ is the total angular momentum operator and $M$ is the projection of $\vec{F}$ onto a chosen quantization axis. At intermediate fields, the eigenbasis is a function of the magnetic field and can be represented as a superposition of either $\ket{IJFM}$ or $\ket{Im_IJm_J}$ states. Note that $M$ is a good quantum number at all field strengths. We call an eigenstate of $\RamHam$ a Ramsey state, which we denote as $\ket{R}$ and which has the energy $E_R$. The number of eigenstates of $\RamHam$ is $N_R$, such that $1 \leq R \leq N_R$.

We treat the apparatus as a one dimensional system and account for the actual geometry by rotating the basis of the hyperfine states at the appropriate locations (see Section \ref{Sec_RotMat}). The total Hamiltonian can thus be written as 
\begin{align}
\hat H(x) = \frac{{\hat p}^2}{2m} + \RamHam\boldsymbol{\left(\right.} \vec{B}(x) \boldsymbol{\left.\right)} \label{Eqn_totalHami}
\end{align}
where $\hat p$ is the centre of mass momentum operator, $m$ is the molecular mass, and $x$ is the position of the molecule in the apparatus. The magnetic field $\vec{B}(x)$ is now spatially dependent, reflecting the magnetic field profiles of the two branches of the apparatus. 

In principle, the total Hamiltonian should incorporate molecule-surface interaction terms, such as the molecule-surface interaction potential. However, instead of treating the molecule-surface interactions explicitly, we include the interactions effectively through the use of a scattering transfer matrix (see Section \ref{Sec_ScatMat}). This allows us to separate the details of the molecule-surface interaction from the propagation of the molecules through the apparatus. We can then treat the molecular propagation analytically while allowing for the scattering matrix to be determined by the level of theory practical for a particular system. Even more importantly, this approach allows us to treat the scattering matrix elements as free parameters that can be determined by fitting the calculated signal to an experimental signal. For the present manuscript, we treat the scattering matrix elements as arbitrary parameters, focussing primarily on the development of a theoretical formalism to describe the molecular propagation. We choose particular values for the scattering matrix elements only when we apply the formalism specifically to \oH~(Section \ref{Sec_orthoHydrogen}). We also assume that the surface is static on timescales relevant to the experiment, such that the scattering matrix is time independent.


The system eigenstates $\ket{ER}$ are defined by the total Hamiltonian (\ref{Eqn_totalHami}) through $\hat H \ket{ER} = E \ket{ER}$. Note that the system eigenstate $\ket{ER}$ is $N_\text{R}$ degenerate and that any linear combination of these states with the same label $E$ is also an eigenstate of $\hat H$. This degeneracy occurs as, while the $N_\text{R}$ different Ramsey states may have different energies, the kinetic energy can always be selected to maintain the same total energy. For the sake of convenience, we choose the orthonormal basis to be that defined by $\RamHam\boldsymbol{\left(\right.}\vec B(x)\boldsymbol{\left.\right)} \ket{ER} = E_R \ket{ER}$  for $x\leq 0^{-}$. The zero of $x$ is defined to be immediately after the magnetic lens, while $y^{\pm} \equiv \lim_{\delta\to y^{\pm}} \delta$. 
We use these limit definitions as we will deal with discontinuities in the magnetic field when working with the transfer matrix formalism (Section \ref{Sec_transferMatrixFormalism}). As an example of the use of this notation, the statement that both one-sided limits are equal at the point $x$ $\left(i.e.~\lim_{a\to x^{-}}f(a)=\lim_{b\to x^{+}}f(b)\right)$ can be written as $f(x^{-})=f(x^{+})$.
%

The above definition of $\ket{ER}$ produces, for all $x$, a unique labelling of the system eigenstate $\ket{ER}$ by the total energy $E$ and the internal state $R$, where $R$ is a Ramsey state in the high magnetic field located immediately after the magnetic lens (i.e. at $x = 0^{-}$). Note that, because of this definition, $\RamHam\boldsymbol{\left(\right.}\vec B(x)\boldsymbol{\left.\right)} \ket{ER} \neq E_R \ket{ER}$  for $x\geq 0^{+}$; that is, the system eigenstates are superpositions of the \textit{local} Ramsey states for $x\geq 0^{+}$.
This unique labelling of the system eigenstates is valid for all $x$ as the eigenstate wavefunctions 
have a well-defined phase relationship throughout the entire apparatus.
See Section \ref{Sec_PropAndDiscontMat} for more details on the specifics of this phase relationship.


\section{Impact of the Magnetic Lens on the Molecular States}
\label{Sec_magLensImpact}

The magnetic lenses are designed to focus molecules with certain hyperfine states either onto the sample or into the detector. The remaining molecules are either defocussed or insufficiently focussed and  contribute significantly less to the experimental signal. Roughly, high-field seeking states are defoccussed, some of the low-field seeking states are well focussed and the rest of the low-field seeking states are partially focussed. The actual proportions of each hyperfine state in the molecular beam must be measured or calculated from simulation. These magnetic lenses typically use large magnetic fields and large magnetic field gradients to perform this focussing \cite{Dworski2004,Jardine2001}.  

In general, magnetic lenses may take different forms, but we will consider lenses that have one key feature: the internal degrees of freedom of the outgoing molecular wavepackets are decohered in the high-magnetic field basis (i.e. $\ket{Im_IJm_J}$). More precisely, we assume that the wavepacket exiting the magnetic lens is a mixed state of the form:
\begin{align}
\rho_0 &= \sum_{R_0} P_{R_0} \ket{\Psi_{R_0k_0}}\bra{\Psi_{R_0k_0}},  \label{Eqn_initialMixedState}
\end{align}
where
\begin{align}
\ket{\Psi_{R_0k_0}} &= \int \text{d}r~\psi_{R_0k_0}(r) \ket{rR_0};  \label{Eqn_PsiInitial}
\end{align}
$\psi_{R_0k_0}(r) \equiv \braket{r|\Psi_{R_0k_0}}$ is the wavefunction of a molecule in state $\ket{R_0}$; $\rho_0$ is the initial (time $t=0$) density matrix;  $\ket{rR_0}\equiv\ket{r}\ket{R_0}$; $\ket{r}$ is an eigenstate of the position operator; $\ket{R_0}$ is an eigenstate of $\RamHam(\vec{B}_\mathrm{lens})$; $\vec{B}_\mathrm{lens}$ is a high magnitude, $z$-aligned magnetic field; $k_0$ is the experimentally-determined mean wavenumber of the wavepacket; and $P_{R_0}$ is the probability that the hyperfine statevector of the molecule is $\ket{R_0}$. Note that $\rho_0$ is diagonal in $\ket{R_0}$ but not in $\ket{r}$ (or $\ket{k}$, the momentum basis). Also, $\vec{B}_\mathrm{lens} = \vec{B}\left(x=0^{-}\right)$ corresponds to the final portion of the auxiliary field (i.e. a strong, $z$ aligned, dipolar field), not the field inside the hexapole magnet itself (see Section \ref{Sec_exptDescription}).

That such a form of the wavepacket is valid follows from the work by Utz et al.~\cite{Utz2015}. The authors show that the two wavepackets arising from a spin--$\frac{1}{2}$ particle passing through a Stern-Gerlach apparatus are quickly decohered with respect to one another, even before they separate spatially. That is, the quantum dynamics themselves cause decoherence between the spin degrees of freedom (but not the spatial); a measurement or coupling to an external bath is not required. This decoherence occurs as the large magnetic field gradients cause a rapid oscillation in the off-diagonal terms of the extended Wigner distribution. That is, the phase relationship between the spin-up and spin-down components oscillates heavily in both the position and momentum bases, destroying coherence. 


Given that the magnetic lenses we consider act like a Stern-Gerlach apparatus for the molecular hyperfine states, it is reasonable to assume that the internal hyperfine degrees of freedom will also decohere. Thus, we need only determine the values of $P_{R_0}$ for a specific magnetic lens. These can be found via semi-classical calculations \cite{Godsi2017,Kruger2018}, may  be measured experimentally \cite{Kruger2018} or may potentially be  determined by solving the full 3D Schr\"odinger equation within the lens.

The mean velocity $v_0$ and velocity spread $\sigma_v$ of the molecules in the molecular beam can be measured experimentally \cite{Godsi2017}. Both of these values are determined from the position and profile of scattering peaks obtained from the scattering of the probe molecules by appropriate sample surfaces \cite{Godsi2017}. We assume that the initial wavefunction of a molecule $\psi_{R_0k_0}(r)$ is Gaussian and is characterized by $k_0\equiv  mv_0/\hbar$ and $\sigma_k\equiv m\sigma_v/\hbar$, where $m$ is the mass of the molecule. More precisely,
\begin{align}
\psi_{R_0k_0}(r) &= \int  \text{d}k~\frac{1}{\left(2\pi\sigma_k^2\right)^{\frac{1}{4}}}e^{-\frac{\left(k-k_0^{R_0}\right)^2}{4\sigma_k^2}}\frac{e^{i kr}}{\sqrt{2\pi}} \nonumber \\
	&= \sqrt{\sigma_k}\left(\frac{2}{\pi}\right)^{\frac{1}{4}}e^{ik_0^{R_0}r}e^{-r^2\sigma_k^2} \label{Eqn_PsiInitialProfile}
\end{align}
where $k_0^{R_0}$ is taken to be $k_0$. Though $k_0^{R_0}$ may in fact depend slightly (on the order of ppm) on $R_0$, we show later that the experimental signal is insensitive to small changes in $k_0^{R_0}$.
%


\section{Wavepacket Propagation and Signal Calculation}
\label{Sec_wavePacketPropagation}

The primary measured value of the experiment is a current that is proportional to the molecular flux entering the detector. This measured current is a function of the magnetic fields, the scattering geometry, and the surface properties. The molecular flux entering the detector  can be calculated as the product of the molecular flux incident to the apparatus and the probability that a molecule entering the apparatus will successfully pass through the apparatus and be detected. 
It is this probability of detection $P_{\text{detection}}$ that is sensitive to the experimental parameters and surface properties. Note that the incident molecular flux could  be either continuous or pulsed, as long as the density is low enough that the molecules can be considered non-interacting.

As the detector has a finite time-response, the probability of detection is given by 
\begin{align}
	P_{\text{detection}} = \frac{1}{\tau}\int_{t_1}^{t_2} \text{d} t \langle \hat C (t) \rangle,
	\end{align}
	where $t_1$ and $t_2$ are the initial and final times of the detection window $\tau = t_2-t_1$, and $\langle \hat C (t) \rangle$ is the expectation value of the detector measurement operator $\hat C$. This expectation value is given by 
	\begin{align}
	\langle \hat C (t) \rangle = \Tr \hat \rho(t) \hat C,
	\end{align}
	where $\hat \rho(t) \equiv \hat U \rho_0 \hat U^\dagger = \sum_{R_0} P_{R_0} \ket{\Psi_{R_0k_0}(t)}\bra{\Psi_{R_0k_0}(t)}$ is the time evolved density matrix, $\hat U \equiv e^{-i\frac{\hat H}{\hbar}t}$ is the time evolution operator, $\rho_0$ is the density matrix  (\ref{Eqn_initialMixedState}) at $t=0$, and $\ket{\Psi_{R_0k_0}(t)} \equiv \hat U \ket{\Psi_{R_0k_0}}$. 

Given that the detector consists of a  magnetic lens  that focusses molecules with particular states into a measuring apparatus, such as an ionization detector \cite{Godsi2017}, and that the internal degrees of freedom of these molecules are decohered by the second magnetic lens (see Section \ref{Sec_magLensImpact}), we can model the detector with a diagonal operator 
\begin{align}
\hat C = \sum_{R_D} \int \text{d}x~c_{R_D}(x)\ket{xR_D}\bra{xR_D}
\end{align}
 The matrix elements of $\hat C$ are the probabilities $c_{R_D}(x)$ of detecting, at position $x$, a molecule whose internal state is a high-field eigenstate $\ket{R_D}$ of $\RamHam$.
Note that $c_{R_D}(x)= 0~\mathrm{for}~x<x_D$, the detector position.
	
Using the time evolution operator, we determine the time-dependence of the density matrix $\rho(t)$ to be
\begin{align}
\rho(t) =  \sum_{R_0RR'} \int \text{d}E~\int \text{d}E'~P_{R_0}e^{-\frac{i}{\hbar}(E-E') t} \alpha_{k_0R_0}^{ER}{\alpha_{k_0R_0}^{*E'R'}}\ket{ER}\bra{E'R'}, \label{Eqn_rhoOfTime}
\end{align}
where $\alpha_{k_0R_0}^{ER} \equiv \int \text{d}r~\psi_{R_0k_0}(r)\Phi_{R_0}^{*ER}(r)$ is the overlap between the initial wavefunction $\psi_{R_0k_0}(r)$ and the system eigenstate wavefunction $\Phi_{R_0}^{ER}(r) \equiv \braket{rR_0|ER}$. 



\begin{figure}[H]
	\centering
	\includegraphics[width=\textwidth]{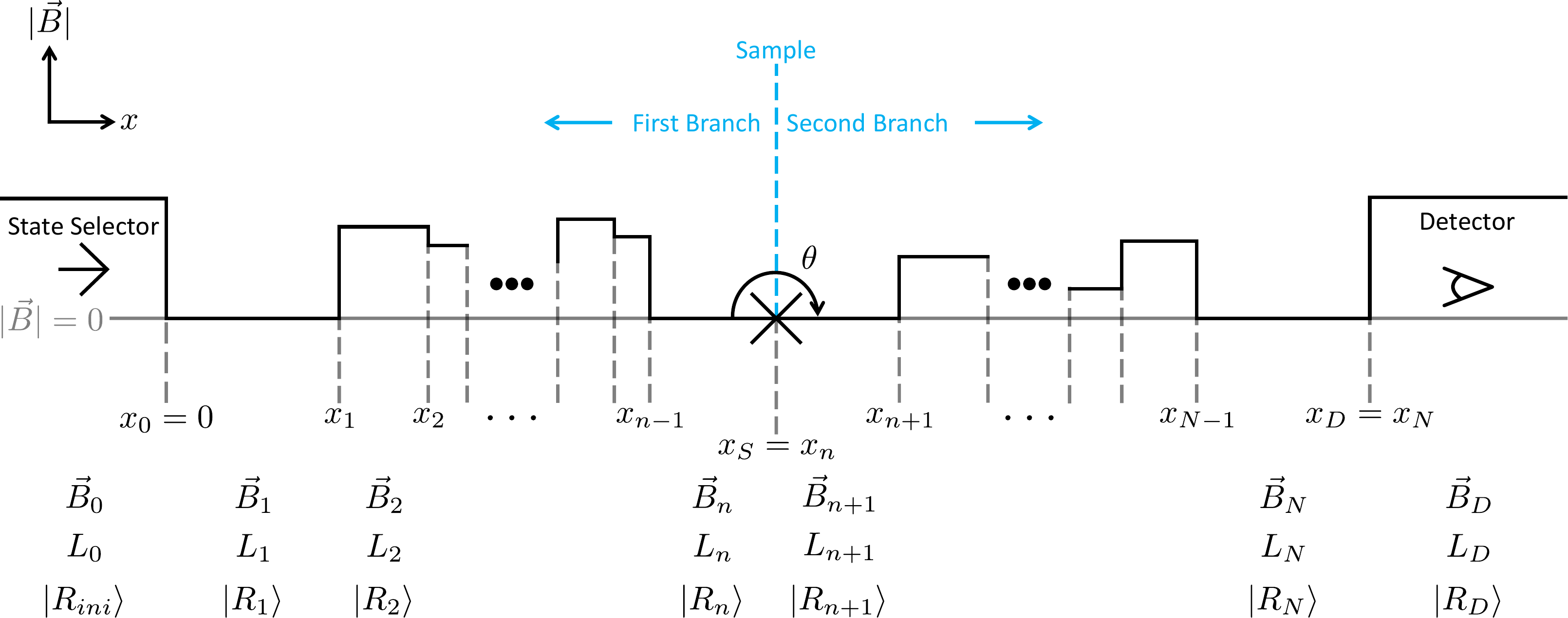}
	\caption{Generic field profile of a molecular hyperfine interferometry experiment. The actual magnetic field profiles of the experiment are approximated by $N+2$ regions of length $L_i$ and constant magnetic field $\vec{B}_i$ (black line). The true field profile is asymptotically approached as $N\to\infty$. We assume large magnetic fields in the regions of the state selector (large arrow) and the detector (eye), which, when combined with the dephasing discussed in Section \ref{Sec_magLensImpact}, allows us to neglect propagation in the selector and detector regions. That is, the exact locations of $x_0$ and $x_D$ are unimportant as long as $x_0$ is in the high-field region of the state selector, $x_D$ is in the high-field region of the detector, and all propagation is treated coherently between the two points. The initially Gaussian wavepacket propagates from $x_0^{-}$ along the first branch to the sample surface (cross) at $x_S$ then, after scattering, propagates along the second branch to $x_D^{+}$. The two branches are separated by an angle $\theta$. The vertical axis indicates the magnitude of the magnetic field $|\vec{B}|$ (the direction is not depicted for clarity), with $|\vec{B}|=0$ indicated by the grey solid line. $\ket{R_i}$ denotes the set of eigenstates of $\RamHam(\vec{B}_i)$, Eqn. (\ref{Eqn_RamHam}), in each region. }
	\label{Fig_ApparatusAbstraction}
\end{figure}

We can evaluate $\langle \hat C (t) \rangle$ by inserting a resolution of the identity $\sum_{R_D}\int \text{d}r~\ket{rR_D}\bra{rR_D}$, where $\RamHam\boldsymbol{\left(\right.}\vec B(x)\boldsymbol{\left.\right)} \ket{R_D} = E_{R_D} \ket{R_D}$  for $x\geq x_D^{+}$ and $x_D$ is the starting location of the detector (see Figure \ref{Fig_ApparatusAbstraction}). In other words, $\ket{R_D}$ is a Ramsey state in the strong dipolar magnetic field of the detector auxiliary field. The result is
\begin{align}
\langle \hat C (t) \rangle = \sum_{R_D,R_D'}\int \text{d}r~\int \text{d}r'~\bra{r'R_D'}\rho(t)\ket{rR_D}\bra{rR_D}\hat C\ket{r'R_D'} \label{Eqn_OperatorExpectValTraceEvald}
\end{align}
where we have evaluated the trace in the $\ket{r'R_D'}$ basis and
\begin{align}
\bra{r'R_D'}\rho(t)\ket{rR_D} =\sum_{R_0RR'} \int \text{d}E~\int \text{d}E'~P_{R_0}e^{-\frac{i}{\hbar}(E-E') t} \alpha_{k_0R_0}^{ER}{\alpha_{k_0R_0}^{*E'R'}}\Phi_{R_D'}^{ER}(r')\Phi_{R_D}^{*E'R'}(r). \label{Eqn_rhoOfTimeMatElements}
\end{align}
We emphasize that $R$ and $R_D$ are indices of \emph{different} sets of Ramsey states, i.e. $\braket{R|R_D}\neq \delta_{RR_D}$, unless the magnetic fields at the first magnetic lens ($x=0^{-}$) and the detector magnetic lens ($x=x_D^{+}$) happen to be identical.

We also have
\begin{align}
\bra{rR_D}\hat C\ket{r'R_D'} &= \sum_{R_D''} \int \text{d}z~c_{R_D''}(z)\delta(r-z)\delta_{R_DR_D''}\delta(r'-z)\delta_{R_D'R_D''} \nonumber \\
	&= c_{R_D}(r)\delta(r'-r)\delta_{R_D'R_D} 
\end{align}
which, when inserted with Eqn. (\ref{Eqn_rhoOfTimeMatElements}) into Eqn. (\ref{Eqn_OperatorExpectValTraceEvald}), results in
\begin{align}
\langle \hat C (t) \rangle = \sum_{R_0RR'} \int \text{d}E~\int \text{d}E'~P_{R_0}e^{-\frac{i}{\hbar}(E-E') t} \alpha_{k_0R_0}^{ER}{\alpha_{k_0R_0}^{*E'R'}}\left(\sum_{R_D}\int \text{d}r~\Phi_{R_D}^{ER}(r)\Phi_{R_D}^{*E'R'}(r) c_{R_D}(r)\right).
\end{align}

The \added[id=JTC_2]{initial} wavepacket is  almost entirely confined to the region $r\leq0^-$, as $\psi_{R_0k_0}(r)$  has a Gaussian profile (\ref{Eqn_PsiInitialProfile}) with spatial width on the order of \SI{10}{\angstrom} (as determined from the measured velocity distribution for \oH~\cite{Godsi2017}). 
Thus, we can evaluate $\alpha_{k_0R_0}^{ER} \equiv \int \text{d}r~\psi_{R_0k_0}(r)\Phi_{R_0}^{*ER}(r)$ if we know $\Phi_{R_0}^{ER}(r)$ for $r\leq0^-$. Given the definition of the eigenstate $\ket{ER}$, discussed in Section \ref{Sec_molecularHami}, we show  in Section \ref{Sec_PropAndDiscontMat} that $\Phi_{R_0}^{ER}(r)=A_{R}e^{irk^{ER}}\delta_{RR_0}$ for $r\leq0^-$ (cf. Eqn.~(\ref{Eqn_coeffKspaceExpansionReduced})), where $k^{ER}\equiv\frac{\sqrt{2m\left(E-E_\text{R}\right)}}{\hbar}$ (cf. Eqn.~(\ref{Eqn_kDefInField})). Combined with the definition (\ref{Eqn_PsiInitial}) of $\psi_{R_0k_0}(r)$,
\begin{align}
\alpha_{k_0R_0}^{ER} &\approx\int \text{d}r~\delta_{RR_0}A_{R}^*\sqrt{\sigma_k}\left(\frac{2}{\pi}\right)^{\frac{1}{4}}e^{i\left(k_0^{R_0}-k^{ER}\right)r}e^{-r^2\sigma_k^2}\nonumber \\
	&= \delta_{RR_0}\Gamma^{ER}_{k_0R_0}
\end{align}
where $\Gamma^{ER}_{k_0R_0}=A_{R}^*\frac{(2\pi)^\frac{1}{4}}{\sqrt{\sigma_k}}e^{-\frac{\left(k^{ER}-k_0^{R_0}\right)^2}  {4\sigma_k^2}}$. 
Thus,
\begin{align}
	\langle \hat C (t) \rangle &=\sum_{R_0} \int \text{d}E~\int \text{d}E'~P_{R_0}e^{-\frac{i}{\hbar}(E-E') t}\Gamma^{ER_0}_{k_0R_0} \Gamma^{*E'R_0}_{k_0R_0}\left(\sum_{R_D}\int \text{d}r~\Phi_{R_D}^{ER_0}(r)\Phi_{R_D}^{*E'R_0}(r) c_{R_D}(r)\right),
\end{align}
where we have performed the sums over $R$ and $R'$.

If the detection window $t_2-t_1$ is large enough that the entire wavepacket passes through the detection region defined by $c_{R_D}(z)$ we have
\begin{align}
P_{\text{detection}} &= \frac{1}{\tau}\int_{t_1}^{t_2} \text{d} t \langle \hat C (t) \rangle \approx \frac{1}{\tau}\int_{-\infty}^{\infty} \text{d} t \langle \hat C (t) \rangle \nonumber \\
	&=  \sum_{R_0} \int \text{d}E~\int \text{d}E'~P_{R_0}\frac{2\pi\hbar}{\tau}\delta\left(E-E'\right)\Gamma^{ER_0}_{k_0R_0} \Gamma^{*E'R_0}_{k_0R_0}\left(\sum_{R_D}\int \text{d}r~\Phi_{R_D}^{ER_0}(r)\Phi_{R_D}^{*E'R_0}(r) c_{R_D}(r)\right) \nonumber \\
	&=  \sum_{R_0} \int \text{d}E~P_{R_0}\left|\Gamma^{ER_0}_{k_0R_0}\right|^2 \left(\sum_{R_D}\int \text{d}r~\frac{2\pi\hbar}{\tau}\left|\Phi_{R_D}^{ER_0}(r)\right|^2  c_{R_D}(r)\right),
\end{align}
where $\frac{2\pi\hbar}{\tau}\delta\left(E-E'\right) = \frac{1}{\tau}\int_{-\infty}^{\infty} \text{d} te^{-\frac{i}{\hbar}(E-E') t}$. 



Physically, one can see that the probability of detection is proportional to the overlap $\left|\Gamma^{ER_0}_{k_0R_0}\right|^2$ of the initial wavepacket and a system eigenstate multiplied by the overlap $\int \text{d}r~\frac{2\pi\hbar}{\tau}\left|\Phi_{R_D}^{ER_0}(r)\right|^2  c_{R_D}(r)$ of the same system eigenstate and the detection region, as expected. 

Substituting for $\left|\Gamma^{ER_0}_{k_0R_0}\right|^2$ and given that 
\begin{align*}
\Phi_{R_D}^{ER_0}(r) &\equiv \braket{rR_0|ER} \\
&= e^{ik_{R_D}r}\braket{R_D|ER_0} \\
&\equiv e^{ik_{R_D}r}\beta_{R_D}^{ER_0}
\end{align*}
 for $r\geq x_D^+$ (cf. Eqn.~(\ref{Eqn_coeffKspaceExpansionReduced})), we have
\begin{align}
	P_{\text{detection}} &=  \sum_{R_0} P_{R_0}\left|A_{R_0}\right|^2\int \text{d}E~\frac{(2\pi)^\frac{1}{2}}{\sigma_k}e^{-\frac{\left(k^{ER_0}-k_0^{R_0}\right)^2}  {2\sigma_k^2}} \sum_{R_D} c_{R_D}\frac{2\pi\hbar}{\tau}\left|\beta_{R_D}^{ER_0}\right|^2
	\label{Eqn_P_DetectPenultimate}
\end{align}
where $c_{R_D} \equiv \int \text{d}r~c_{R_D}(r)$ and $\beta_{R_D}^{ER_0}\equiv\braket{R_D|ER_0}$, the projection of the system eigenstate $\ket{ER_0}$ onto the detector eigenstate $\ket{R_D}$ at $x_D^{+}$. For the purposes of comparing to experiment, only the dependence of $P_{\text{detection}}$ on the experimental parameters is needed, not its absolute value. Also,  the value of $A_{R_0}=1$ as $A_{R}e^{irk^{ER}}\equiv\braket{rR_0|ER_0}=e^{irk^{ER}} (\mathrm{for}~r\leq0^-)$  because of the specific definition of the system eigenstates (see Section \ref{Sec_molecularHami}).
Additionally, one can see that $P_{\text{detection}}$ is not sensitive to minor (on the order of ppm) changes in $k_0^{R_0}$ as $\sigma_k\propto k_0$ in experiment \cite{Godsi2017}. Finally, in Eqn.~(\ref{Eqn_P_DetectPenultimate}), only $\beta_{R_D}^{ER_0}$ is dependent on the magnetic fields, the scattering geometry, and the surface properties. It is thus sufficient to work with the following equation:
\begin{align}
P_{\text{detection}} &\propto  \sum_{R_0} P_{R_0}\int \text{d}E~e^{-\frac{\left(k^{ER_0}-k_0\right)^2}  {2\sigma_k^2}} \sum_{R_D} c_{R_D}\left|\beta_{R_D}^{ER_0}\right|^2 \label{Eqn_P_DetectFinal}
\end{align}

To determine the values of $\beta_{R_D}^{ER_0}$, 
we derive and apply the transfer matrix method with internal degrees of freedom (Section \ref{Sec_transferMatrixFormalism}).

\section{Transfer Matrix Formalism with Internal Degrees of Freedom}
\label{Sec_transferMatrixFormalism}


The transfer matrix method as applied in quantum transport turns the solution of the time-independent Schr\"odinger equation of a 1D system into a product of matrices \cite{Walker1992}. Pedagogical introductions can be found in Refs.~\cite{Walker1992,sanchez2012,mello2004}. The present problem 
has two unique features: (i) the propagating molecules have many internal degrees of freedom which may be mixed as the molecule transitions from one local field to another and (ii) molecules change their propagation direction after scattering by the surface. Problem (i) is addressed in Section \ref{Sec_PropAndDiscontMat}, while (ii) is addressed in Section  \ref{Sec_RotMat}. 
The impact of scattering on the internal degrees of freedom is accounted for by using a scattering transfer matrix (Section \ref{Sec_ScatMat}). 

The transfer matrix formalism we present in Section \ref{Sec_PropAndDiscontMat} is similar to the mixed multicomponent transfer matrix formalism described in \cite{diago2006} and can be viewed as an extension and application of the transfer matrix formalism used in the study of molecular tunnelling \cite{saito1994,jarvis1998}. The formalism combines transfer matrices that incorporate the internal molecular degrees of freedom of a composite particle \cite{saito1994,jarvis1998} with eigenbasis changes between regions of the external potential. Similar eigenbasis changes have been  employed in the transfer matrix formalism used in the envelope function approximation, which is used to calculate electronic properties in abrupt semi-conductor heterostructures \cite{foreman1996}. We further extend the transfer matrix formalism in Sections \ref{Sec_RotMat} and \ref{Sec_ScatMat} to account for the impact of scattering on the molecules and their relevant internal degrees of freedom.


\subsection{Propagation and Discontinuity Matrices}
\label{Sec_PropAndDiscontMat}

We first break up the arbitrary magnetic field profiles of the apparatus into rectangular regions of constant field, as shown in Figure \ref{Fig_ApparatusAbstraction}. 
We then solve the Schr\"odinger equation for a single eigenstate in a single region of constant field. Subsequently, we determine the impact of the boundary conditions that exist at the discontinuity between two regions of constant field. Using these solutions, we determine matrices that describe the spatial dependence of the eigenstate wavefunction coefficients within a region of constant field (propagation matrices) and matrices that describe how these coefficients change across the discontinuity between two regions of constant field (discontinuity matrices). Note that while we derive these matrices for molecules whose internal degrees of freedom are described by the Ramsey Hamiltonian (\ref{Eqn_RamHam}), the formalism is not limited to this Hamiltonian.


 Within a region of uniform magnetic field, the Ramsey Hamiltonian $\RamHam$ is constant, which allows us to derive the propagation matrix that includes the internal degrees of freedom. We begin by expanding a system eigenstate $\ket{E\tilde R}$ as
\begin{align}
\ket{E\tilde R} = \sum_{R}\int\mathrm{d}x~ \Phi_{R}^{E\tilde R}(x) \ket{xR}, \label{Eqn_eigstateExpanDef}
\end{align}
where $\Phi_{R}^{E\tilde R}(x)\equiv\braket{xR|E\tilde R}$, we define $\ket{xR}\equiv\ket{x}\ket{R}$, and $\ket{R}$ is one of the  $N_R$ Ramsey states of a molecule in some magnetic field $\vec{B}$. Note that $\vec{B}$ is not necessarily the local magnetic field $\vec{B}_\mathrm{loc}$ of the current region and thus $\ket{R}$ is not necessarily an eigenstate of $\RamHam(\vec{B}_\mathrm{loc})$ at this point.
Also, the eigenstates $\ket{E \tilde{R}}$ are labelled by their energy $E$ and a particular Ramsey index $\tilde R$, such that $\RamHam(\underaccent{\tilde} {\vec B})\ket{E\tilde R}=E_{\tilde R}\ket{E\tilde R}$, with $\underaccent{\tilde} {\vec B}$ an arbitrarily chosen magnetic field. 

Using Eqn. (\ref{Eqn_eigstateExpanDef}), the Schr\"odinger equation with the total Hamiltonian (\ref{Eqn_totalHami}) can be shown to be (Appendix \ref{App_schroedingEqn}):
%
\begin{align}
-\frac{\hbar^2}{2m}\frac{\partial^2}{\partial x^2}\Phi_{R_0}^{E\tilde R}(x) &=  \Phi_{R_0}^{E\tilde R}(x)E-\sum_{R}\RamHamEl_{R_0R}\Phi_{R}^{E\tilde R}(x), \label{Eqn_coeffSchroedEqn}
\end{align}
where $ \RamHamEl_{R_0R} = \bra{R_0}\RamHam(\vec{B}_\mathrm{loc}) \ket{R}$. Eqn.~(\ref{Eqn_coeffSchroedEqn}) is in general difficult to solve because of the coupling of the internal degrees of freedom by $\RamHam(\vec{B}_\mathrm{loc})$. However, if we choose the eigenbasis of the internal degrees of freedom to satisfy $\RamHam(\vec{B}_\mathrm{loc})\ket{R}=E_\text{R}\ket{R}$ (that is, $\ket{R}$ is now a Ramsey state of a molecule in the local magnetic field $\vec{B}_\mathrm{loc}$), the equations decouple and we obtain
\begin{align}
\frac{\partial^2}{\partial x^2}\Phi_{R}^{E\tilde R}(x) &= -\frac{2m}{\hbar^2}\left(  E-E_\text{R}\right) \Phi_{R}^{E\tilde R}(x). \label{Eqn_coeffSchroedEqnDiagd}
\end{align}
The solution is
\begin{align}
\Phi_{R}^{E\tilde R}(x) = A_{R}e^{ik_Rx}+B_{R}e^{-ik_Rx}, \label{Eqn_coeffKspaceExpansion}
\end{align}
where $A_R$ and $B_R$ are $R$-dependent coefficients and
\begin{align}
k_R\equiv\frac{\sqrt{2m\left(E-E_\text{R}\right)}}{\hbar}. \label{Eqn_kDefInField}
\end{align}

As per the single channel transfer matrix method \cite{Walker1992}, given that $\Phi_{R}^{E\tilde R}\left(x+\Delta x\right) = A_{R}e^{ik_Rx}e^{ik_R\Delta x}+B_{R}e^{-ik_Rx}e^{-ik_R\Delta x}$, we can collect the $A_R$ and $B_R$ coefficients into a $2N_R$-dimensional coefficient vector $\vec\phi_x = \left(A_1,A_2,...,A_{N_R,}B_1,B_2,...,B_{N_R}\right)^T$ and write
\begin{align}
\vec\phi_{x_2} = \mathbf{\Pi}_{x_2-x_1}\vec\phi_{x_1},
\end{align}
where $\mathbf{\Pi}_{x}$ is the $2N_R\times2N_R$ propagation matrix
\begin{align}
\mathbf{\Pi}_{x} &\equiv \left[\bigoplus_R e^{ik_Rx}\right] \oplus \left[\bigoplus_R e^{-ik_Rx}\right] \nonumber \\
&= \begin{pmatrix}
e^{ik_1x} & & & & & \\
& \ddots & &  &  \bigzero & \\
& & e^{ik_{N_R}x} & & & \\
& &  & e^{-ik_1x} & & \\
& \bigzero& & \ddots & & \\
& & & & & e^{-ik_{N_R}x}\\
\end{pmatrix}, \label{Eqn_fullPropMat}
\end{align}
where $\oplus$ denotes the direct sum. 

Following the derivation of Ref.~\cite{Walker1992}, we can determine how the coefficients transform across a step discontinuity in the magnetic field. Using the propagation matrix (\ref{Eqn_fullPropMat}) and a relabelling of the coordinate system, we can always set the discontinuity to appear at $x=0$. Given that Eqn.~(\ref{Eqn_coeffSchroedEqn}) applies everywhere, the coefficients $\Phi_{R}^{E\tilde R}(x)$ and their derivatives are continuous across the discontinuity (i.e. $\Phi_{R}^{E\tilde R}(x)\in C^1(x)$), for each value of $R$. However, the coefficients are only known when $\ket{R}$ is an eigenstate of $\RamHam(\vec{B}_\mathrm{loc})$, which differs on each side of the discontinuity (that is, $\vec{B}(0^{-})\neq\vec{B}(0^{+})$). Note that the wavevector $\ket{E\tilde R}$ is the same everywhere in the system. Thus, by writing the wavevector $\ket{E\tilde R}$ in the two different bases corresponding to the eigenstates of $\RamHam$ on each side of the field, we see that the coefficients at a specific value of $x$ are related by a basis transformation:
\begin{align}
\ket{E\tilde R^{-}}&=\ket{E\tilde R^+} \nonumber \\
\sum_{R^{-}}\int\mathrm{d}x~ \Phi_{R^{-}}^{E\tilde R}(x) \ket{xR^{-}}   &=\sum_{R^{+}}\int\mathrm{d}x~ \Phi_{R^{+}}^{E\tilde R}(x) \ket{xR^{+}} \nonumber \\
\sum_{R^{-}R^{+}}\int\mathrm{d}x~ \Phi_{R^{-}}^{E\tilde R}(x) \braket{R^{+}|R^{-}} \ket{xR^{+}}  &=\sum_{R^{+}}\int\mathrm{d}x~ \Phi_{R^{+}}^{E\tilde R}(x) \ket{xR^{+}} \nonumber \\
\implies \Phi_{R^{+}}^{E\tilde R}(x) = &\sum_{R^{-}} \Phi_{R^{-}}^{E\tilde R}(x)\braket{R^{+}|R^{-}}, \label{Eqn_coeffBasisTransform}
\end{align} 
where $\ket{E\tilde R^{\pm}}$ is the wavevector written in the basis of $\ket{R^{\pm}}$, $\ket{R^{\pm}}$ are the eigenstates of $\RamHam\boldsymbol{\left(\right.}\vec B(0^{\pm})\boldsymbol{\left.\right)}$ on the left $(-)$ and right $(+)$ sides of the discontinuity at $x=0$, respectively, and $\sum_{R^{+}} \ket{R^{+}}\bra{R^{+}}$ was inserted in the third line (recall that $\ket{xR^{-}}\equiv\ket{x}\ket{R^{-}}$). The values $\braket{R^{-}|R^{+}}$ are 
recognized as the matrix elements $S_{R^{-}R^{+}}$ of the matrix $\mathbf{S}^{R^{+}}_{R^{-}}$ whose columns are the eigenstates of $\RamHam\boldsymbol{\left(\right.}\vec B(0^{+})\boldsymbol{\left.\right)}$ written in the $\ket{R^{-}}$ basis.

Since $\Phi_{R}^{E\tilde R}(x)\in C^1(x)$ for each value of $R$ separately, we can equate the two limits $\lim_{x\to 0^{\mp}} \Phi_{R^{+}}^{E\tilde R}(x)$ and the two limits of the derivative $\lim_{x\to 0^{\mp}} \frac{\partial}{\partial x} \Phi_{R^{+}}^{E\tilde R}(x)$. Solving the resultant equations for the coefficients $A_{R^{+}}$ and $B_{R^{+}}$, we obtain (Appendix \ref{App_discontCoeff}):

\begin{align}
A_{R^{+}} &=  \sum_{R^{-}} S_{R^{-}R^{+}}^* \Delta_{R^{+}R^{-}}^{+} A_{R^{-}} + \sum_{R^{-}} S_{R^{-}R^{+}}^*\Delta_{R^{+}R^{-}}^{-} B_{R^{-}} \label{Eqn_Acoeff}\\
B_{R^{+}} &=  \sum_{R^{-}} S_{R^{-}R^{+}}^*\Delta_{R^{+}R^{-}}^{-} A_{R^{-}} + \sum_{R^{-}} S_{R^{-}R^{+}}^*\Delta_{R^{+}R^{-}}^{+} B_{R^{-}} \label{Eqn_Bcoeff}
\end{align}
where $S_{R^{-}R^{+}}^* \equiv \braket{R^{+}|R^{-}}$,  $\Delta_{R^{+}R^{-}}^{\pm} \equiv \frac{1}{2}\left(1\pm\frac{k_{R^{-}}}{k_{R^{+}}}\right)$, $k_{R^{\pm}} \equiv \frac{\sqrt{2m \left(E-E_{\text{R}^{\pm}}\right) }}{\hbar}$, and $E_{\text{R}^{\pm}}\equiv \braket{R^{\pm}|\RamHam\boldsymbol{\left(\right.}\vec B(0^{\pm})\boldsymbol{\left.\right)}|R^{\pm}}$. There are $N_R$ such sets of equations, one for each value of $R^{+}$. Working again with $\vec\phi_x = \left(A_1,A_2,...,A_{N_R,}B_1,B_2,...,B_{N_R}\right)^T$, one can write the matrix equation
\begin{align}
\vec\phi_{x^{+}} = \mathbf{K}\vec\phi_{x^{-}},
\end{align}
where $x^{\mp}$ indicates the location just before $({-})$ or just after $({+})$ the discontinuity located at $x$ and $\mathbf{K}$ is the $2N_R\times2N_R$ discontinuity matrix
\begin{align}
\mathbf{K} &\equiv \begin{pmatrix}
{\mathbf{S}^{R^{+}}_{R^{-}}}^\dagger\circ\mathbf{\Delta^{+}}& {\mathbf{S}^{R^{+}}_{R^{-}}}^\dagger\circ\mathbf{\Delta^{-}}\\
{\mathbf{S}^{R^{+}}_{R^{-}}}^\dagger\circ\mathbf{\Delta^{-}}&{\mathbf{S}^{R^{+}}_{R^{-}}}^\dagger\circ\mathbf{\Delta^{+}} 
\end{pmatrix}  \label{Eqn_fullDiscontMat}
\end{align}
where $\circ$ denotes the element-wise Hadamard product, such that $({\mathbf{S}^{R^{+}}_{R^{-}}}^\dagger\circ\mathbf{\Delta^{\pm}})_{R^{+}R^{-}}\equiv S_{R^{-}R^{+}}^* \Delta_{R^{+}R^{-}}^{\pm}$. This matrix allows one to calculate the coefficients of the wavefunction as one moves from one region of constant magnetic field to another through a discontinuity. Thus, if one breaks up any magnetic field profile into a series of constant regions separated by discontinuities, one can systematically approach a perfect description of the propagation of a molecule with internal degrees of freedom through a magnetic field of arbitrary profile through repeated application of $\mathbf{K}$ and $\mathbf{\Pi}_{x}$. Furthermore, this approach is not restricted to molecules moving through magnetic fields. Many other types of quantum objects moving in a single dimension with internal degrees of freedom that couple to an external static potential can also be analyzed in this way.

The above analysis indicates that one needs to keep track of $2N_R$ components to build up the eigenstates of the system exactly. However, for the current application in mind, one only needs $N_R$ components as the magnetic fields typically change the linear molecular momentum by such a small amount that the amplitudes $B_R$ of the reflected part of the wavefunction are negligible. That is, any backscattering of the  molecules by the magnetic fields is negligible and can be ignored.

For example, a typical velocity of the \oH~molecules in the experiment of Ref. \cite{Godsi2017} is $v_{H_2}=\SI{1450}{m/s}$. This corresponds to the kinetic energy $E_{H_2} = \frac{1}{2}m_{H_2} v_{H_2}^2 = 5.31\times10^9 \text{ kHz}$. The data reported by Ramsey \cite{Ramsey1952} indicates that the maximum energy change for the hyperfine states of \oH~at 500G is approximately -2550 kHz. The experiment of Ref. \cite{Godsi2017} has magnetic fields up to about 1000G. For such fields, the energy changes are approximately linear, so we expect the maximum change in energy  to be $\Delta E \approx -5100$ kHz.  In the field-free region before the discontinuity, $k_{R^{-}} \approx m_{H_2} v_{H_2} / \hbar$ and after the discontinuity in the field, $k_{R^{+}} \approx \sqrt{2m_{H_2}(E - \Delta E)}/\hbar$, as per Eqn.~(\ref{Eqn_kDefInField}). Then, $|\Delta_{R^{+}R^{-}}^{-}| \approx 2.4 \times 10^{-7}$ and $|\Delta_{R^{+}R^{-}}^{+}| \approx 1$, making $\mathbf{K}$ approximately diagonal and illustrating the decoupling of the forward and backward channels under typical experimental conditions.

We thus only need to keep track of the $A_{R}$ components, which correspond to the forward-propagating momenta. We can define a new coefficient vector
\begin{align}
\vec\psi_x \equiv \left(A_1,A_2,...,A_{N_R}\right)^T \label{Eqn_coeffVectorDef}.
\end{align}
The corresponding $N_R\times N_R$ propagation $\mathbf{P}_{x}$ and discontinuity $\mathbf{D}$ matrices are
\begin{align}
\mathbf{P}_{x} &\equiv \bigoplus_R e^{ik_Rx} \nonumber \\
&= \begin{pmatrix}
e^{ik_1x} & &\bigzero  \\
& \ddots &  \\
\bigzero &  & e^{ik_{N_R}x}  \\
\end{pmatrix} \label{Eqn_propMatDef} \\
\nonumber \\
\mathbf{D} &\equiv {\mathbf{S}^{R^{+}}_{R^{-}}}^\dagger\circ\mathbf{\Delta^{+}} \nonumber \\
&\approx {\mathbf{S}^{R^{+}}_{R^{-}}}^\dagger, \label{Eqn_discontMatDef}
\end{align}
where the matrix elements of ${\mathbf{S}^{R^{+}}_{R^{-}}}^\dagger$ are $ S_{R^{-}R^{+}}^* \equiv \braket{R^{+}|R^{-}}$, 
$\RamHam\boldsymbol{\left(\right.}\vec B(0^{\pm})\boldsymbol{\left.\right)}\ket{R^{\pm}}=E_{R^{\pm}}\ket{R^{\pm}}$, $0^{\pm}$ indicates the position just to the left $({-})$ or right $({+})$ of the discontinuity, and $k_R$ is defined as in Eqn.~(\ref{Eqn_kDefInField}). Specifically, $\mathbf{D}$ changes the basis of the vector $\vec\psi_x$ from $\ket{R^{-}}$ to $\ket{R^{+}}$. 
That is, $\vec\psi_x$ is always in the eigenbasis of $\RamHam\boldsymbol{\left(\right.}\vec B(x)\boldsymbol{\left.\right)}$. Finally, given that $B_R\approx0$, the eigenstate coefficients are now
\begin{align}
\Phi_{R}^{E\tilde R}(x) = A_{R}e^{ik_Rx}. \label{Eqn_coeffKspaceExpansionReduced}
\end{align}

Given that a generic transfer matrix $\mathbf{M}$ has the property $\mathbf{M}\mathbf{\sigma}_z\mathbf{M}^\dagger=\mathbf{\sigma}_z$\cite{Walker1992}, the decoupling of the forward and backward channels implies that the forward channel matrix $\mathbf{M}_F$ (composed of a product of $\mathbf{P}_{x}$ and $\mathbf{D}$ matrices) is now unitary.

\subsection{Rotation Matrices}
\label{Sec_RotMat}
Scattering by the sample surface changes both the propagation direction and the internal states of the molecule.
To take into account the change in the direction of the propagation path when applying the transfer matrix formalism, we need only change the orientation of the quantization axis. However, to address the impact  of scattering on the internal states, we need to apply a scattering matrix that is written with respect to a particular reference frame (which is often a sample-fixed frame, see Section \ref{Sec_ScatMat}). Thus, instead of just rotating the quantization axis from the first branch to the second branch (to account for the change in the direction of propagation), we need to first rotate from the initial reference frame ($xyz$ in Figure \ref{Fig_experimentSchematic}) to the reference frame of the scattering matrix. Then, after applying the scattering matrix, we need to rotate from the scattering matrix reference frame to the final reference frame ($x'y'z'$ in Figure \ref{Fig_experimentSchematic}). To perform these rotations coherently, we apply  $N_\mathrm{R}\times N_\mathrm{R}$ rotation matrices $\mathbf{R}(\phi,\Theta,\chi)$ to $\vec\psi_x$, where $\phi$, $\Theta$, and $\chi$ are the Euler angles in the $ZYZ$ convention (with $Y$ and $Z$ being the axes of a space-fixed frame; see Ref. \cite{Zare1988}). In this way, we can account for both specular and non-specular scattering geometries and for various orientations of the sample surface.

To change the orientation of the quantization axis, we perform passive rotations on the state vector $\vec\psi_x$. These passive rotations modify the basis of $\vec\psi_x$, but leave the physical state unchanged. For example, if we were to assume that the only impact of scattering was to change the propagation direction, we would need to perform a passive rotation of the state vector about the $y$ axis by the angle $\theta$ to account for a change of angle $\theta$ in the propagation direction (for the definition of the axes shown in Figure \ref{Fig_experimentSchematic}). We would perform this rotation by applying the equivalent active rotation of angle $-\theta$ to $\vec\psi_x$; that is, by using the matrix $\mathbf{R}(0,-\theta,0)$.

For the general case, we work with the rotation matrices $\mathbf{R}(\phi,\Theta,\chi)$, whose matrix elements, when written in the $\ket{R}$ eigenbasis of $\RamHam(\vec{B}_\mathrm{loc})$ where $\vec{B}_\mathrm{loc}$ is the local magnetic field, are:

\begin{align}
\mathbf{R}_R(\phi,\Theta,\chi) &\equiv \left[\bra{R'}\hat{R}(\phi,\Theta,\chi)\ket{R}\right] \nonumber \\
&= \left[\sum_{FMF'M'}\braket{R'|F'M'}\braket{F'M'|\hat{R}(\phi,\Theta,\chi)|FM}\braket{FM|R} \right]\nonumber \\ 
&= {\mathbf{S}^{R}_{FM}}^\dagger\mathbf{R}_{FM}(\phi,\Theta,\chi){\mathbf{S}^{R}_{FM}}, 
\end{align}
where $\ket{FM}\equiv\ket{IJFM}$ is an angular momentum state with total angular momentum $F$, $z$ axis projection $M$, total nuclear spin  angular momentum $I$ and total rotational angular momentum $J$; the subscripts of $\mathbf{R}_R$ and $\mathbf{R}_{FM}$ denote the basis of the matrix representation, $\ket{R}$ and $\ket{FM}$, respectively; $\mathbf{S}^{R}_{FM}$ is the matrix whose columns are the eigenstates $\ket{R}$ written in the $\ket{FM}$ basis,
$\hat{R}(\phi,\Theta,\chi)$ is the rotation operator (with the same $ZYZ$ convention mentioned above) and 
\begin{align}
\mathbf{R}_{FM}(\phi,\Theta,\chi) &= \left[\delta_{FF'} D^F_{M'M}(\phi,\Theta,\chi)\right]  \nonumber \\ 
&= \left[\delta_{FF'}e^{-i\phi M'}d^F_{M'M}(\Theta)e^{-i\chi M}\right]   , \label{Eqn_RotMatFM}
\end{align}
where $ D^F_{M'M}(\phi,\Theta,\chi)$ are the Wigner D-matrices and $d^F_{M'M}(\Theta)$ are the Wigner small d-matrices~\cite{Zare1988}. Note that $\mathbf{R}_{FM}(\phi,\Theta,\chi)$ is diagonal in $F$, because of conservation of angular momentum, but not diagonal in $M$ \cite{Zare1988}. 
Thus, one must be careful to also perform a passive rotation on the local magnetic field vector if rotations are performed in a region with non-zero field. Typically, however, the sample chamber is magnetically shielded. 

\added[id=JTC_2]{We also note that the rotation may impact how to appropriately match the boundary conditions between the eigenstate immediately after rotation and the eigenstate at the start of the second branch. As the propagation matrices $\mathbf{P}_{x}$ (\ref{Eqn_propMatDef}) are defined with respect to the momentum, which may be positive or negative, it is important to choose the sign of the momentum that results in the probability current flowing in the same direction as the molecular propagation. For example, using the axis definitions in Figure \ref{Fig_experimentSchematic}, $+k_R$ is chosen for the first branch and $-k_R$ for the second branch.}

\subsection{Scattering Transfer Matrices}
\label{Sec_ScatMat}
Scattering by the sample surface can involve many complex phenomena that may change the internal state, the momentum, and the total energy of the scattering molecule. For the present manuscript, we focus on scattering processes that conserve the total energy of the molecules.  Energy-conserving scattering may, however, include transfer of energy between the internal and translational degrees of freedom. Such scattering processess are described by a general, non-diagonal scattering matrix in the basis of the molecular states.


The interactions of the molecules with the sample surface can be phenomenologically described with the total scattering \emph{transfer} matrix. This matrix is the $2N_\mathrm{R}\times 2N_\mathrm{R}$ matrix $ {\mathbf{\tilde\Sigma}}$ that relates the wavefunctions on the ``left'' side of the scattering event to those on the ``right'' (as opposed to the scattering matrix, which relates the incoming wavefunctions to the outgoing). However, because the initial wavepacket (\ref{Eqn_PsiInitialProfile}) does not contain any negative momentum states, the magnetic fields of the solenoids do not cause significant backscattering (Section \ref{Sec_PropAndDiscontMat}), and the detector only detects molecular flux in the forward scattering direction, we need only work with the $N_\mathrm{R}\times N_\mathrm{R}$ matrix $\mathbf{\Sigma}\equiv\mathbf{P}_{\mathrm{fwd}}\mathbf{\tilde\Sigma}{\mathbf{P}^\dagger_{\mathrm{fwd}}}$, where $\mathbf{P}_{\mathrm{fwd}}$ is an $N_\mathrm{R}\times 2N_\mathrm{R}$ projection matrix onto the forward scattering states. 
We define $\mathbf{\Sigma}$ in the $\ket{Im_IJm_J}$ basis, where the $\ket{Im_IJm_J}$ states are themselves defined with respect to the quantization axis that is normal to the surface sample. We choose this basis to relate to scattering calculations, which are frequently carried out in the $\ket{Jm_J}$ basis with a quantization axis normal to the sample surface. In principle, however, any suitable set of Ramsey states $\ket{R_\Sigma}$ could be chosen as the basis for the scattering transfer matrix and any suitable quantization axis could be chosen, to take advantage  of relevant symmetries.


In general, the scattering transfer matrix elements $\Sigma_{Im_IJm_JI'm'_IJ'm'_J}$  
are functions of the incident energy $E$, the outgoing energy $E'$, the incident momentum $\vec{k}$, and the outgoing momentum $\vec{k}'$. As we are restricting ourselves to iso-energetic processes, $E=E'$. Also, Eqn.~(\ref{Eqn_kDefInField}) defines the magnitudes of the momentum before and after the scattering event. This leaves the scattering transfer matrix elements as functions of only energy and the four angles that define the scattering geometry. These angles are: (1) the angle between the two branches, (2) the angle between the surface normal and the scattering plane,  
(3)  the angle between the first branch and the projection of the surface normal on the scattering plane, and (4) the azimuthal angle of the sample. The scattering plane is the plane defined by the two branches of the apparatus.

Given that the experiment only probes a single scattering direction at a time (see Section \ref{Sec_exptDescription} and Figure \ref{Fig_experimentSchematic}), the scattering transfer matrix will not, in general, be unitary. This incorporates state-dependent loss channels into the formalism.
Additionally, the scattering transfer matrix is, in general, time-dependent. Here, we assume that the time-scales of the surface dynamics are significantly different from the molecule-surface or wavepacket-surface interaction time-scales and assume $\mathbf{\Sigma}$ to be time-independent.


Because $\mathbf{\Sigma}$ is defined with respect to the surface normal, we use rotation matrices to appropriately change the basis of $\vec\psi$ before and after applying the scattering transfer matrix. We ensure that the total rotation corresponds to the change in propagation direction induced by scattering off of the sample surface and that the quantization axis is again coplanar with the two branches of the apparatus.

The scattering transfer matrix elements for a specific molecule-surface interaction can be determined from scattering calculations \cite{Godsi2017,Diaz2009}. Alternatively, they can be treated as free parameters and determined from the experimental measurements by solving the inverse scattering problem. Such a problem can potentially be solved efficiently using machine learning based on Bayesian optimization \cite{Vargas2017,BML}.



\subsection{Calculation of Eigenstate Coefficients}
To determine the dependence of the probability of detection (\ref{Eqn_P_DetectFinal}) on the magnetic fields and the surface properties, we must determine the coefficients $\beta_{R_D}^{ER_0}$. This can be done by multiplying the initial coefficient vector $\vec{\psi}^{ER_0}_{x_0}$ (\ref{Eqn_coeffVectorDef}) of a system eigenstate $\ket{ER_0}$ by a succession of transfer matrices to obtain the final coefficient vector $\vec{\psi}^{ER_0}_{x_D}\equiv\left(\beta_{1}^{ER_0},\beta_{2}^{ER_0},\cdots,\beta_{N_\mathrm{R}}^{ER_0}\right)^T$:
\begin{align}
\vec{\psi}^{ER_0}_{x_D}&= {\mathbf{S}^{R_D}_{R_N}}^\dagger\mathbf{M}_2\mathbf{M}_\Sigma\mathbf{M}_1\vec{\psi}^{ER_0}_{x_0}, \label{Eqn_finalCoeffVector}
\end{align}
where \replaced[id=JTC_2]{$\mathbf{M}_{1}$ and $\mathbf{M}_{2}$ describe the propagation through the  first and second branches of the apparatus, respectively}{$\mathbf{M}_{1/2}$ describes the propagation through the  first/second branch of the apparatus}, $\mathbf{M}_{\Sigma}$ describes the scattering, and ${\mathbf{S}^{R_D}_{R_N}}^\dagger$ changes the basis of the coefficient vector  to the eigenbasis $\ket{R_D}$ of $\RamHam\boldsymbol{\left(\right.}\vec{B}(x_D^{+})\boldsymbol{\left.\right)}$ at the location of the detector $x_D$. The $\mathbf{M}$ matrices are defined as
\begin{align}
\mathbf{M}_1 &= \mathbf{P}_{L_n} {\mathbf{S}^{R_n}_{R_{n-1}}}^\dagger\cdots\mathbf{P}_{L_2}{\mathbf{S}^{R_2}_{R_1}}^\dagger \mathbf{P}_{L_1} {\mathbf{S}^{R_1}_{R_{ini}}}^\dagger \label{Eqn_mMatrix1} \\
\mathbf{M}_\Sigma &= {\mathbf{S}^{R_n}_{FM}}^\dagger\mathbf{R}_{FM}(\alpha',\beta',\gamma')\mathbf{\Sigma}_{FM}\mathbf{R}_{FM}(\alpha,\beta,\gamma){\mathbf{S}^{FM}_{R_{n}}}^\dagger \label{Eqn_mMatrixSigma} \\
\mathbf{M}_2 &= \mathbf{P}_{L_N} {\mathbf{S}^{R_N}_{R_{N-1}}}^\dagger\cdots\mathbf{P}_{L_{n+1}} {\mathbf{S}^{R_{n+1}}_{R_n}}^\dagger \label{Eqn_mMatrix2}
\end{align}
where $R_{ini}$ refers to the eigenbasis $\ket{R_{ini}}$ of $\RamHam\boldsymbol{\left(\right.}\vec{B}(0^{-})\boldsymbol{\left.\right)}$ at the initial location of the wavepacket; 
$R_i$ refers to the eigenbasis $\ket{R_i}$ of $\RamHam(\vec{B}_i)$ in region $i$ of the apparatus, as depicted in Figure \ref{Fig_ApparatusAbstraction}; 
$FM$ refers to the $\ket{IJFM}$ basis where $\vec{F}\equiv\vec{I}+\vec{J}$ and $M$ is the projection on the local $z$ axis; 
$\alpha$, $\beta$, and $\gamma$ are the Euler angles that rotate the reference frame of the first branch ($xyz$ in Figure \ref{Fig_experimentSchematic}) onto the reference frame of the scattering transfer matrix, whose quantization axis is normal to the sample surface (see Section \ref{Sec_RotMat} and Section \ref{Sec_ScatMat}); 
$\alpha'$, $\beta'$, and $\gamma'$ are the Euler angles that rotate the scattering transfer matrix reference frame onto the reference frame of the second branch ($x'y'z'$ in Figure \ref{Fig_experimentSchematic}); 
$L_i$ is the \added[id=JTC_2]{signed} length of region $i$, as depicted in Figure \ref{Fig_ApparatusAbstraction}; \added[id=JTC_2]{the sign of $L_i$ indicates the direction of propagation with respect to the local $x$ or $x'$ axis;}  
$N$ is the total number of regions between $x_0$ and $x_D$ (see Figure \ref{Fig_ApparatusAbstraction});
$n$ is the number of regions between the initial position of the wavepacket $x_0=0$ and the sample position $x_S$;
$\mathbf{\Sigma}_{FM}\equiv{\mathbf{S}^{FM}_{R_{IJ}}}^\dagger\mathbf{\Sigma}{\mathbf{S}^{R_{IJ}}_{FM}}^\dagger$ is the scattering transfer matrix  written in the $\ket{IJFM}$ basis; $R_{IJ} \equiv Im_IJm_J$; and $\mathbf{\Sigma}$ is the scattering transfer matrix in the  $\ket{Im_IJm_J}$ basis. 
Note the product of the two rotation matrices $\mathbf{R}_{FM}(\alpha',\beta',\gamma')\cdot\mathbf{R}_{FM}(\alpha,\beta,\gamma) = \mathbf{R}_{FM}(\phi,\Theta,\chi)$, where $\phi$, $\Theta$, and $\chi$ are the Euler angles that rotate the reference frame $xyz$ onto the frame $x'y'z'$ (see Figure \ref{Fig_experimentSchematic}). All of the Euler angles mentioned above are in the $ZYZ$ convention, with $Y$ and $Z$ being the axes of a space-fixed frame and as per the convention defined in Ref. \cite{Zare1988}. 
Note also that while the scattering transfer matrix $\mathbf{\Sigma}$ is  written here in the $\ket{Im_IJm_J}$ basis, other suitable bases $\ket{R_\Sigma}$ may be used (see Section \ref{Sec_ScatMat}), where $R_\Sigma$ refers to an arbitrary set of Ramsey states. In such a case, $\mathbf{\Sigma}_{FM}\equiv{\mathbf{S}^{FM}_{R_\Sigma}}^\dagger\mathbf{\Sigma}{\mathbf{S}^{R_\Sigma}_{FM}}^\dagger$.  \added[id=JTC_2]{Also, note that the propagation matrices $\mathbf{P}_{L_i}$ (\ref{Eqn_propMatDef}) are defined with momentum $+k_R$ if the molecular propagation is in the direction of the local $x$ or $x'$ axis or, conversely, with the momentum $-k_R$ if the molecular propagation is in the opposite direction of the local $x$ or $x'$ axis (see Section \ref{Sec_RotMat}).}

By defining a matrix \replaced[id=JTC_2]{$\mathbf{\Psi}^{E}_{x_i}\equiv\left(\vec{\psi}^{E1}_{x_i},\vec{\psi}^{E2}_{x_i},\cdots,\vec{\psi}^{EN_\mathrm{R}}_{x_i}\right)$}{$\mathbf{\Psi}^{E}_{x_i}\equiv\left(\vec{\psi}^{E1}_{x_0},\vec{\psi}^{E2}_{x_0},\cdots,\vec{\psi}^{EN_\mathrm{R}}_{x_0}\right)$}, all $N_\mathrm{R}\times N_\mathrm{R}$ coefficients $\beta_{R_D}^{ER_0}$ can be simultaneously obtained from
\begin{align}
\mathbf{\Psi}^{E}_{x_D}&= {\mathbf{S}^{R_D}_{R_N}}^\dagger\mathbf{M}_2\mathbf{M}_\Sigma\mathbf{M}_1\mathbf{\Psi}^{E}_{x_0} \nonumber \\
&= {\mathbf{S}^{R_D}_{R_N}}^\dagger\mathbf{M}_2\mathbf{M}_\Sigma\mathbf{M}_1\mathbb{1}_{N_\mathrm{R}},
\label{Eqn_finalCoeffMatrix} 
\end{align}
where $\mathbf{\Psi}^{E}_{x_0}\equiv\mathbb{1}_{N_\mathrm{R}}$ because of the specific definition of the system eigenstates (see Section \ref{Sec_molecularHami}).
Using Eqns.~(\ref{Eqn_mMatrix1}--\ref{Eqn_finalCoeffMatrix}), we can obtain $\beta_{R_D}^{ER_0}$, and thus $P_{\text{detection}}$ (\ref{Eqn_P_DetectFinal}), as functions of the magnetic field profile, the scattering matrix elements, and the scattering geometry.

\section{Application to \emph{ortho}-Hydrogen}
\label{Sec_orthoHydrogen}

\begin{figure}[H]
	\centering
	\includegraphics[width=\textwidth]{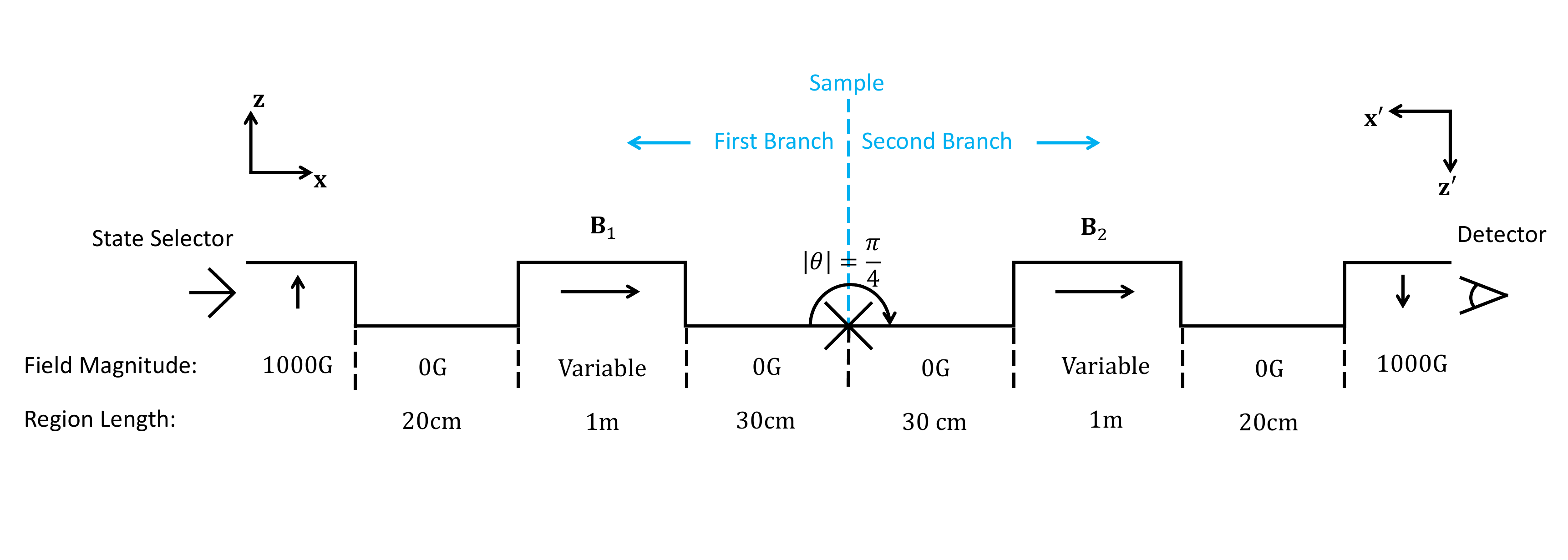}
	\caption{A magnetic field profile that approximates the true magnetic field profile of an experiment using \oH. We combine this approximate field profile with the transfer matrix formalism to calculate the observed signal.
	$\textbf{B}_i$ refers to the different magnetic field vectors of the control fields. $z'$ and $x'$ refer to the new coordinate system defined to align with the second branch of the apparatus (see Figure \ref{Fig_experimentSchematic}). The sample is located at the cross in the centre of the diagram. The surface normal of the sample is assumed to bisect the angle between the two branches of the apparatus. The propagation direction is $x$ before scattering and $-x'$ after scattering. The angle between $x$ and $-x'$ (i.e. the angle between the two arms of the apparatus) is $\theta = \SI{45}{\degree}$. $\textbf{B}_1$ is directed along $x$ and $\textbf{B}_2$ is directed along $-x'$, as per the arrows. The fields just after the state selector and just before the detector are directed toward the $z$ and $z'$ directions, respectively and as per the arrows. Additional computational parameters not shown above can be found in Appendix \ref{App_compParam}. 
}
	\label{Fig_oH2experimentAbstraction}
\end{figure}

The theoretical framework described in Sections \ref{Sec_exptDescription} through \ref{Sec_transferMatrixFormalism} connects the scattering transfer matrix elements $\Sigma_{Im_IJm_JI'm'_IJ'm'_J}$ to the experimentally observed signal, which is proportional to $P_{\text{detection}}$ (\ref{Eqn_P_DetectFinal}). By changing the magnetic field profiles in the two arms of the apparatus, one can obtain information about how the scattering affects various hyperfine states. 
To illustrate our theoretical framework and to demonstrate the impact of the scattering transfer  matrix on the experimentally observed signal, we consider a beam of rotationally cold \oH~and a simplified apparatus that contains only a few regions of constant magnetic field, as depicted in Figure \ref{Fig_oH2experimentAbstraction}. 

\subsection{Rotationally Cold \textit{Ortho}-Hydrogen Hyperfine Hamiltonian}
The Hamiltonian describing the relevant internal degrees of freedom of rotationally cold \oH~is \cite{Ramsey1952}:
\begin{align}
\frac{\RamHam_\oHmath(\vec{B})}{h} &=  - \alpha \hat I \cdot \vec B - \beta \hat J \cdot \vec B -c \hat I \cdot \hat J + \frac{5d}{(2J-1)(2J+3)} [ 3(\hat I \cdot \hat J)^2 +\frac{3}{2} \hat I \cdot \hat J - \hat I^2 \hat J^2 ] 
\end{align}
where, for simplicity, we have neglected magnetic shielding of the nuclear and rotational magnetic moments by the molecule and diamagnetic interactions of the molecule with the magnetic field; 
$\vec B$ is the local magnetic field;
$\hat{I}$ is the nuclear spin operator; 
$\hat{J}$ is the rotational angular momentum operator; 
$\alpha\equiv\frac{\mu_{I}}{hI}\approx\SI{4.258}{kHz}$; $\beta\equiv\frac{\mu_J}{hJ}\approx\SI{0.6717}{kHz}$; $c\approx\SI{113.8}{kHz}$; $d\approx\SI{57.68}{kHz}$; 
$I=1$ is the total nuclear spin angular momentum in units of $\hbar$; $J=1$ is the total rotational angular momentum in units of $\hbar$; 
$\mu_{I}$ is the nuclear magnetic moment of a \emph{single} nucleus;
and $\mu_{J}$ is the magnetic moment due to molecular rotation. The first two terms describe the interaction of the nuclear and rotational magnetic moments with the external magnetic field, the third term describes the nuclear spin-rotational magnetic interaction \cite{Ramsey1952,Kellogg1939,Kellogg1940}, and the terms proportional to $d$ describe the magnetic spin-spin interaction of the two nuclei \cite{Ramsey1952,Kellogg1939,Kellogg1940}.

\begin{figure}[H]
	\centering
	\begin{minipage}{.32\linewidth}
		\centering
		\hspace{2.75em}\textbf{Theory}
		
		\hspace{2.75em}\textbf{Single Velocity}
%


		\xincludegraphicsB[width=\textwidth,label={\bf (a)},pos=ne,fontsize=\normalsize]{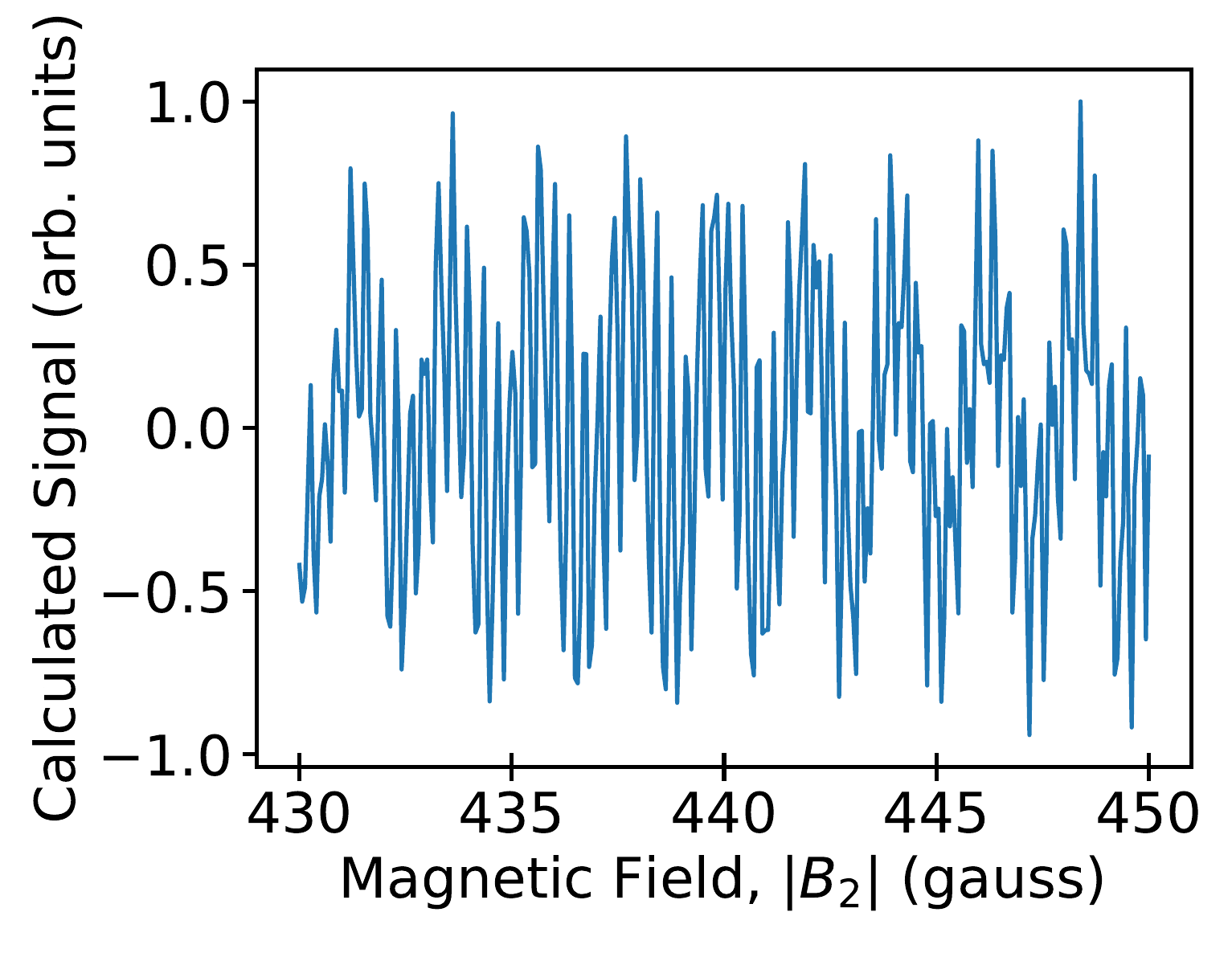}

		\xincludegraphicsB[width=\textwidth,label={\bf (d)},pos=ne,fontsize=\normalsize]{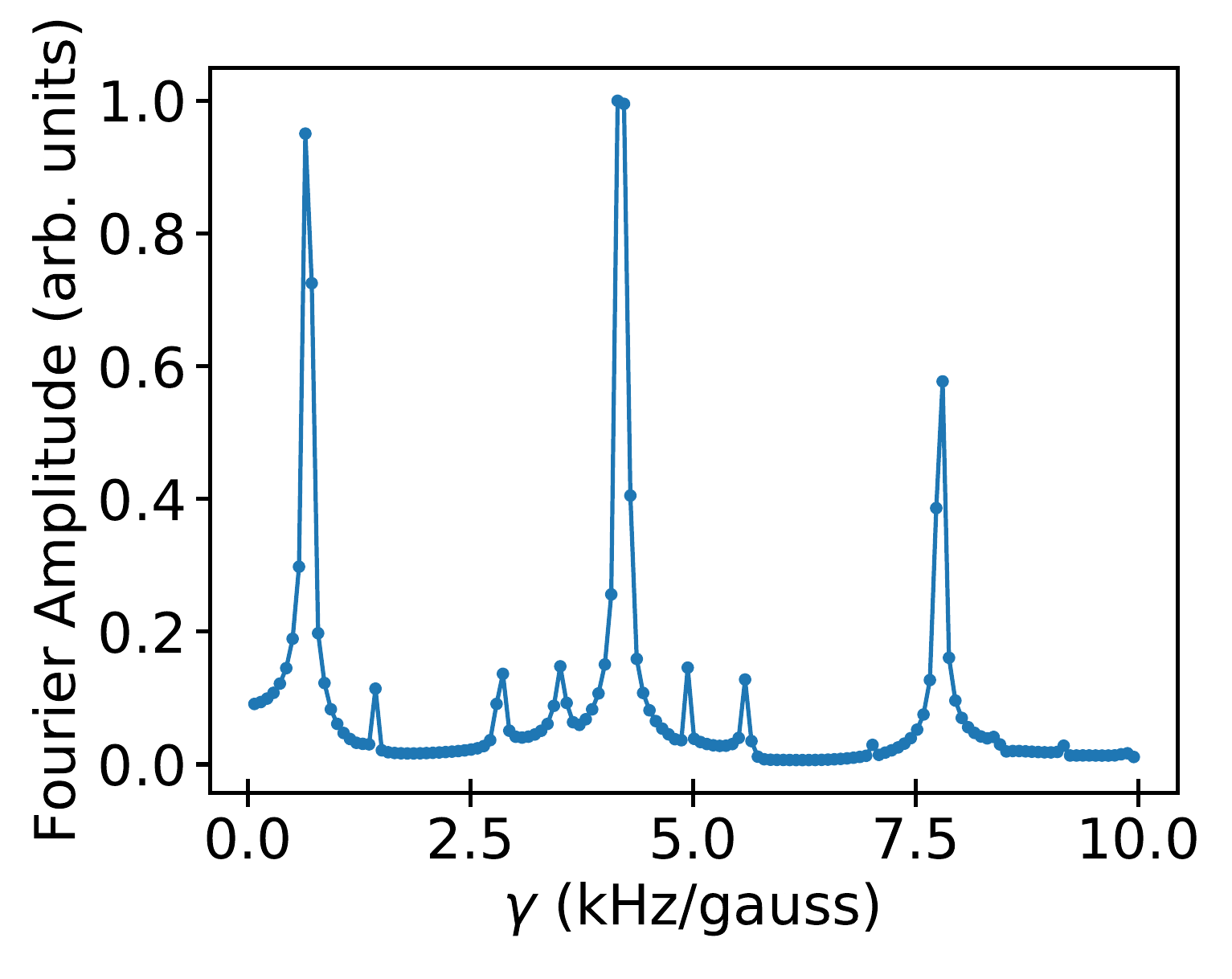}
	\end{minipage}
	\begin{minipage}{.32\linewidth}
		\centering
		\hspace{2.75em}\textbf{Theory} 
		
		\hspace{2.75em}\textbf{Integrated Over Velocity}

		\xincludegraphicsB[width=\textwidth,label={\bf (b)},pos=ne,fontsize=\normalsize]{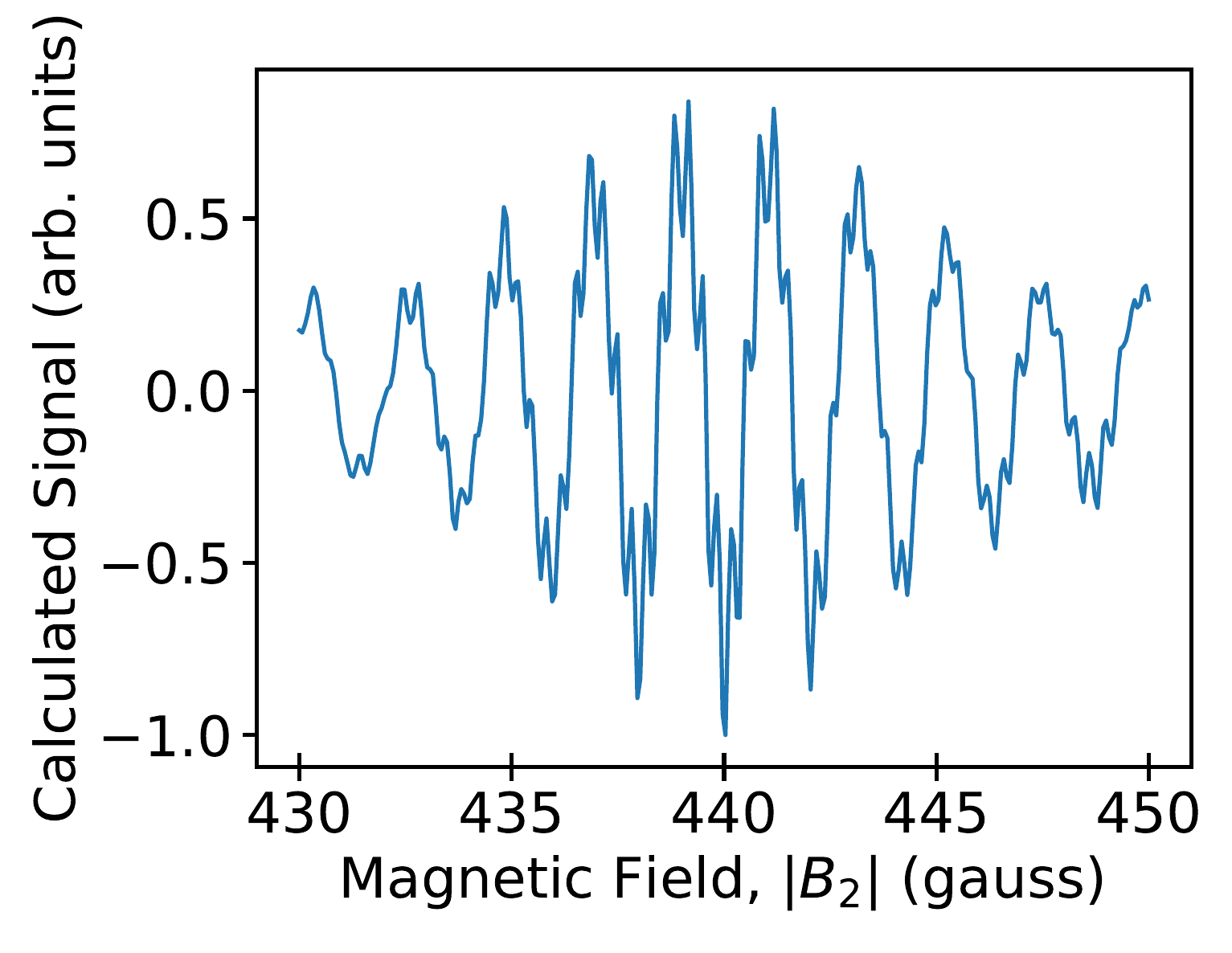}

		\xincludegraphicsB[width=\textwidth,label={\bf (e)},pos=ne,fontsize=\normalsize]{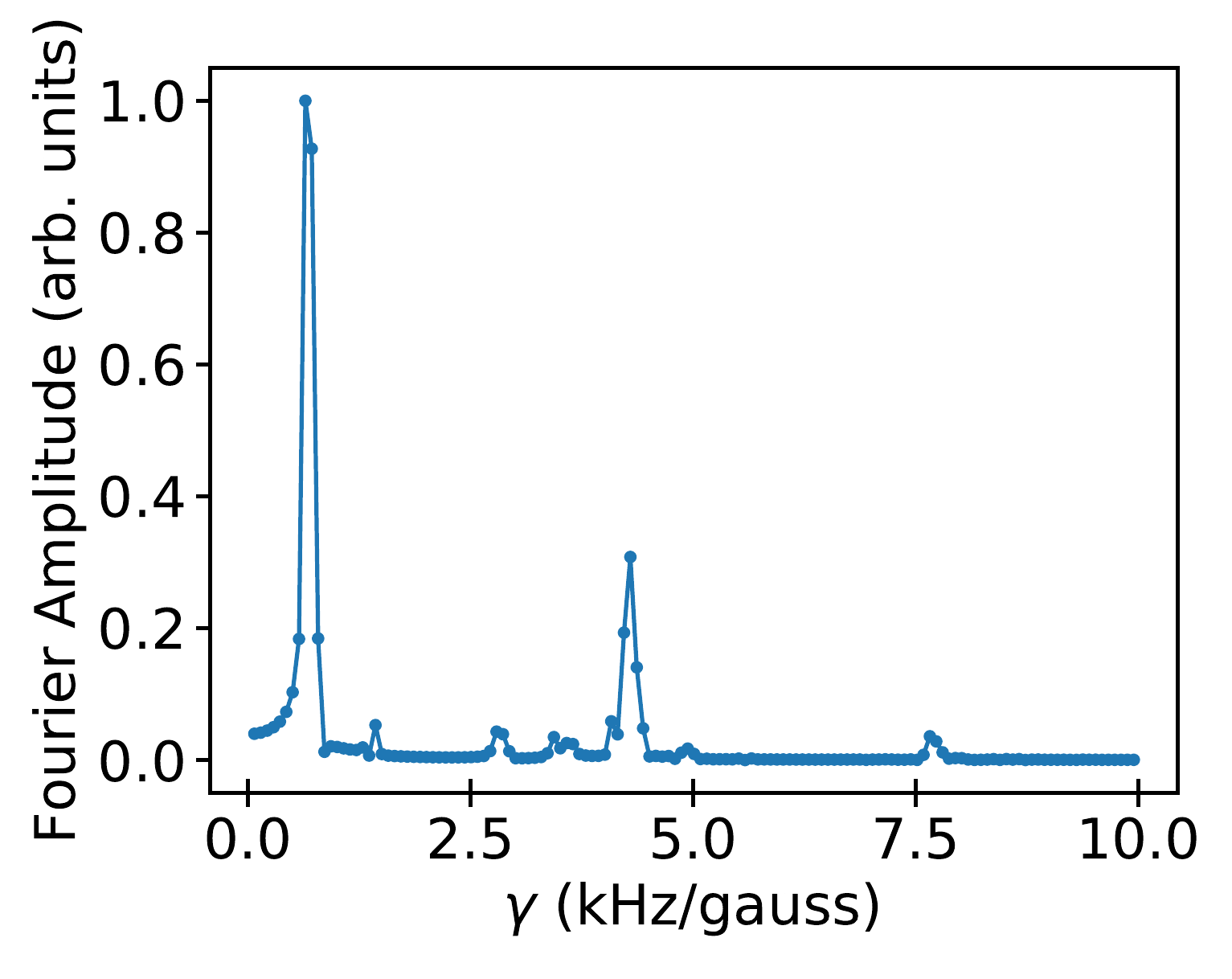}
	\end{minipage}
	\begin{minipage}{.32\linewidth}
		\centering
		\vspace{\baselineskip}
		\hspace{2.5em}\textbf{Experiment}

		\xincludegraphicsB[width=\textwidth,label={\bf (c)},pos=ne,fontsize=\normalsize]{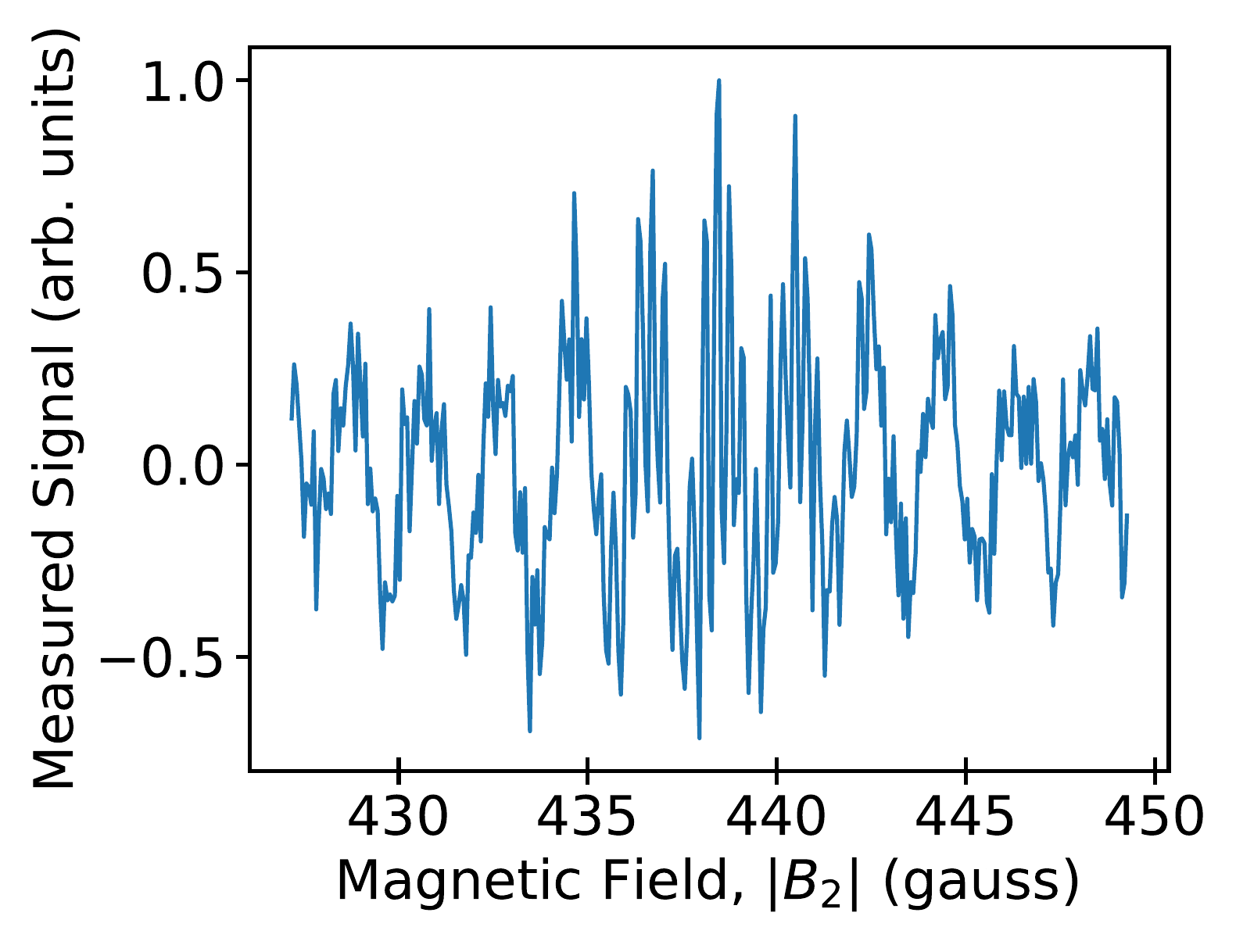}
		
		
		
		\xincludegraphicsB[width=\textwidth,label={\bf (f)},pos=ne,fontsize=\normalsize]{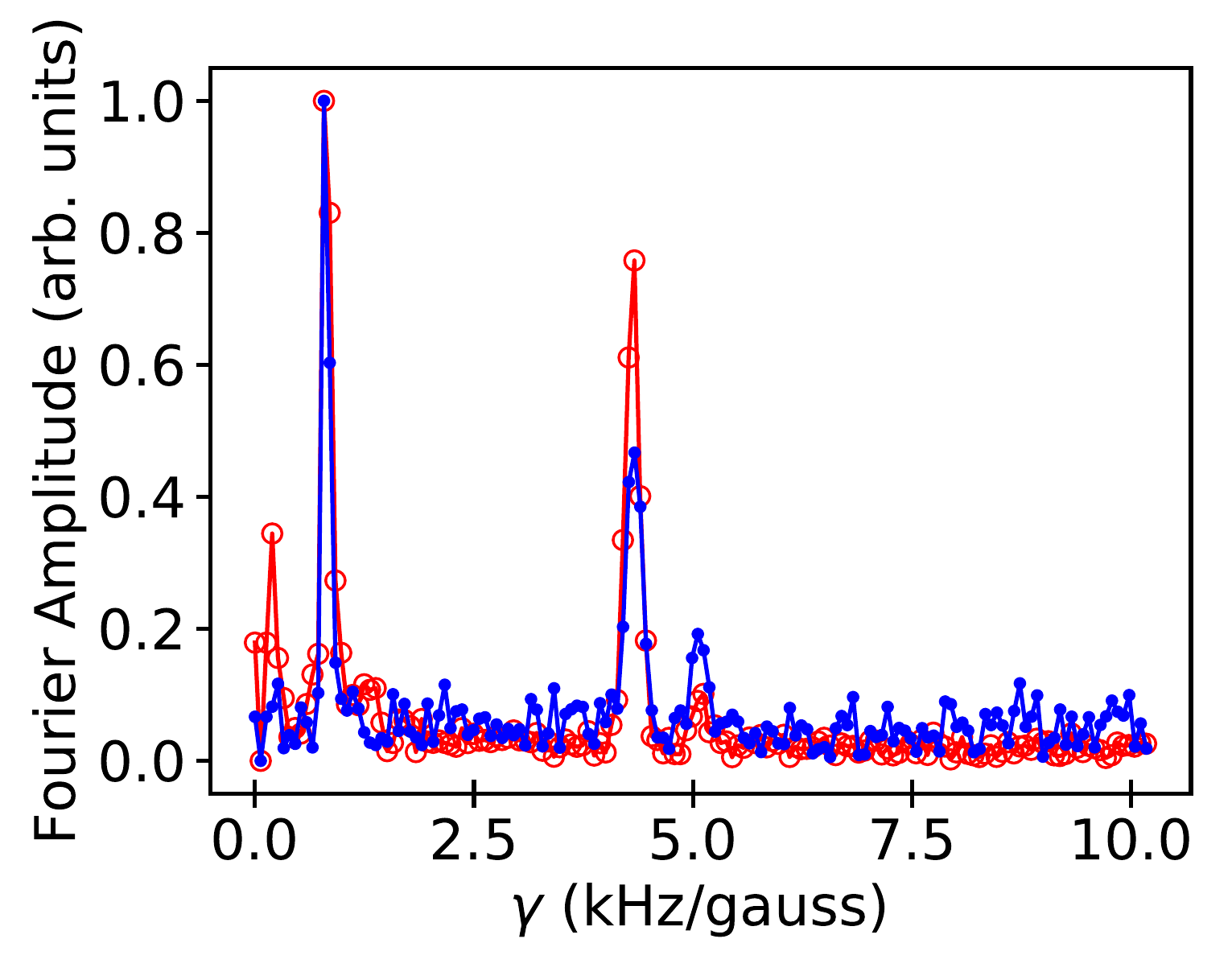}

	\end{minipage}
	\caption{Upper Panels: Calculated and experimental signals close to the spin echo condition versus the magnetic field of the second coil $|B_2|$. Lower Panels: Fourier amplitudes of the upper panels versus the generalized gyromagnetic ratio $\gamma$. For panels (a), (b), (d), and (e), the field profile is depicted in Figure \ref{Fig_oH2experimentAbstraction}; $B_1=\SI{440}{\gauss}$; the scattering transfer matrix $\mathbf{\Sigma} = \mathbb{1}_9$ and is constant for all energies; and the signal is sampled at a rate of 300 points per \SI{20}{\gauss}. Panels (a) and (d) only include a single velocity (or, equivalently, a single value of energy) in Eqn. (\ref{Eqn_P_DetectFinal}) while panels (b) and (e) include the full integral. For the experimental data shown in panel (c), $B_1=\SI{437}{\gauss}$, the sample was the (111) surface of Cu and the signal was sampled every \SI{0.065}{\gauss} (a sampling rate of ${\sim}308$ points per \SI{20}{\gauss}).  Panel (f) shows data for \oH~scattering off of Cu(111) (blue full circles) and Cu(115) (red open circles). All experimental data was obtained from Godsi et al.~\cite{Godsi2017}. }
	\label{Fig_wiggleCurves}
\end{figure}

\subsection{Experiment and Observables}

While there are many possible experimental protocols, we focus on the full interferometer mode used by Godsi et al.~\cite{Godsi2017}. The experiment is performed by initiating a continuous flux of \oH~molecules through the apparatus and measuring the current of the ionization detector while varying the first and second control fields ($B_1$ and $ B_2$ in Figure \ref{Fig_oH2experimentAbstraction}).

In particular, $B_1$ is set to a specific value while $B_2$ is varied around the point ${-}B_1$ (i.e. about the spin-echo condition). In principle, $B_2$ could also be set to vary around ${+}B_1$, where spin echoes have also been observed \cite{Litvin2019}, but we choose to vary $B_2$ about $-B_1$ to match the relevant experiment by Godsi et al.~\cite{Godsi2017}. This variation of the magnetic fields results in  oscillatory curves of the detector current versus $B_2$, as shown in Figure \ref{Fig_wiggleCurves} (a--c). These oscillations reflect the interference pattern that occurs when the various wavepackets recombine after passing through the final control field (see Section \ref{Sec_exptDescription}). This interference pattern contains information about how the individual hyperfine states of the molecule interact with the sample surface. 
%
%
%

The $x$ directed magnetic fields of a solenoid changes the energies of all of the $N_\mathrm{R} = 9$ hyperfine states and induces all ${N_\mathrm{R}\choose2} = 36$ possible transitions. The frequencies of these transitions depend on the magnitude of the magnetic fields. By changing the magnitude of the second magnetic field, we are able to probe the rates of change of these transition frequencies with the magnetic field: the (generalized) gyromagnetic ratios $\gamma_{ij}(B)=\left|\frac{\mathrm{d}f_{ij}(B)}{\mathrm{d}B}\right|$, where $f_{ij}\equiv\frac{1}{h}\mathrm{\Delta}E_{ij} = \frac{E_i-E_j}{h}$, and $E_i$ is the energy of Ramsey state $i$ \cite{Godsi2017}. The Fourier transforms of the oscillatory curves that give these gyromagnetic ratios  are shown in Figure \ref{Fig_wiggleCurves} (d--f). To obtain these results, we assumed that the surface normal of the sample lies in the scattering plane defined by the two branches and bisects the angle defined by the same two branches, such that $\alpha'=\alpha=\gamma'=\gamma=0$, $\beta=3\pi/8$ and $\beta'=-5\pi/8$, where $\beta+\beta'=-\pi/4=-\theta$ (see  Eqn. \ref{Eqn_mMatrixSigma} and Figure \ref{Fig_oH2experimentAbstraction}). \added[id=JTC_2]{Given this geometry and the axis definitions (Figure \ref{Fig_oH2experimentAbstraction}), the propagation matrices are defined with $+k_R$ in the first branch and $-k_R$ in the second.} We also assume that the scattering transfer matrix is the identity matrix and is independent of energy, i.e.~we assume for the present calculation that the only impact of scattering is the change of propagation direction, as modelled with rotation matrices (Section \ref{Sec_RotMat}).

\begin{figure}[H]
	\centering
	\begin{minipage}{.3\linewidth}
		\centering
		\textbf{Theory} 
		
		\textbf{Single Velocity}
		\includegraphics[width=\textwidth]{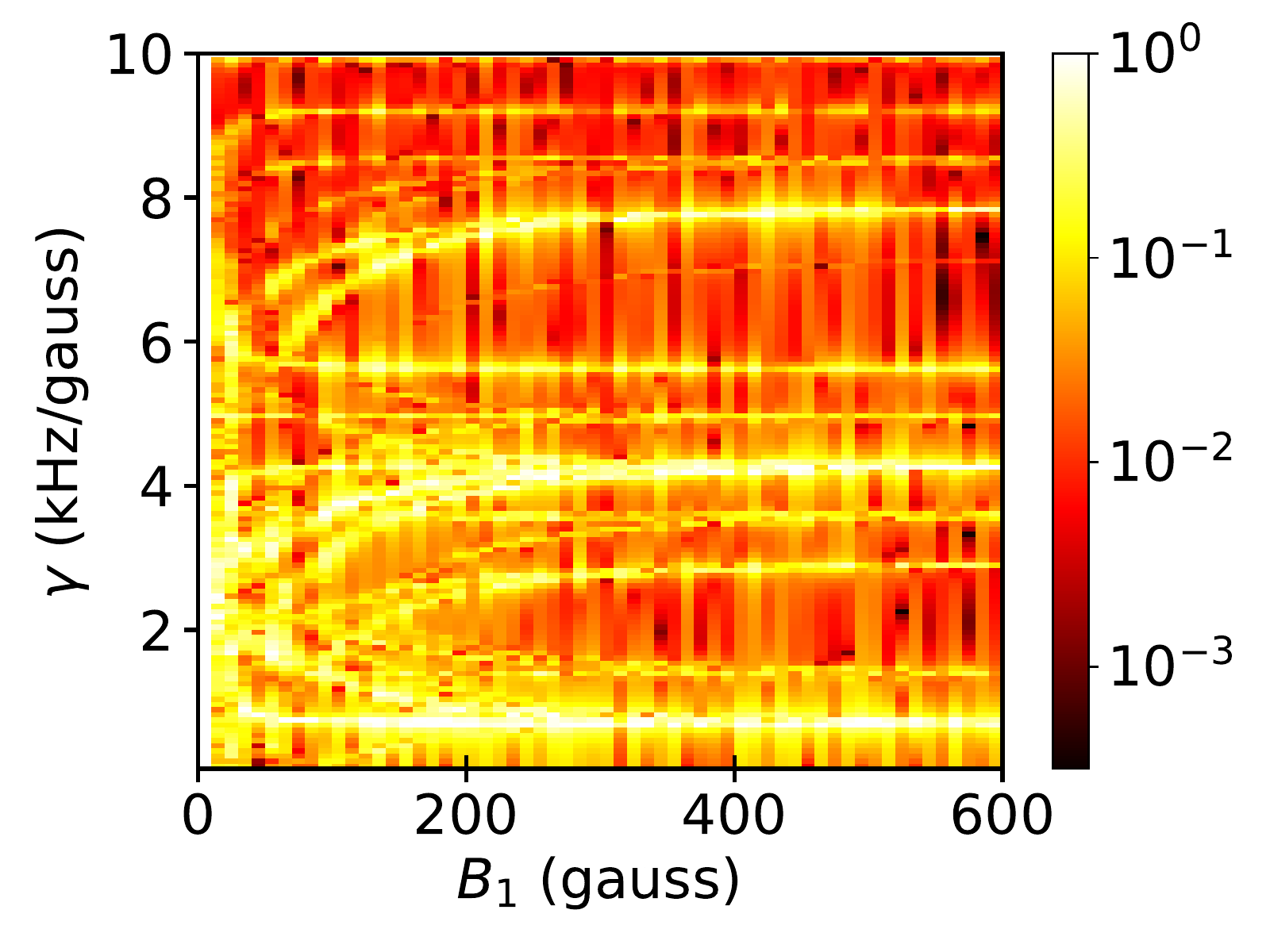}
	\end{minipage}
	\begin{minipage}{.3\linewidth}
		\centering
		\textbf{Theory} 
		
		\textbf{Many Velocities}
		\includegraphics[width=\textwidth]{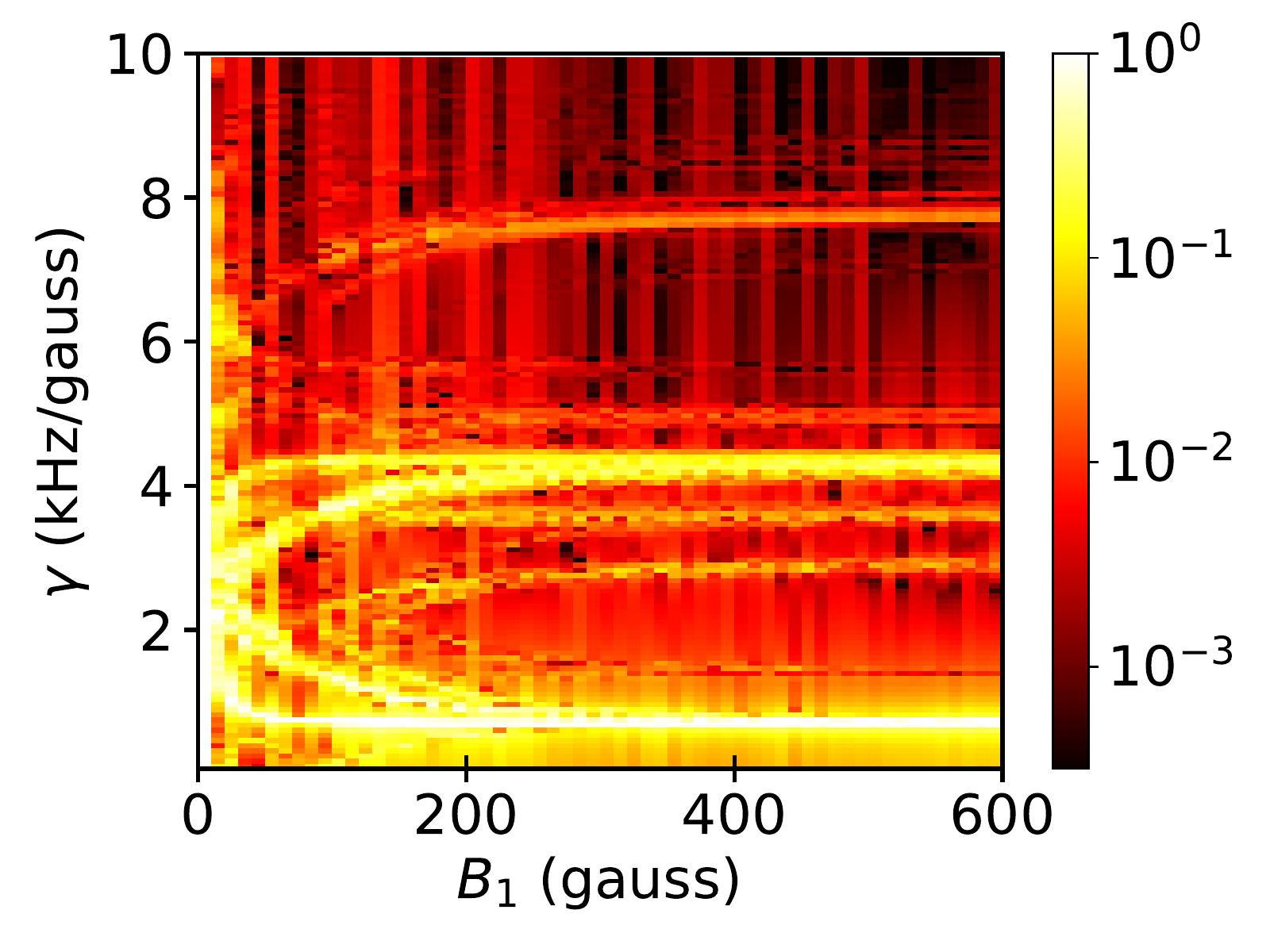}
	\end{minipage}
	\begin{minipage}{.3\linewidth}
		\centering
		\vspace{1\baselineskip}
		
		\textbf{Experiment}
				
		\includegraphics[width=\textwidth]{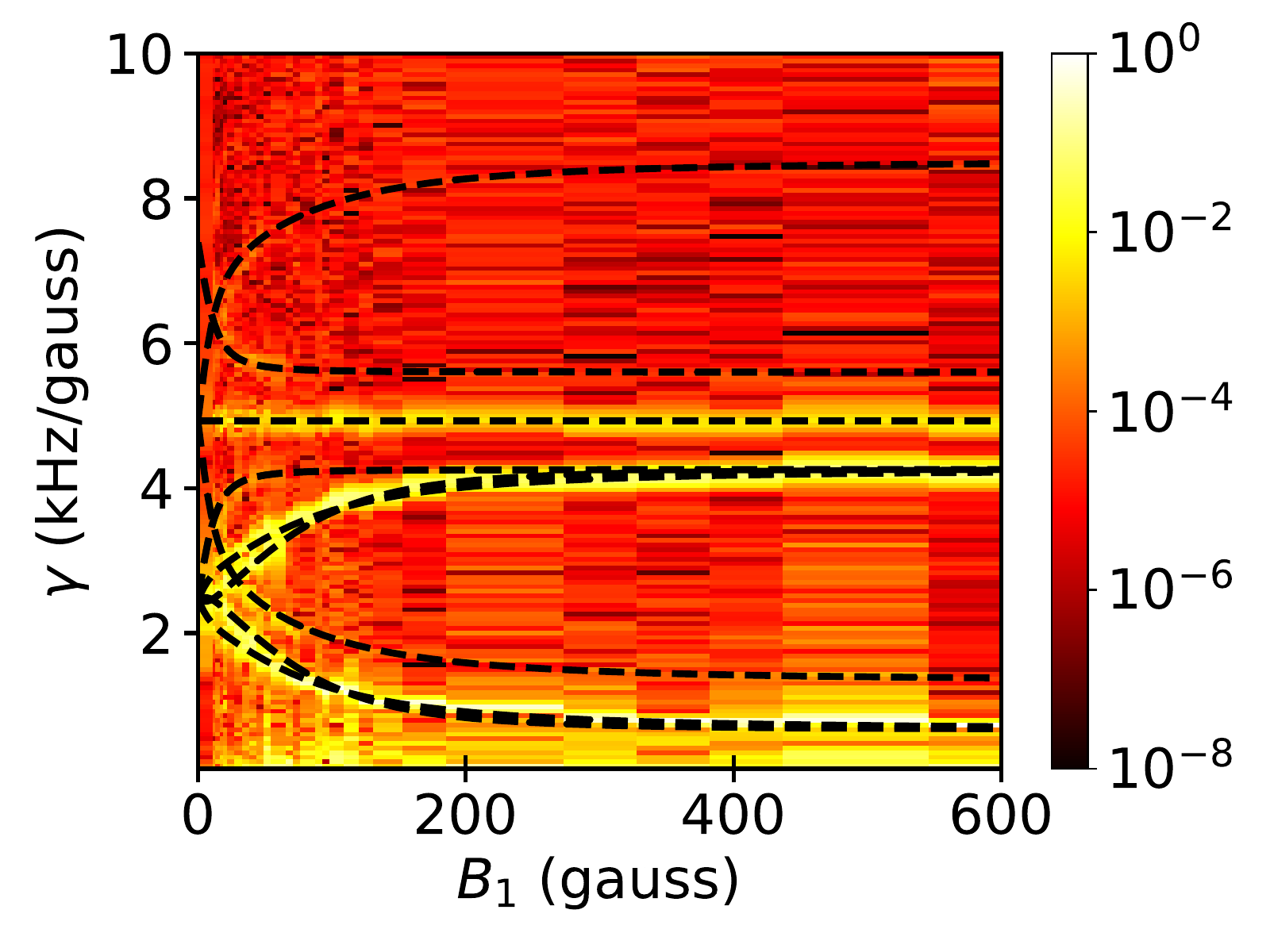}
	\end{minipage}

	%
	\caption{2D Fourier amplitude plots formed by the concatenation of spectra plots (such as Figure \ref{Fig_wiggleCurves} (d-f)) for various values of the magnetic field of the first solenoid $B_1$. Colour indicates the Fourier amplitude. For the theory plots, the field profile is depicted in Figure \ref{Fig_oH2experimentAbstraction}; the scattering transfer matrix $\mathbf{\Sigma} = \mathbb{1}_9$ and is constant for all energies; $B_2$ was varied from $-(B_1-\SI{10}{\gauss})$ to $-(B_1+\SI{10}{\gauss})$; the signal was sampled at a rate of 300 points per \SI{20}{\gauss}; and all data with a value less than $10^{-3.5}$ has been replaced with $10^{-3.5}$ for clarity.  For the experimental plot, the sample was the (111) surface of Cu; all data with a value of less then $10^{-8}$ has been replaced with $10^{-8}$ for clarity; the dashed lines indicate  transitions identified by Godsi et al.~\cite{Godsi2017}; and the data was obtained from Godsi et al.~\cite{Godsi2017}.} 
	\label{Fig_2DFFT_PlotPlusComparison}
\end{figure}

\setstcolor{blue}
%
The location of each feature in the spectra is reflective of a gyromagnetic ratio and is \emph{independent} of the molecule-surface interactions, being only a function of the hyperfine energy level structure of \oH.  The relative height of each feature, however, is dependent on the molecule-surface interactions, as exemplified in the experimental spectrum shown in Figure \ref{Fig_wiggleCurves} (f). From Figure \ref{Fig_wiggleCurves}, one can see that integrating over the velocity distribution is important to produce the spin-echo effect and to bring the observed signal closer to experiment. 


%
%

A different spectrum can be obtained for every possible value of $B_1$ and then combined to form a 2D map of the generalized gyromagnetic ratios and their contributing amplitudes as a function of $B_1$, as shown in Figure \ref{Fig_2DFFT_PlotPlusComparison}. This protocol is equivalent to observing the scattering of molecules with different internal hyperfine states as different values of the magnetic field in the first branch produce different superpositions of the hyperfine states. One can clearly see both the magnetic field-dependence of the gyromagnetic ratios, the impact of integrating over the velocity distribution, and the stark similarities and differences between the experimental and theory plots.

\begin{figure}[H]
	\centering
	
	\begin{minipage}{.3\linewidth}
		\centering
		\xincludegraphicsC[width=\textwidth,label={\bf RP},pos=ne,fontsize=\normalsize]{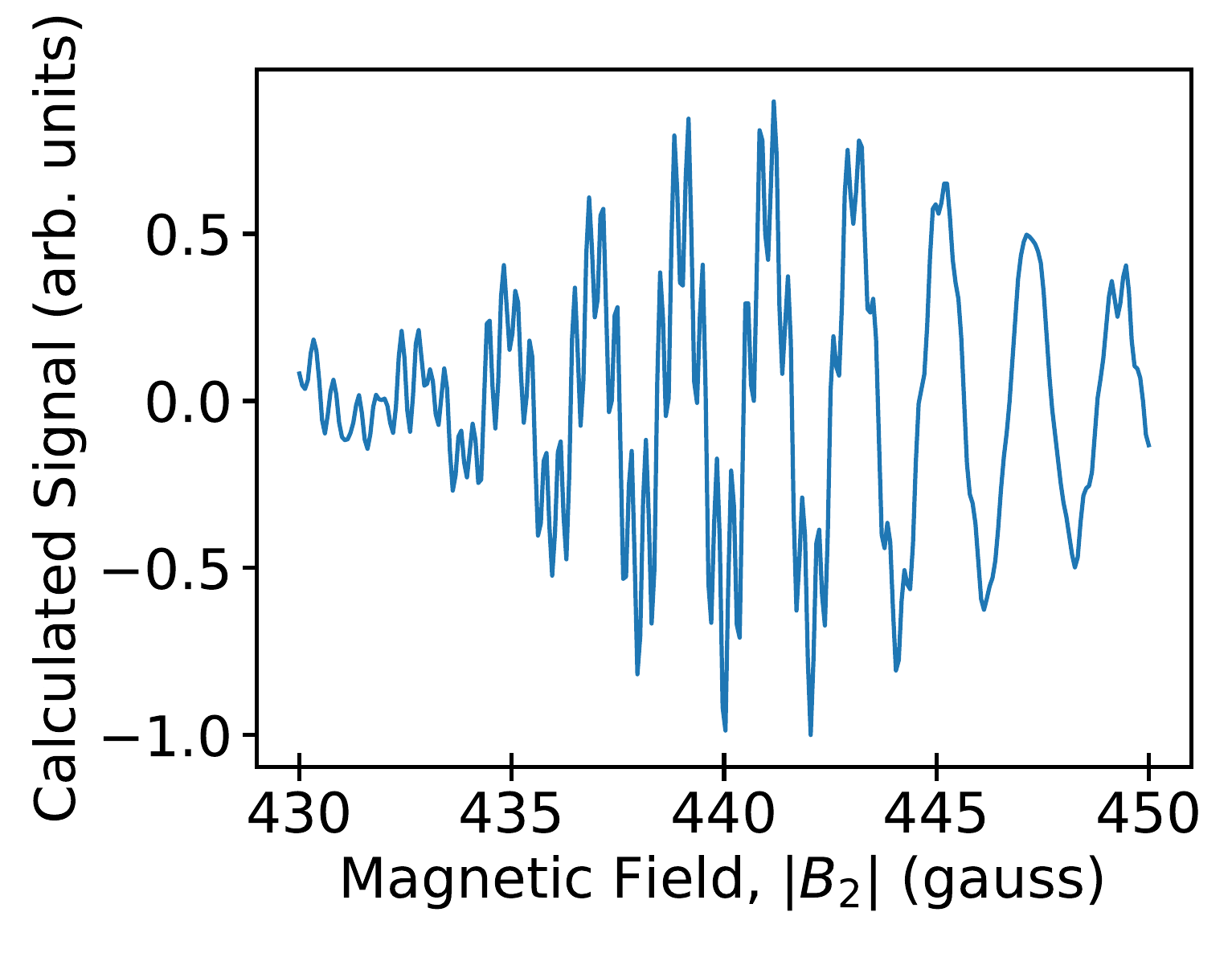}
	\end{minipage}
	\begin{minipage}{.3\linewidth}
		\centering
		\xincludegraphicsC[width=\textwidth,label={\bf RP},pos=ne,fontsize=\normalsize]{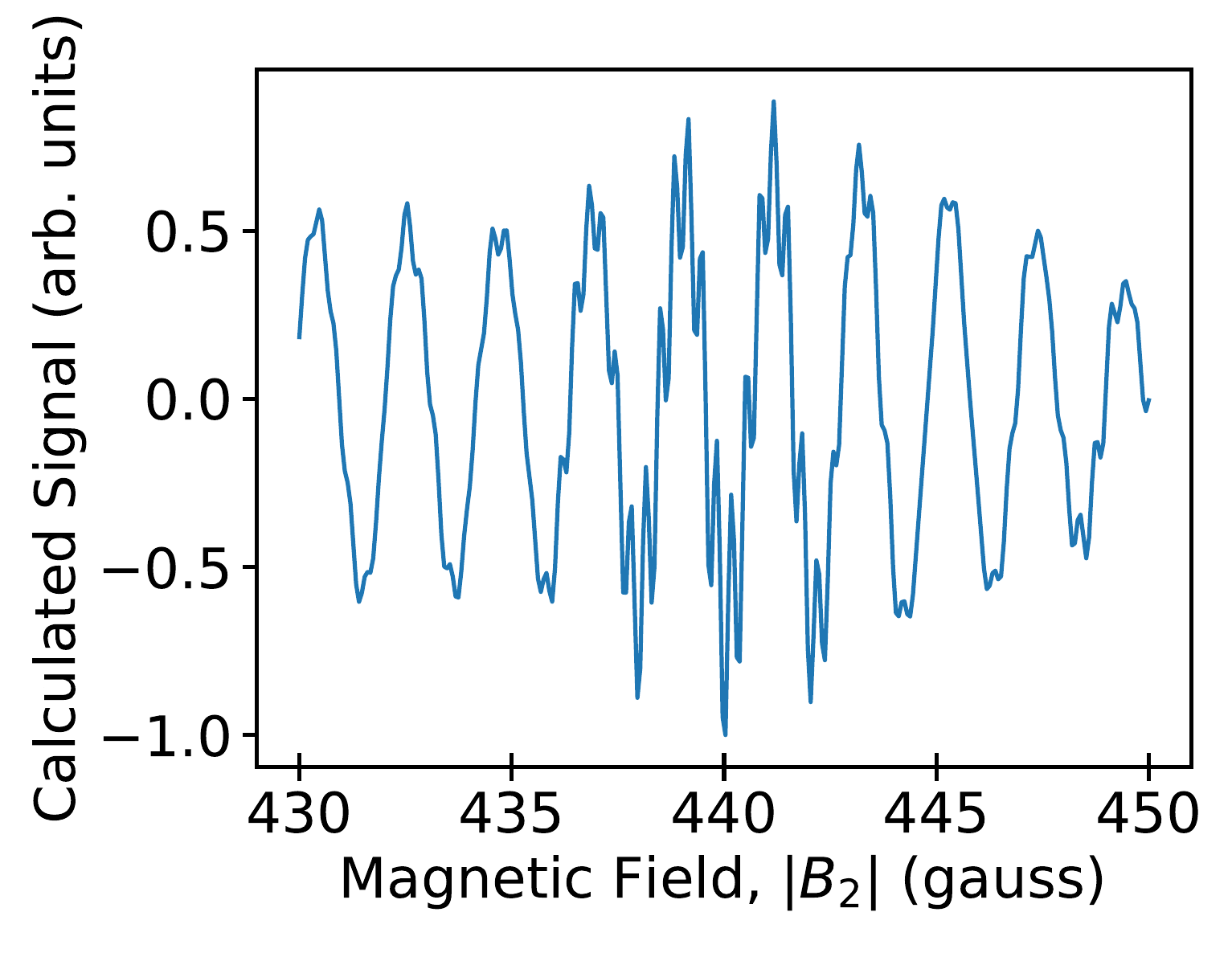}
	\end{minipage}
	\begin{minipage}{.3\linewidth}
		\centering
		\xincludegraphicsC[width=\textwidth,label={\bf RP},pos=ne,fontsize=\normalsize]{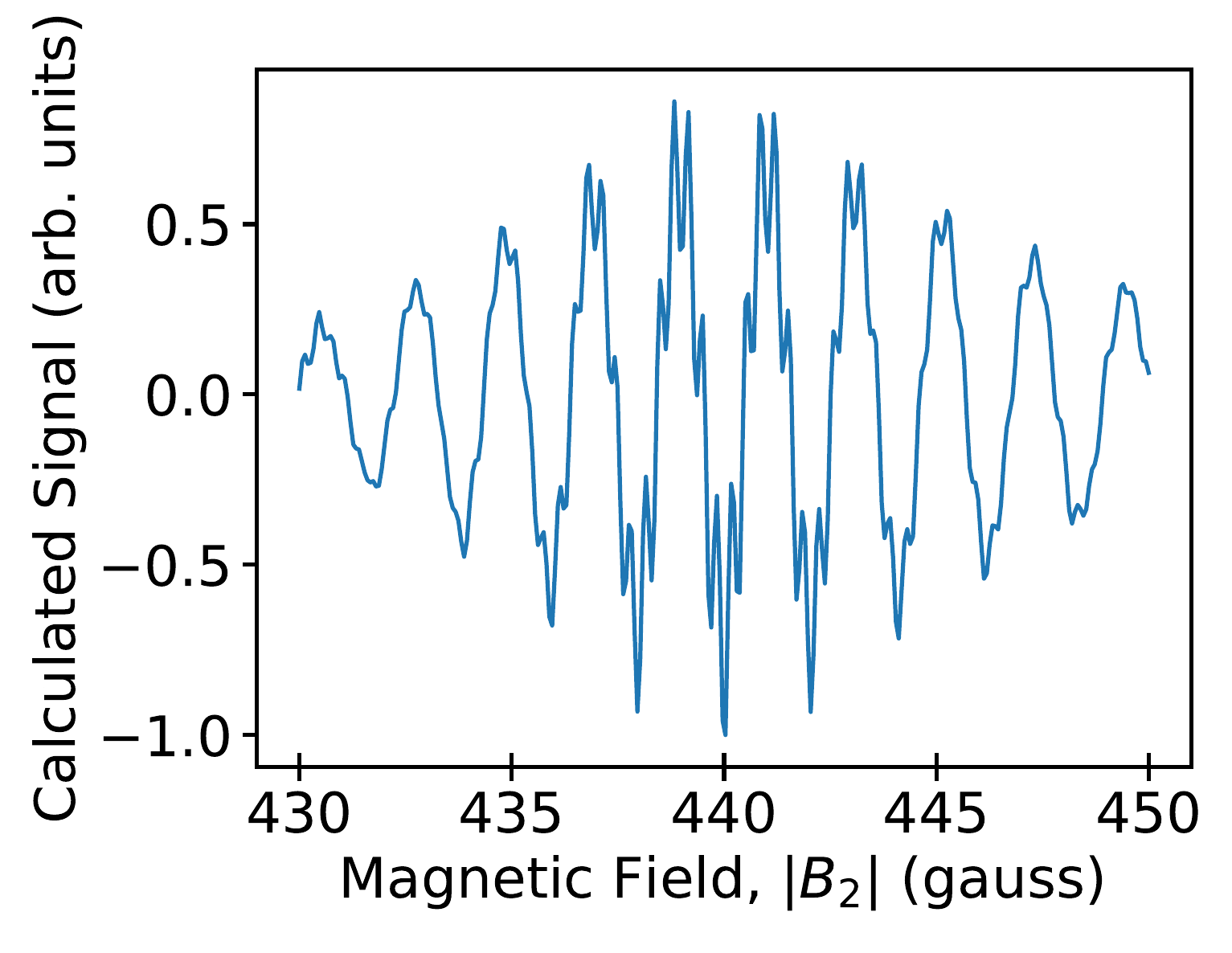}
	\end{minipage}
	
	
	\begin{minipage}{.3\linewidth}
		\centering
		\xincludegraphicsC[width=\textwidth,label={\bf RDA},pos=ne,fontsize=\normalsize]{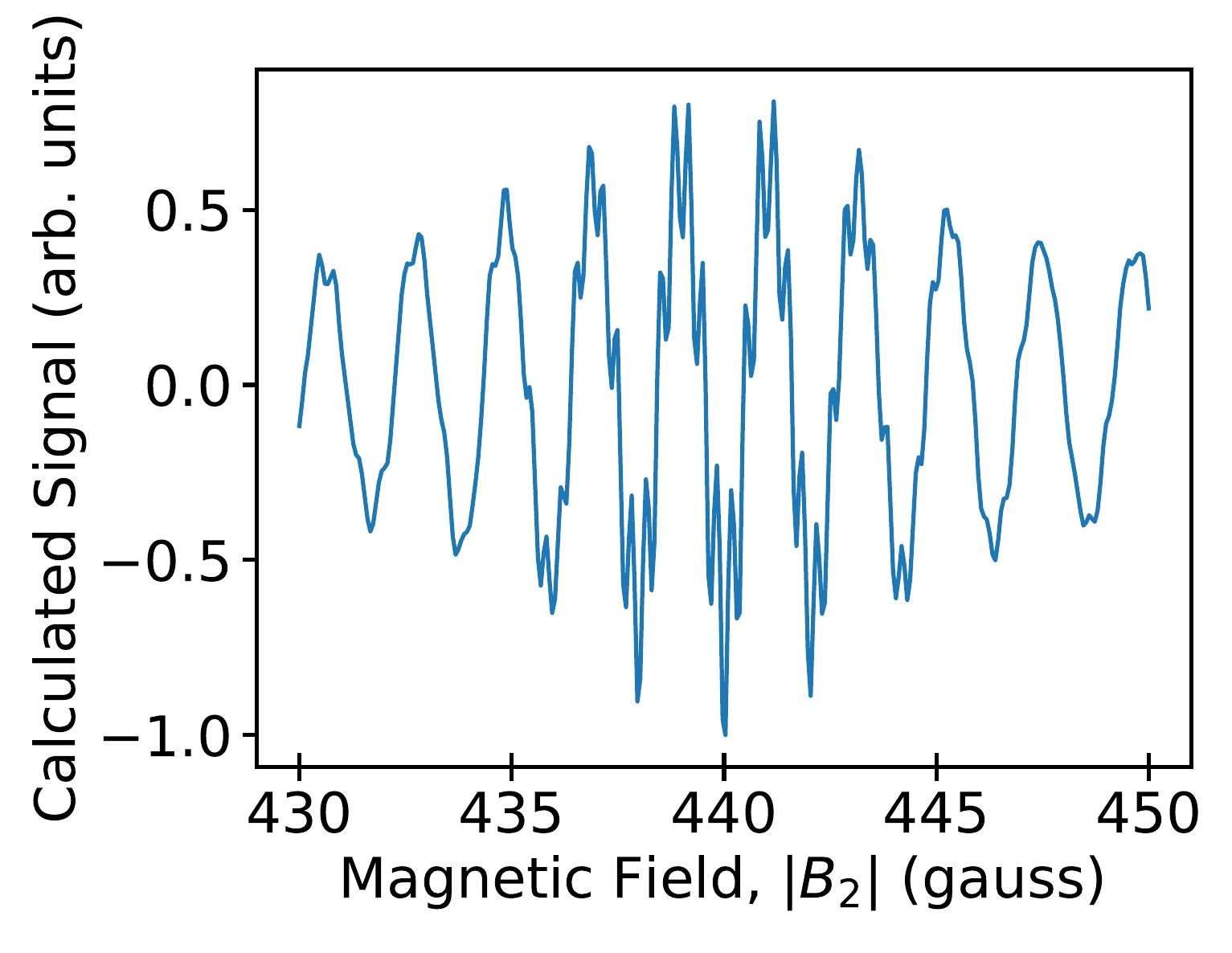}
	\end{minipage}
	\begin{minipage}{.3\linewidth}
		\centering
		\xincludegraphicsC[width=\textwidth,label={\bf RDA},pos=ne,fontsize=\normalsize]{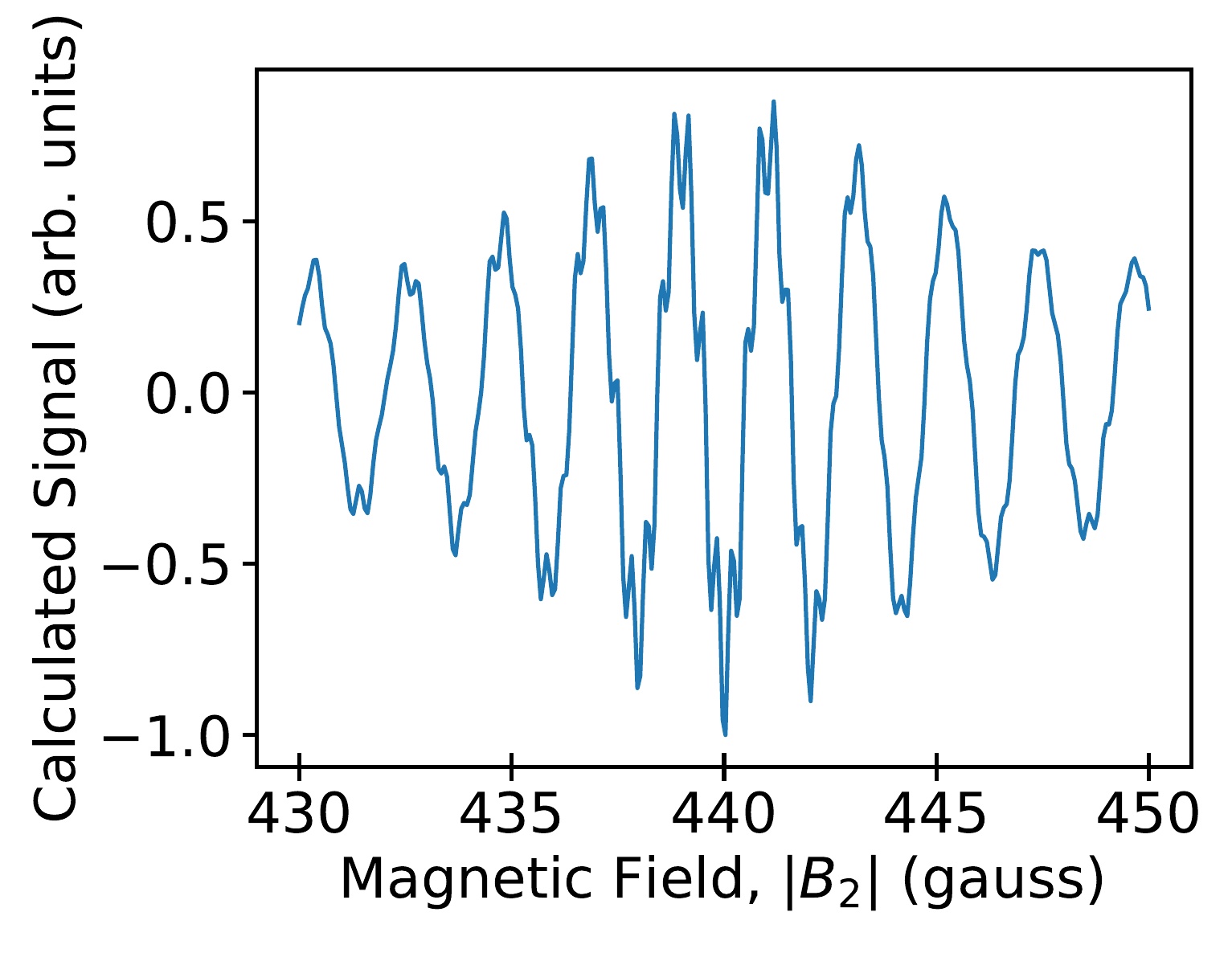}
	\end{minipage}
	\begin{minipage}{.3\linewidth}
		\centering
		\xincludegraphicsC[width=\textwidth,label={\bf RDA},pos=ne,fontsize=\normalsize]{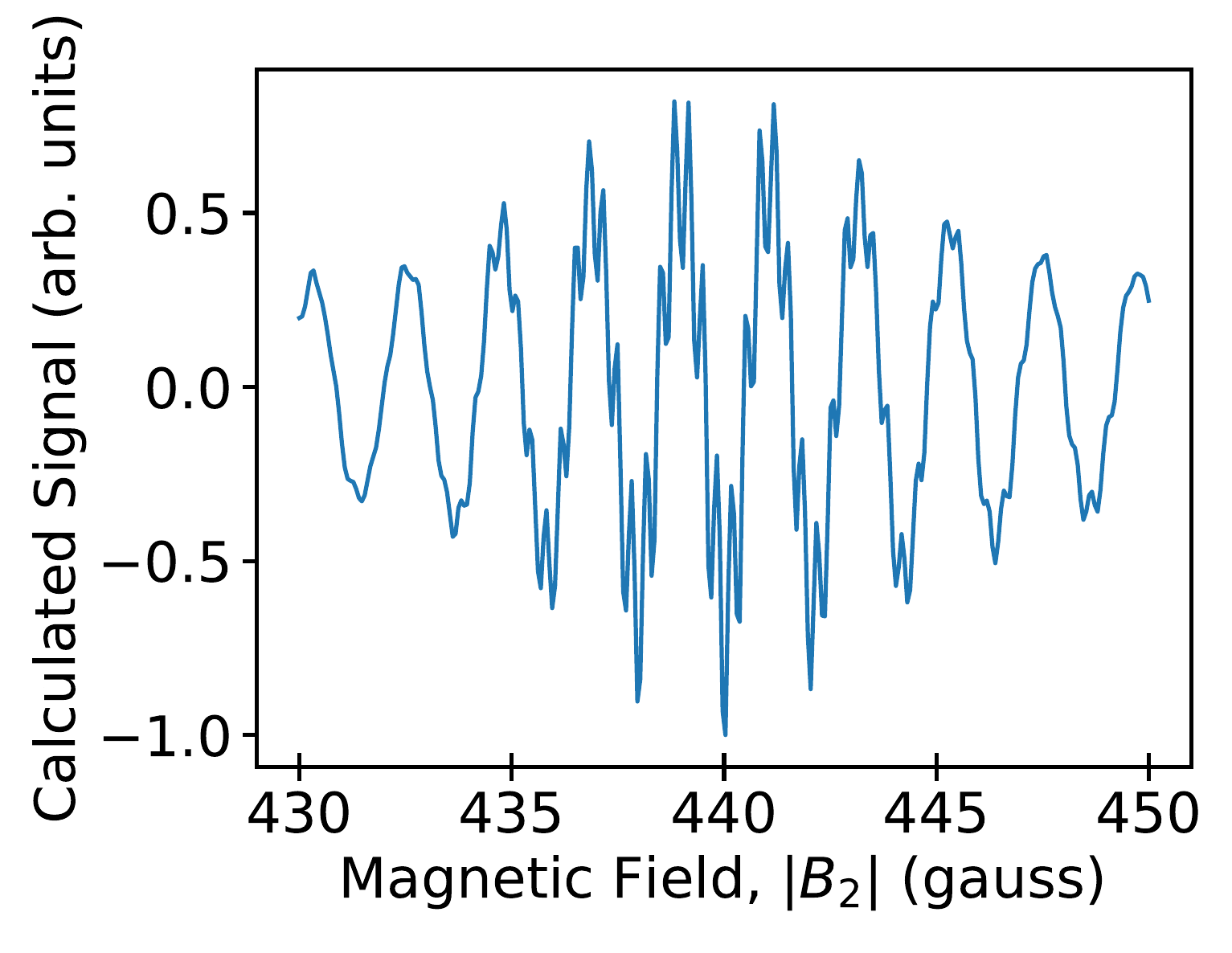}
	\end{minipage}
	
	
	\begin{minipage}{.3\linewidth}
		\centering
		\xincludegraphicsC[width=\textwidth,label={\bf ROM},pos=ne,fontsize=\normalsize]{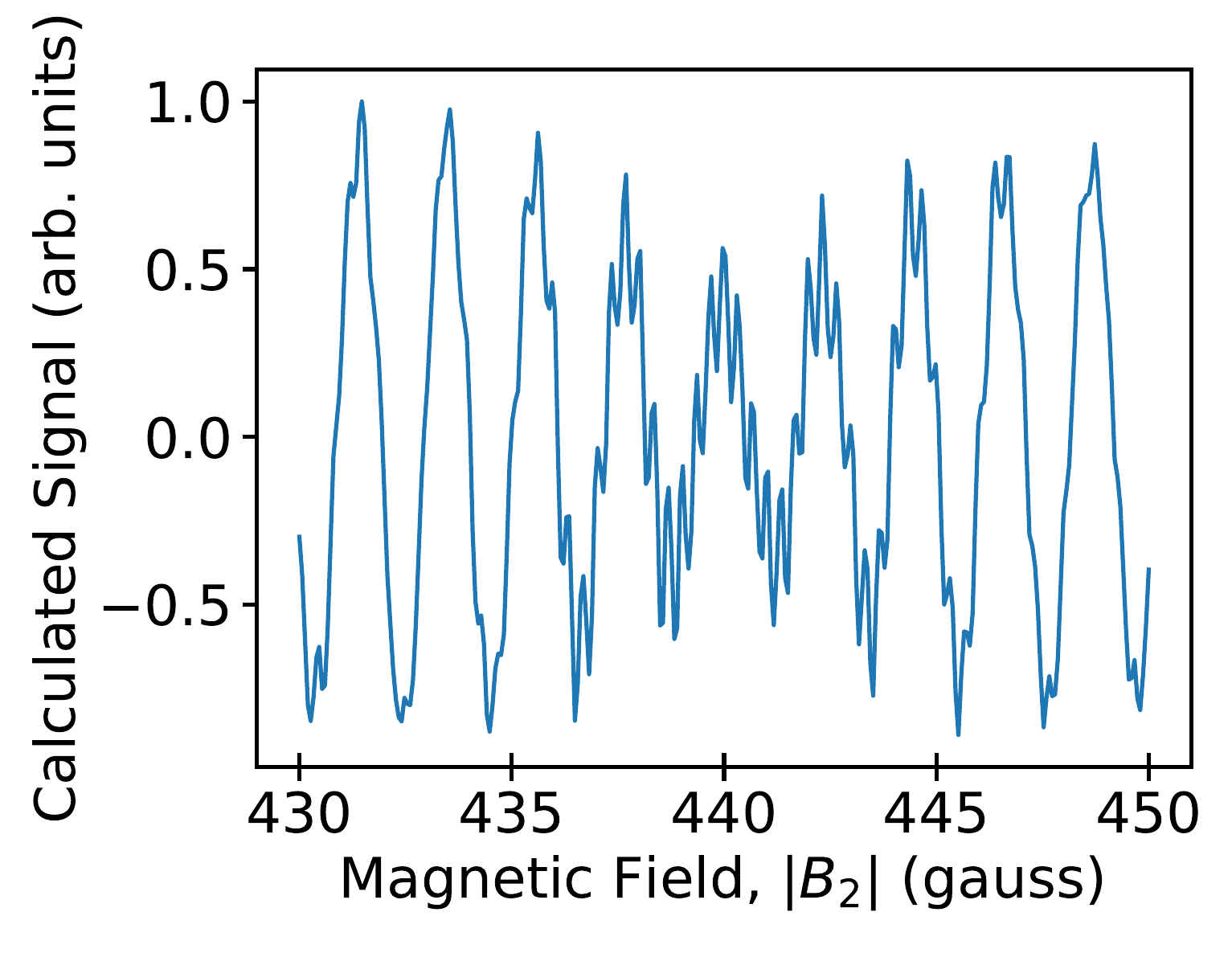}
	\end{minipage}
	\begin{minipage}{.3\linewidth}
		\centering
		\xincludegraphicsC[width=\textwidth,label={\bf ROM},pos=ne,fontsize=\normalsize]{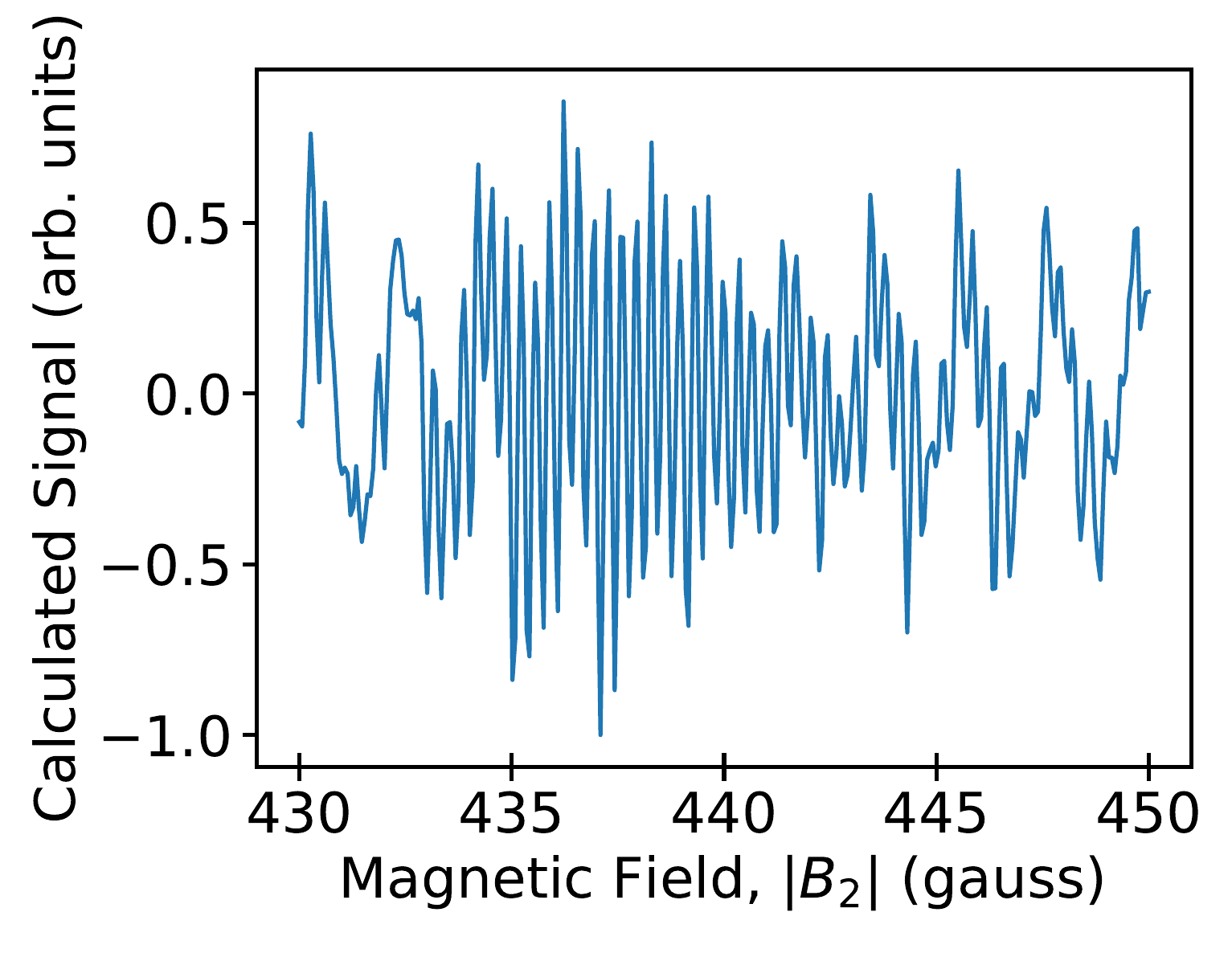}
	\end{minipage}
	\begin{minipage}{.3\linewidth}
		\centering
		\xincludegraphicsC[width=\textwidth,label={\bf ROM},pos=ne,fontsize=\normalsize]{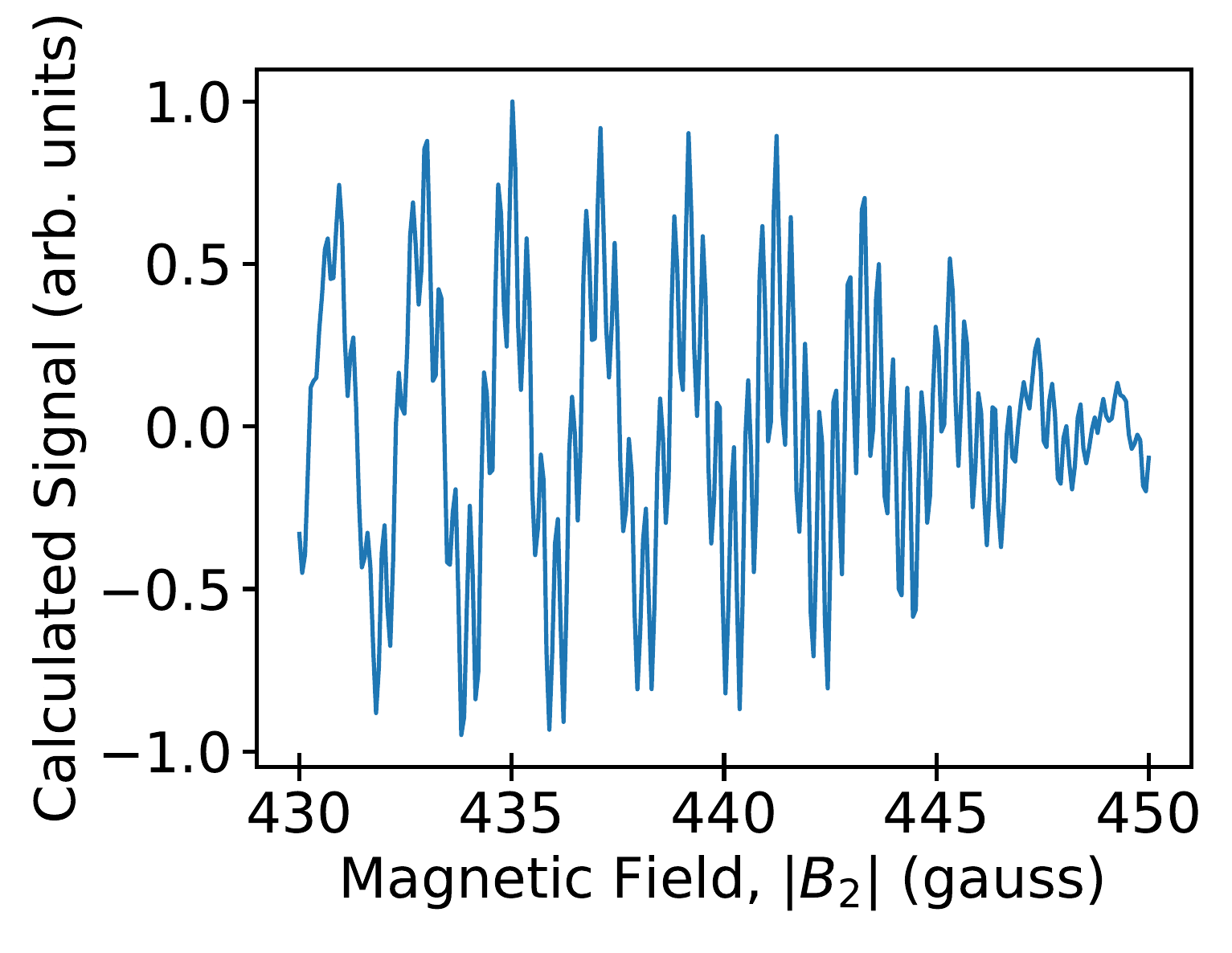}
	\end{minipage}
	
	
	\begin{minipage}{.3\linewidth}
		\centering
		\xincludegraphicsC[width=\textwidth,label={\bf RUM},pos=se,fontsize=\normalsize]{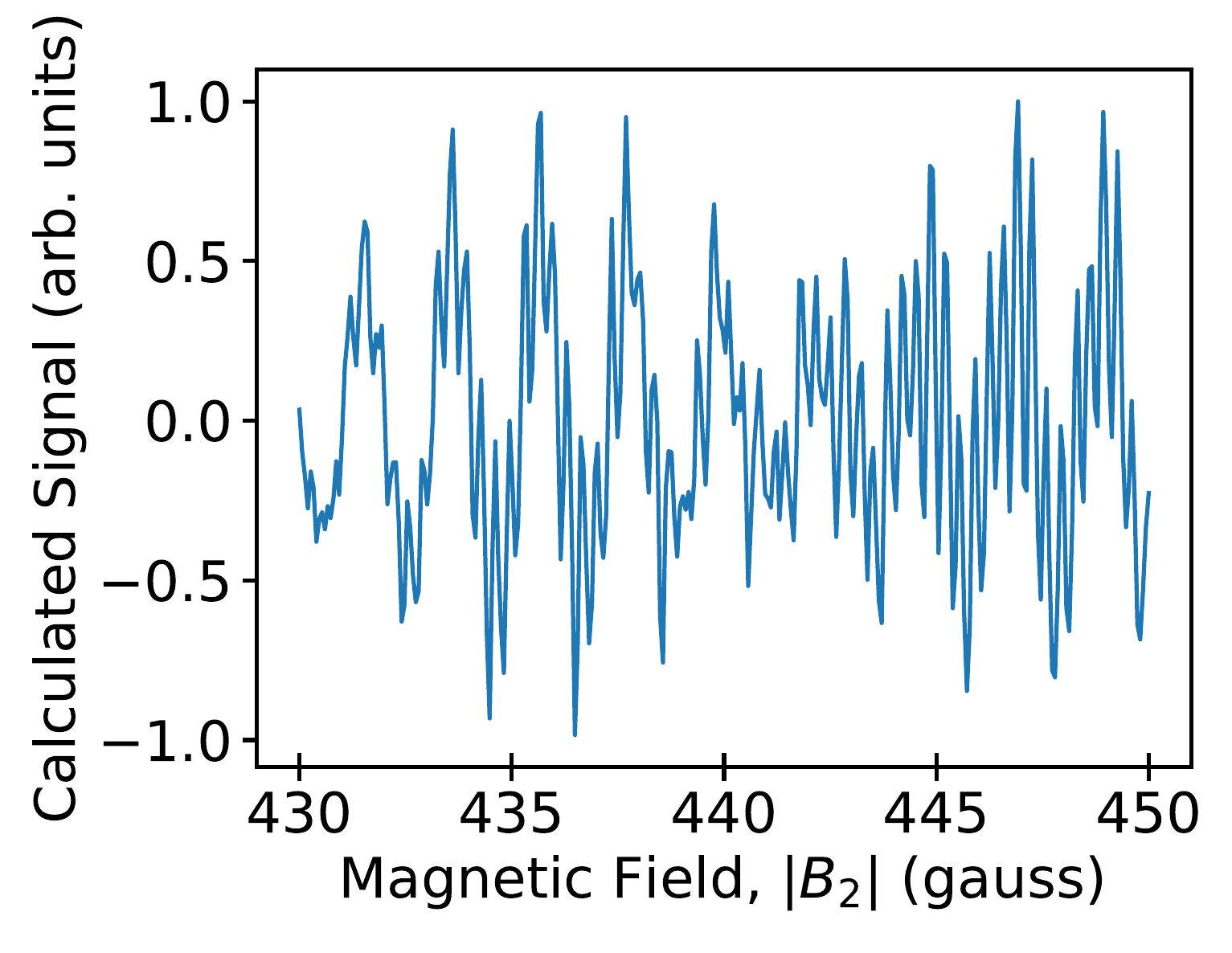}
	\end{minipage}
	\begin{minipage}{.3\linewidth}
		\centering
		\xincludegraphicsC[width=\textwidth,label={\bf RUM},pos=ne,fontsize=\normalsize]{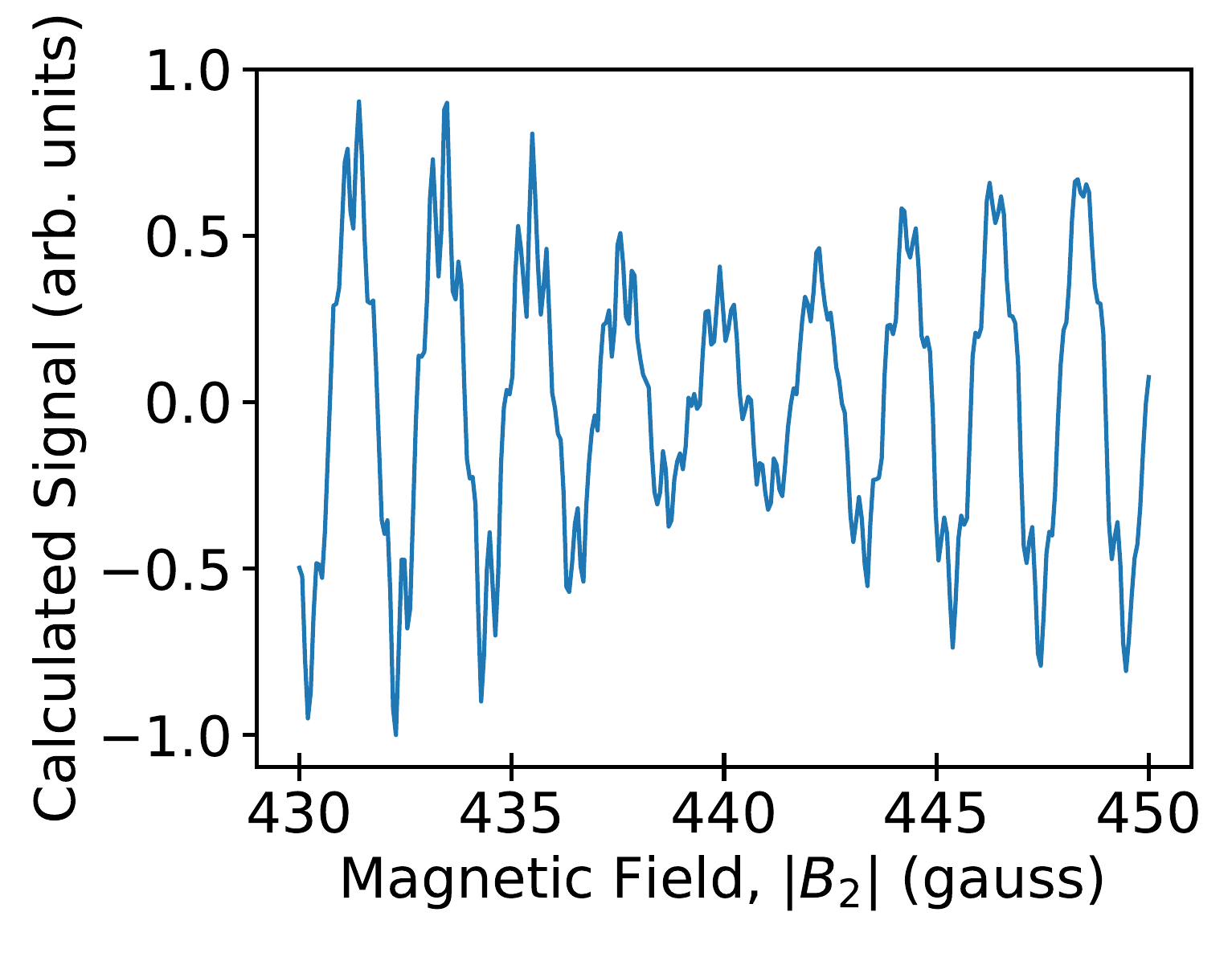}
	\end{minipage}
	\begin{minipage}{.3\linewidth}
		\centering
		\xincludegraphicsC[width=\textwidth,label={\bf RUM},pos=ne,fontsize=\normalsize]{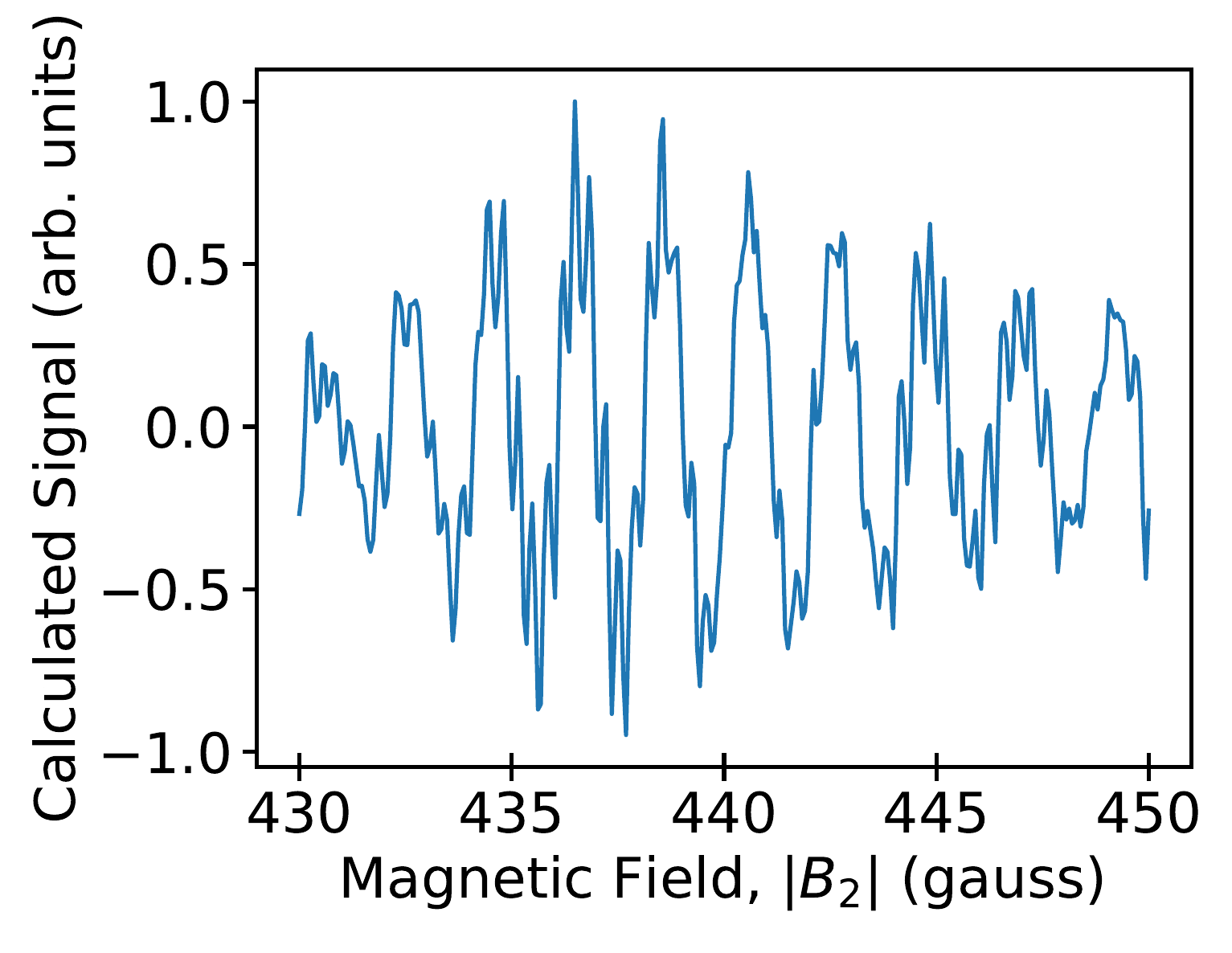}
	\end{minipage}

	\caption{Calculated signals close to the spin echo condition as functions of the magnetic field of the second coil $|B_2|$. The field profile is depicted in Figure \ref{Fig_oH2experimentAbstraction}; $B_1=\SI{440}{\gauss}$; and the signal was sampled at a rate of 300 points per \SI{20}{\gauss}. Each of the plots was created with identical parameters, except for the scattering transfer matrices $\mathbf{\Sigma}$. The scattering transfer matrices are identical for all energies and are randomly chosen for each plot as follows. First row -- Random Phases (\textbf{RP}): $\mathbf{\Sigma} = \bigoplus_{i=1}^9 e^{i\theta_i}$ is a diagonal unitary matrix whose nine phases $\theta_i$ are randomly chosen from a uniform distribution of width $2\pi$. Second row -- Random Diagonal Amplitudes (\textbf{RDA}): $\mathbf{\Sigma} = \bigoplus_{i=1}^9 A_i$ is a diagonal matrix whose diagonal elements are randomly chosen from a uniform distribution on the interval $\left[0,1\right)$. Third Row -- Random Orthogonal Matrices (\textbf{ROM}): $\mathbf{\Sigma}$ is an orthogonal matrix randomly drawn according to the Haar measure on $O(9)$. Fourth Row -- Random Unitary Matrices (\textbf{RUM}): $\mathbf{\Sigma}$ is a unitary matrix randomly drawn according to the Haar measure on $U(9)$. Here, randomly drawing according to the Haar measure can be understood as analogous to drawing from the ``uniform distribution'' over the space of possible matrices \cite{Mezzadri2007}.}
	\label{Fig_wigglePlots_differentMatrices}
\end{figure}

\begin{figure}[H]
	\centering
	
	\begin{minipage}{.3\linewidth}
		\centering
		\xincludegraphics[width=\textwidth,label={\bf RP},pos=ne,fontsize=\normalsize]{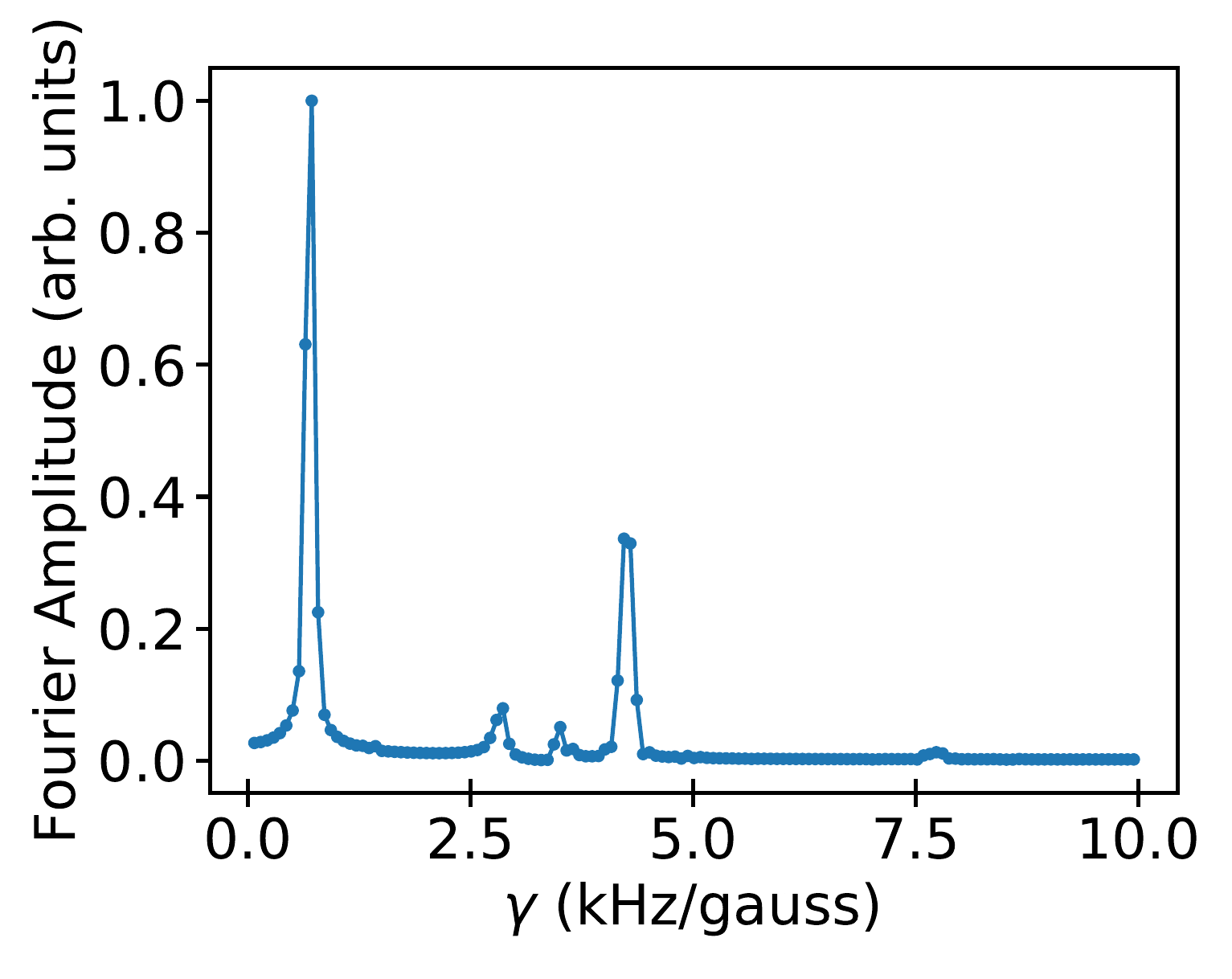}
	\end{minipage}
	\begin{minipage}{.3\linewidth}
		\centering
		\xincludegraphics[width=\textwidth,label={\bf RP},pos=ne,fontsize=\normalsize]{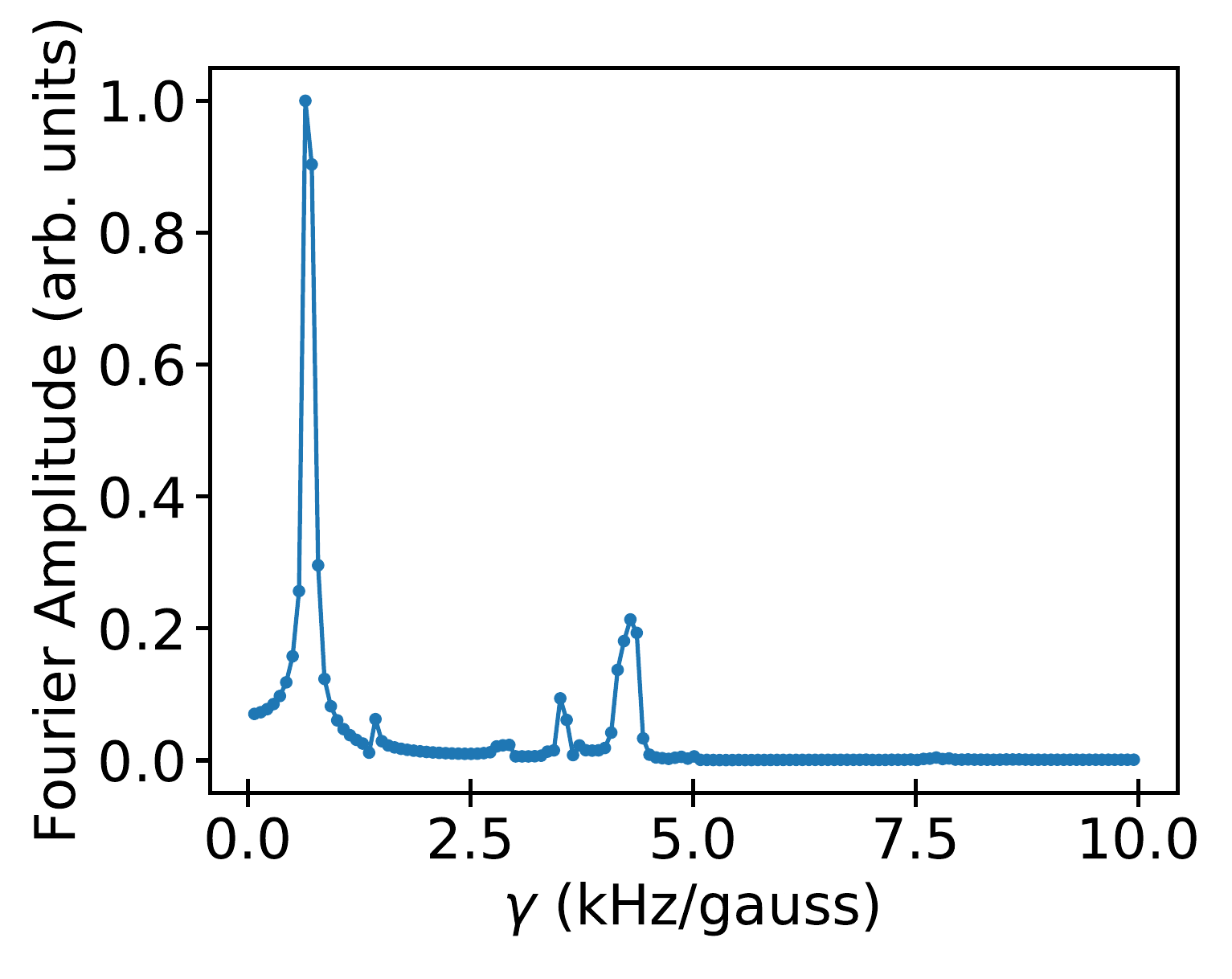}
	\end{minipage}
	\begin{minipage}{.3\linewidth}
		\centering
		\xincludegraphics[width=\textwidth,label={\bf RP},pos=ne,fontsize=\normalsize]{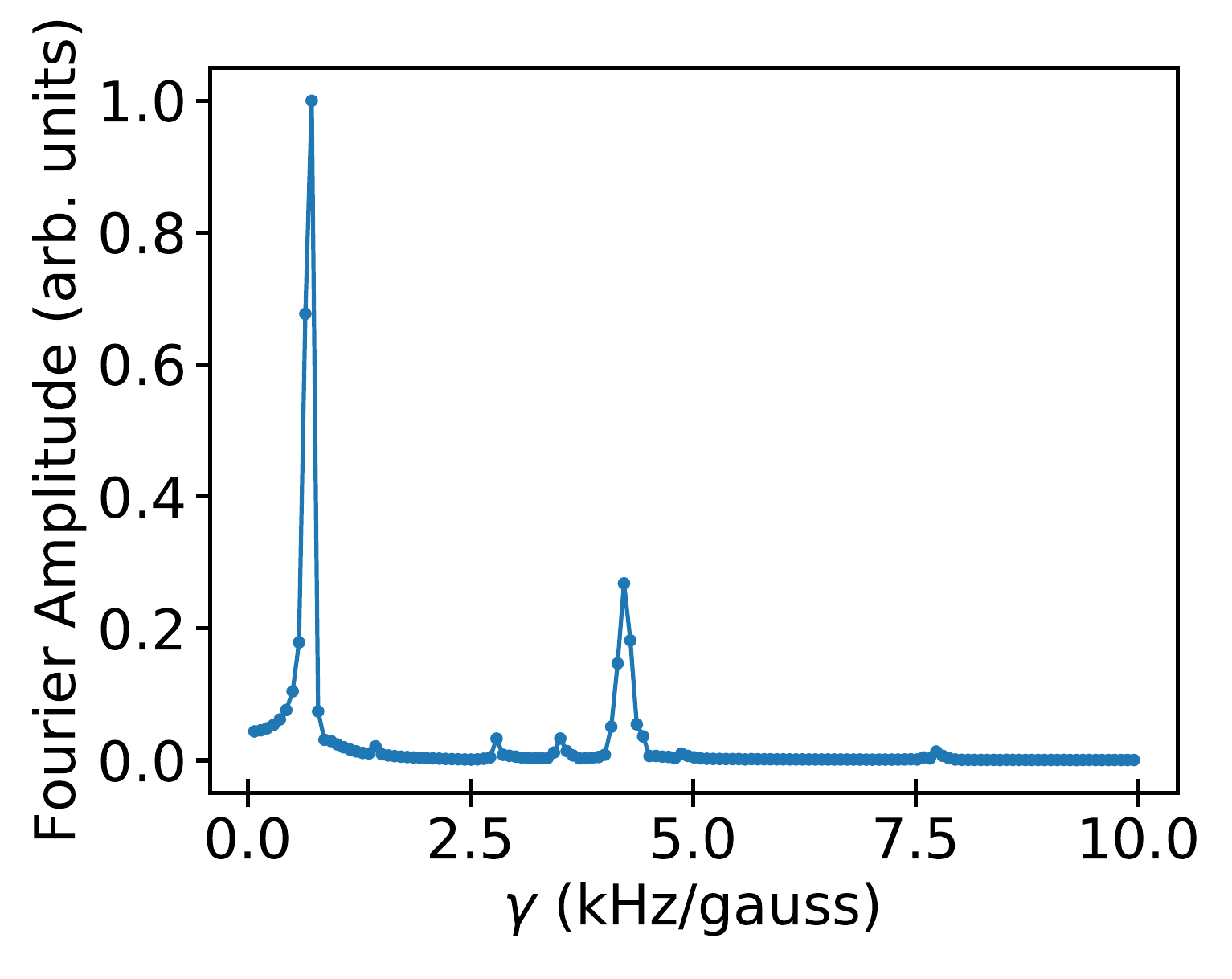}
	\end{minipage}
	
	
	\begin{minipage}{.3\linewidth}
		\centering
		\xincludegraphics[width=\textwidth,label={\bf RDA},pos=ne,fontsize=\normalsize]{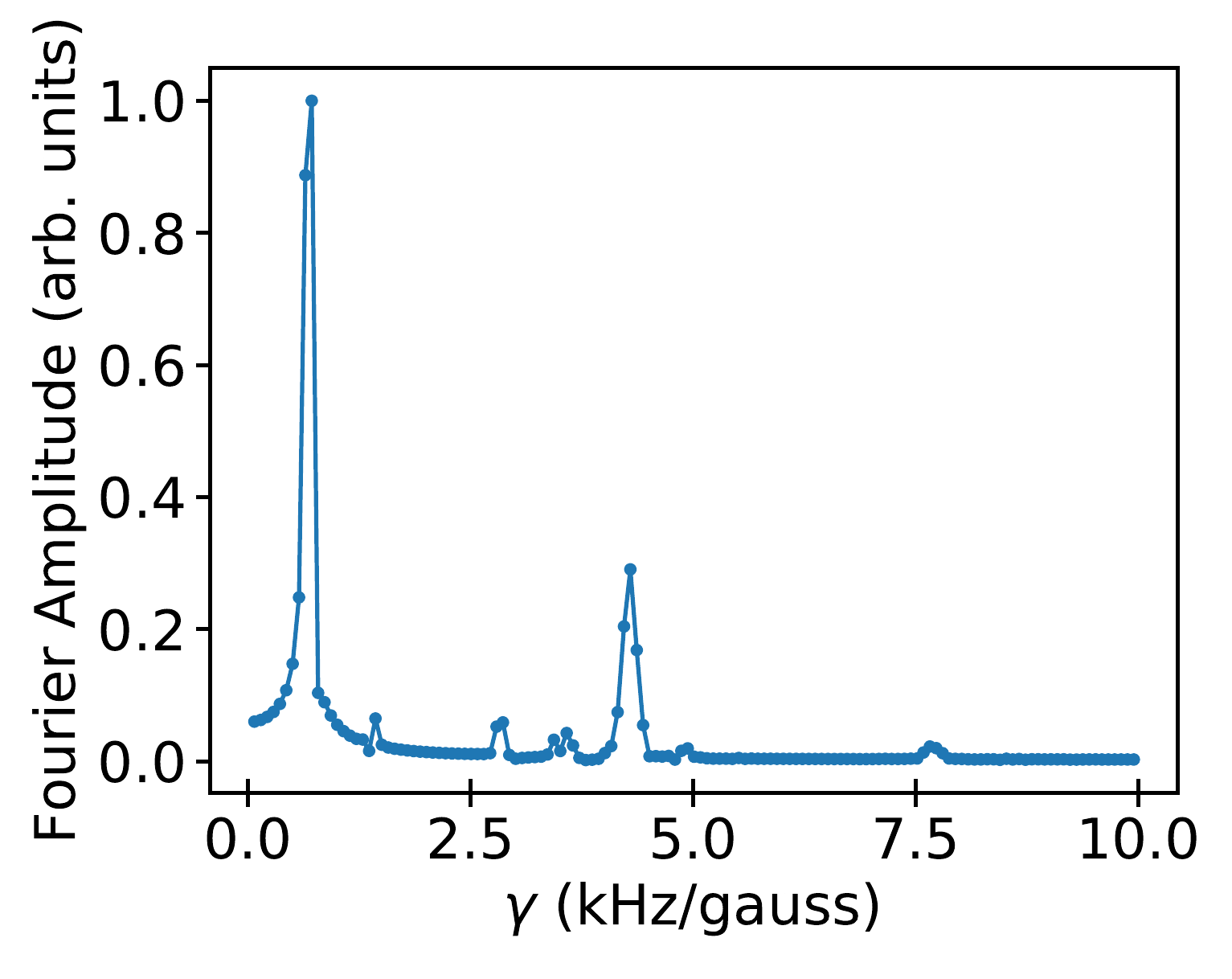}
	\end{minipage}
	\begin{minipage}{.3\linewidth}
		\centering
		\xincludegraphics[width=\textwidth,label={\bf RDA},pos=ne,fontsize=\normalsize]{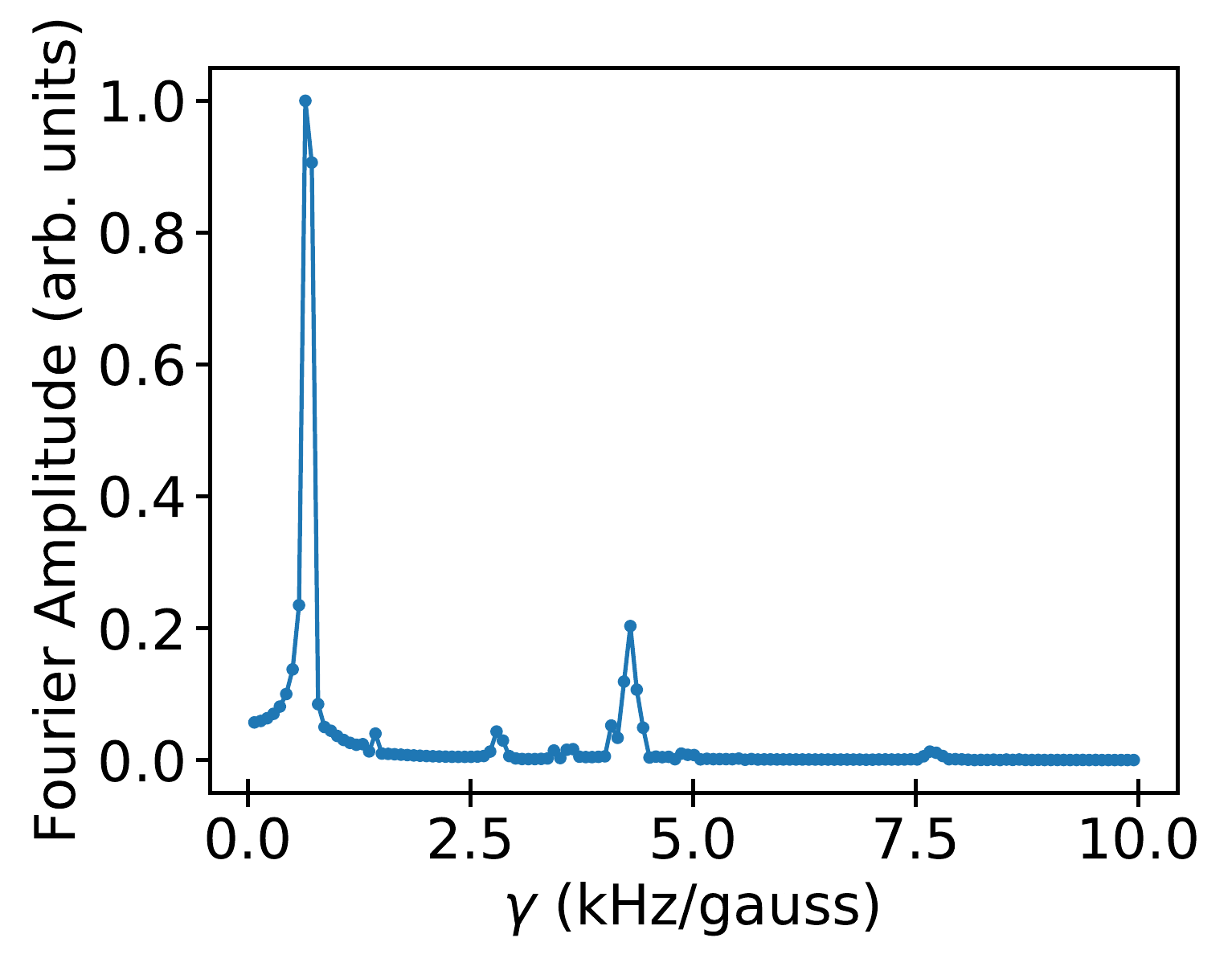}
	\end{minipage}
	\begin{minipage}{.3\linewidth}
		\centering
		\xincludegraphics[width=\textwidth,label={\bf RDA},pos=ne,fontsize=\normalsize]{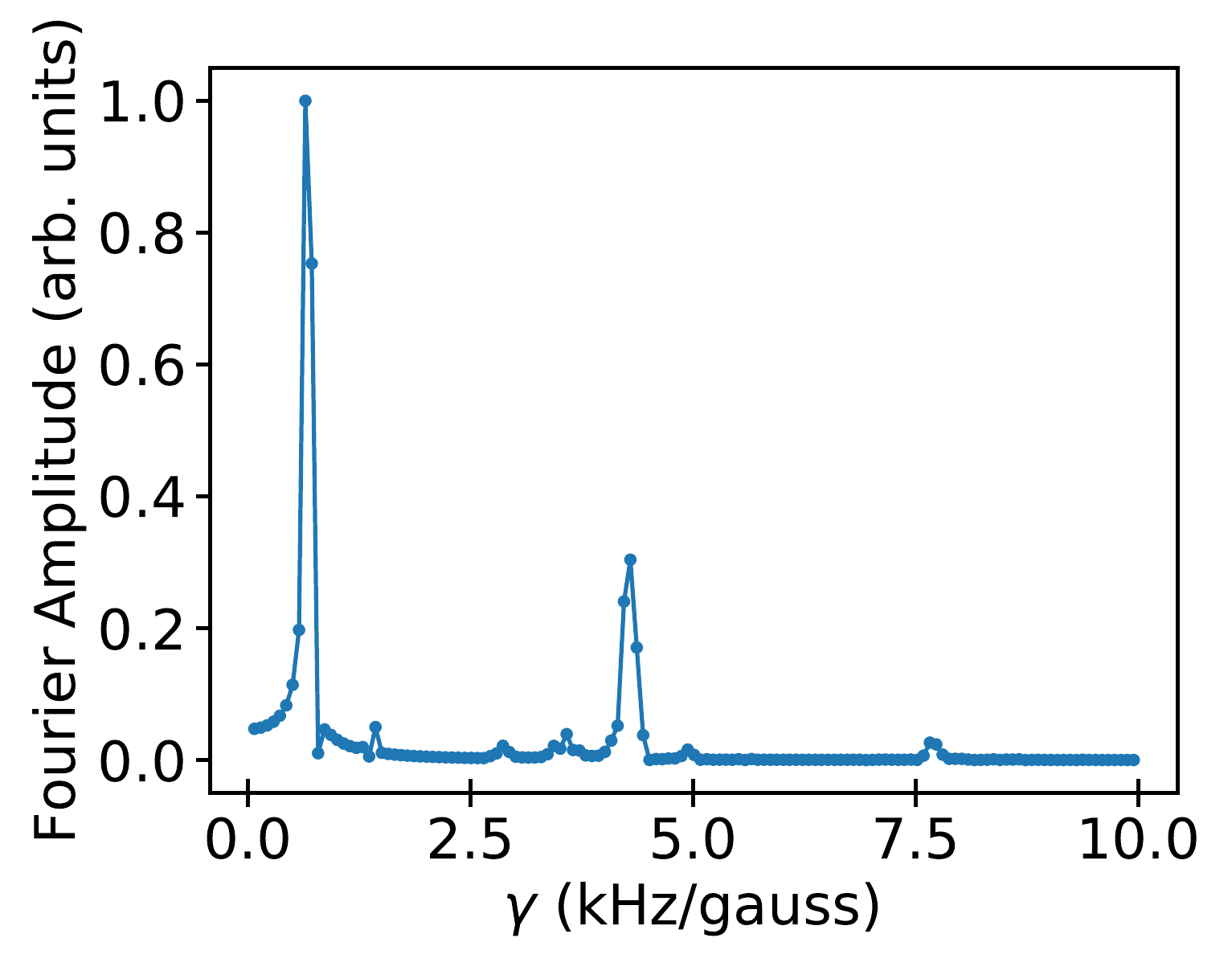}
	\end{minipage}
	
	
	\begin{minipage}{.3\linewidth}
		\centering
		\xincludegraphics[width=\textwidth,label={\bf ROM},pos=ne,fontsize=\normalsize]{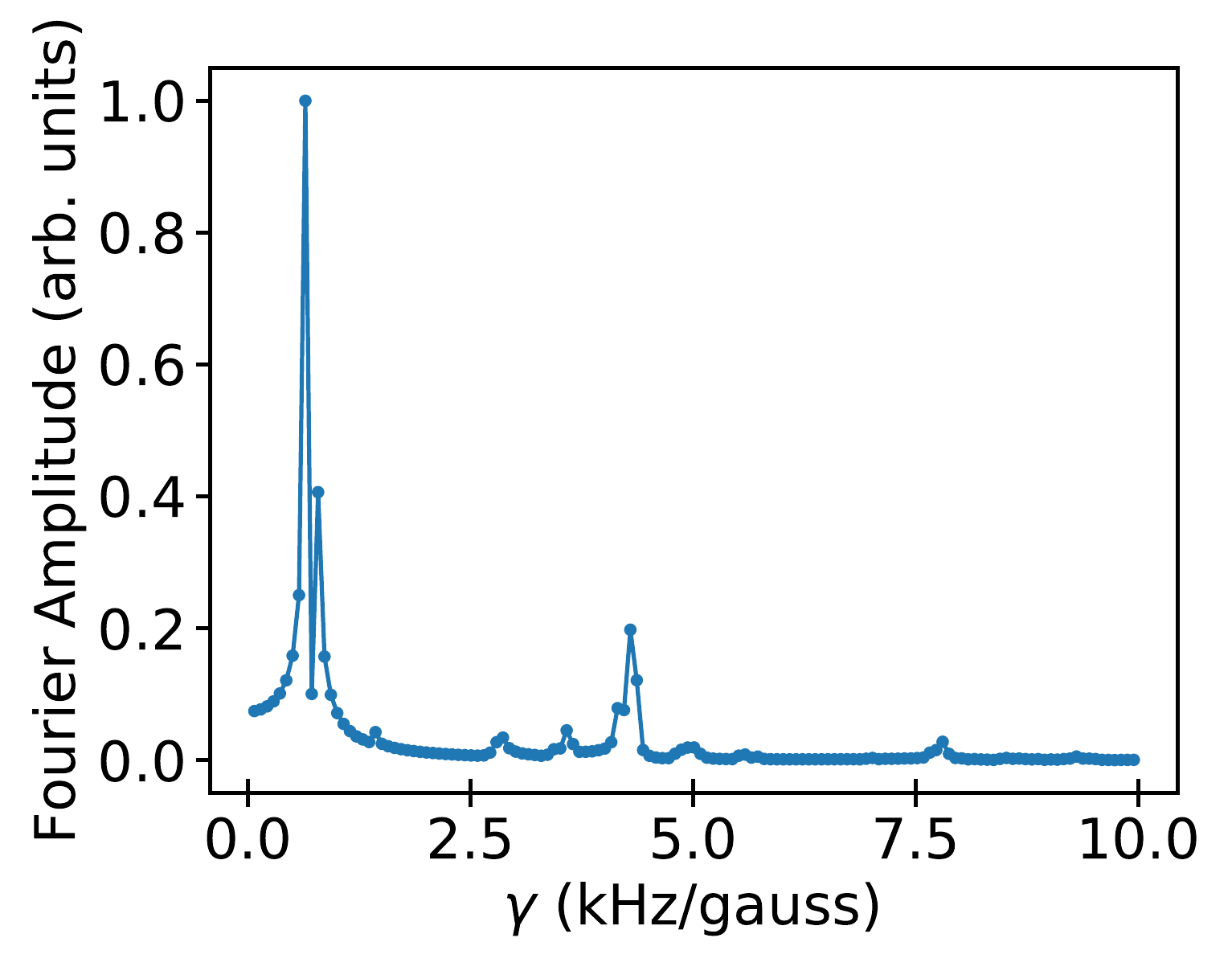}
	\end{minipage}
	\begin{minipage}{.3\linewidth}
		\centering
		\xincludegraphics[width=\textwidth,label={\bf ROM},pos=ne,fontsize=\normalsize]{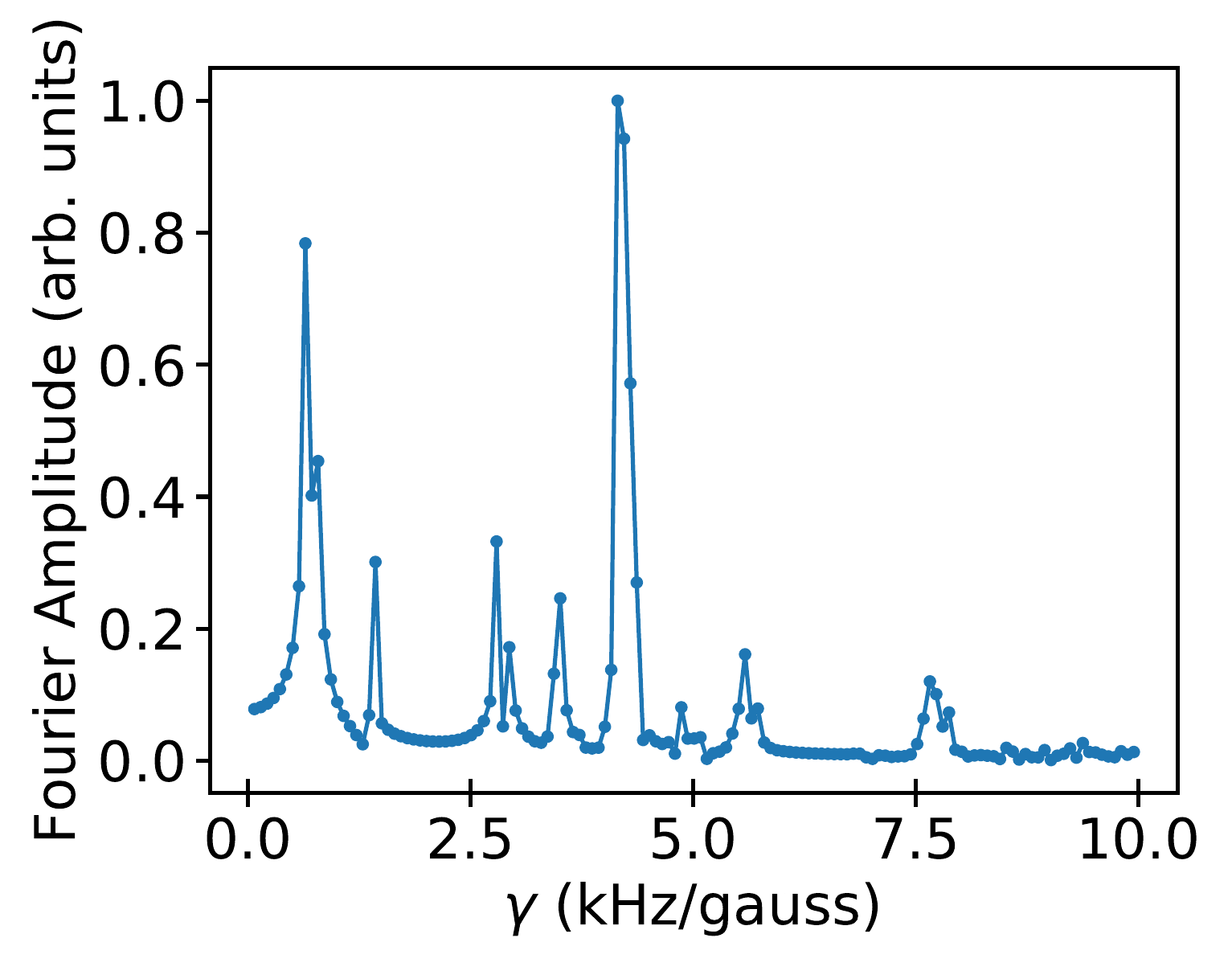}
	\end{minipage}
	\begin{minipage}{.3\linewidth}
		\centering
		\xincludegraphics[width=\textwidth,label={\bf ROM},pos=ne,fontsize=\normalsize]{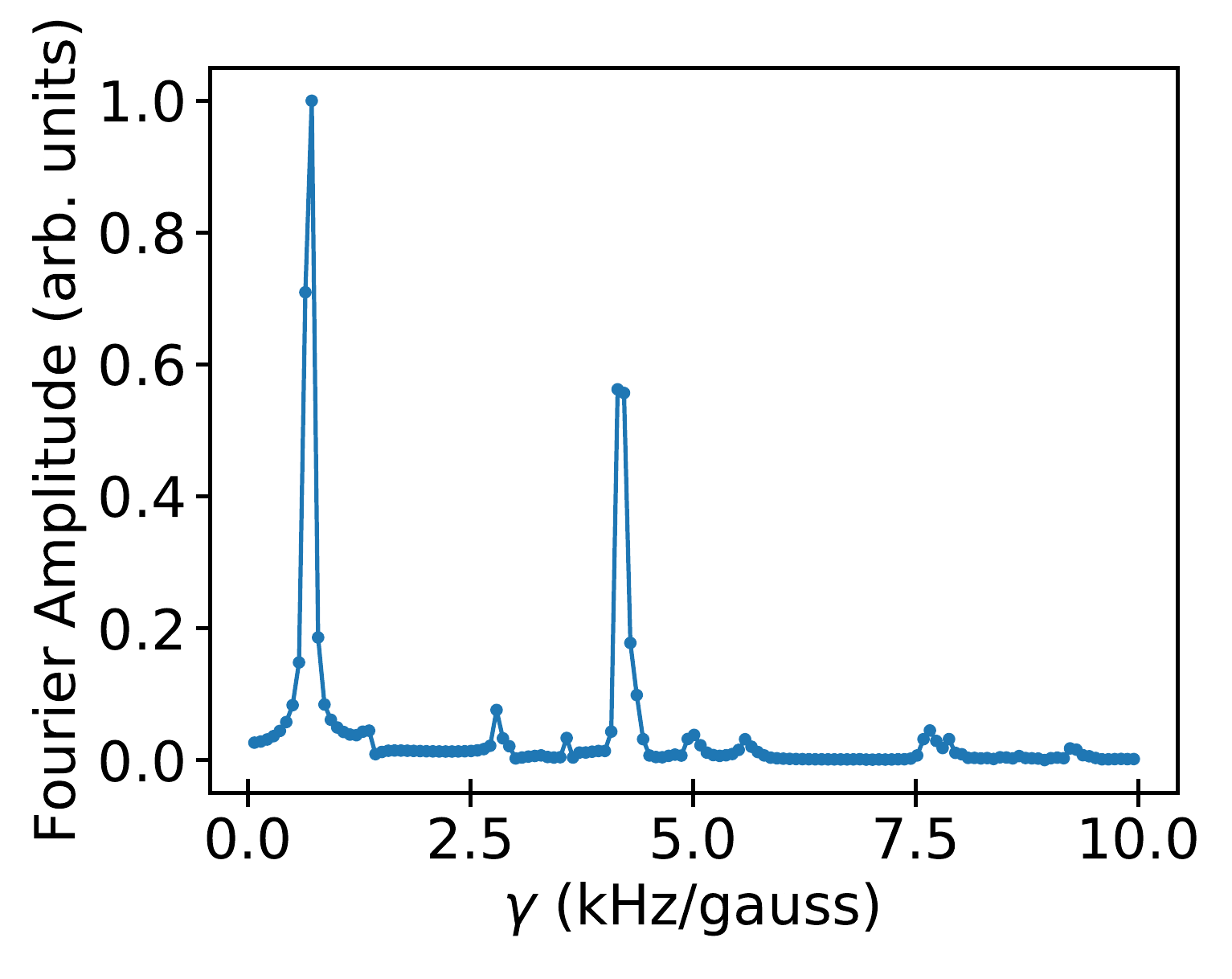}
	\end{minipage}
	
	
	\begin{minipage}{.3\linewidth}
		\centering
		\xincludegraphics[width=\textwidth,label={\bf RUM},pos=ne,fontsize=\normalsize]{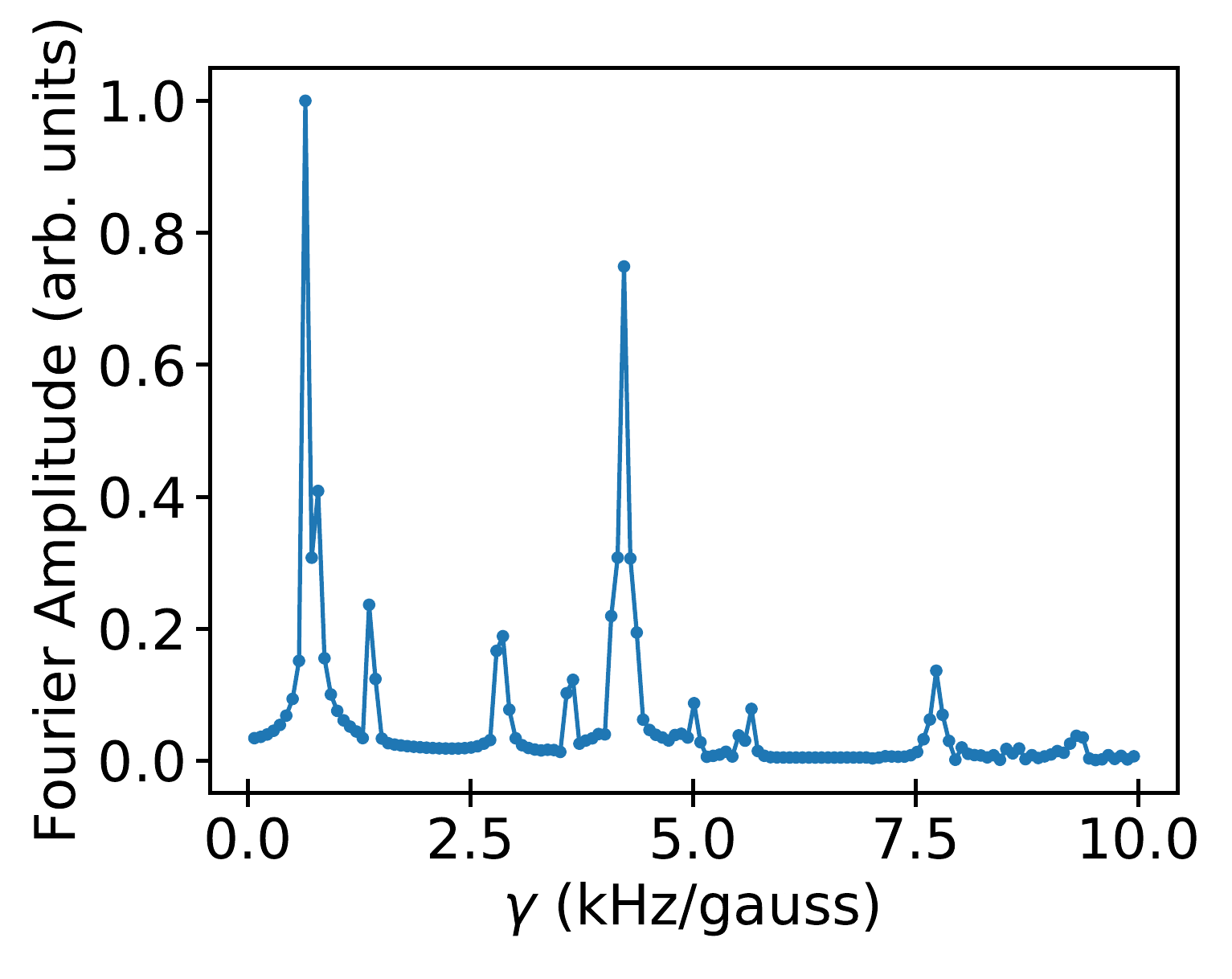}
	\end{minipage}
	\begin{minipage}{.3\linewidth}
		\centering
		\xincludegraphics[width=\textwidth,label={\bf RUM},pos=ne,fontsize=\normalsize]{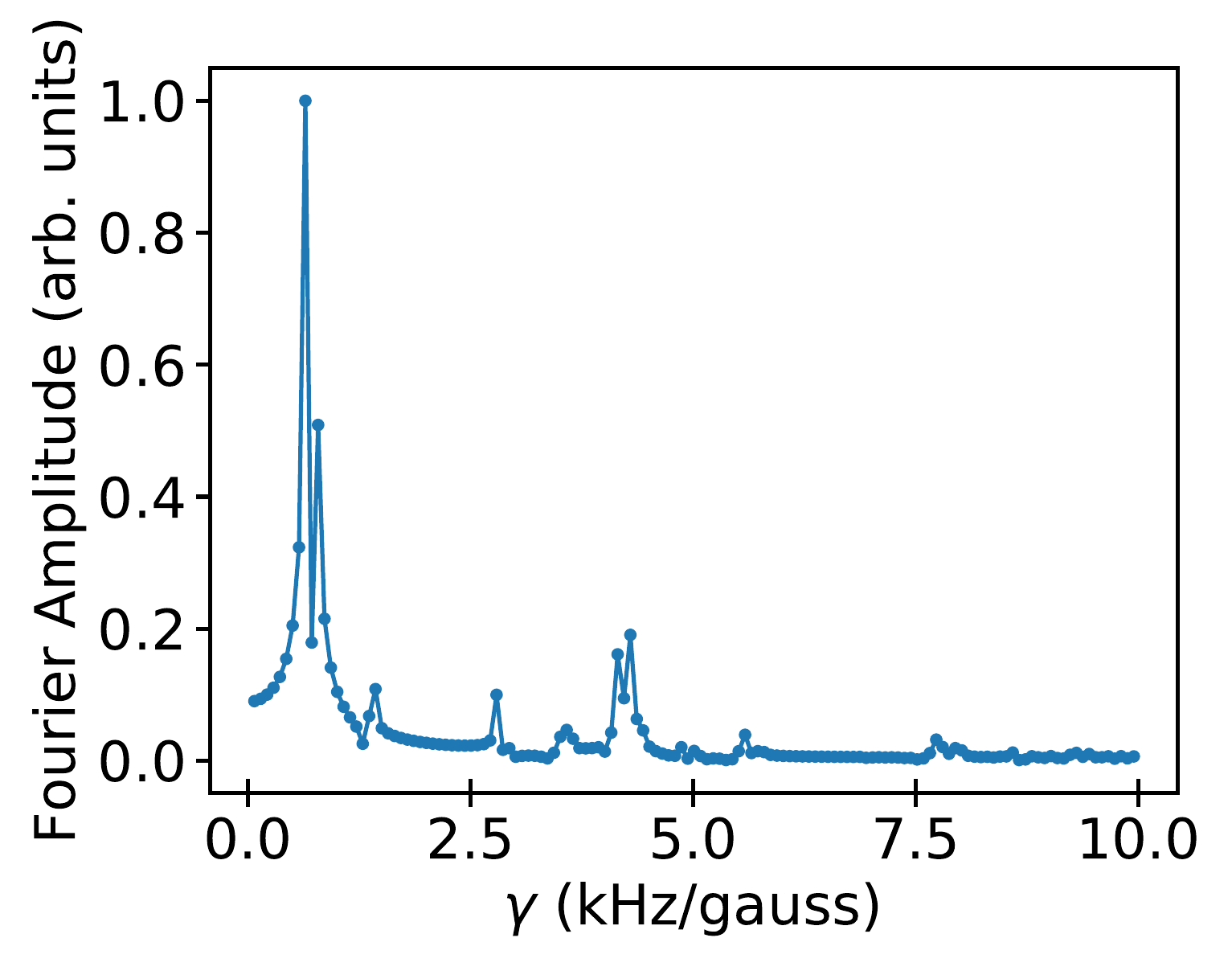}
	\end{minipage}
	\begin{minipage}{.3\linewidth}
		\centering
		\xincludegraphics[width=\textwidth,label={\bf RUM},pos=ne,fontsize=\normalsize]{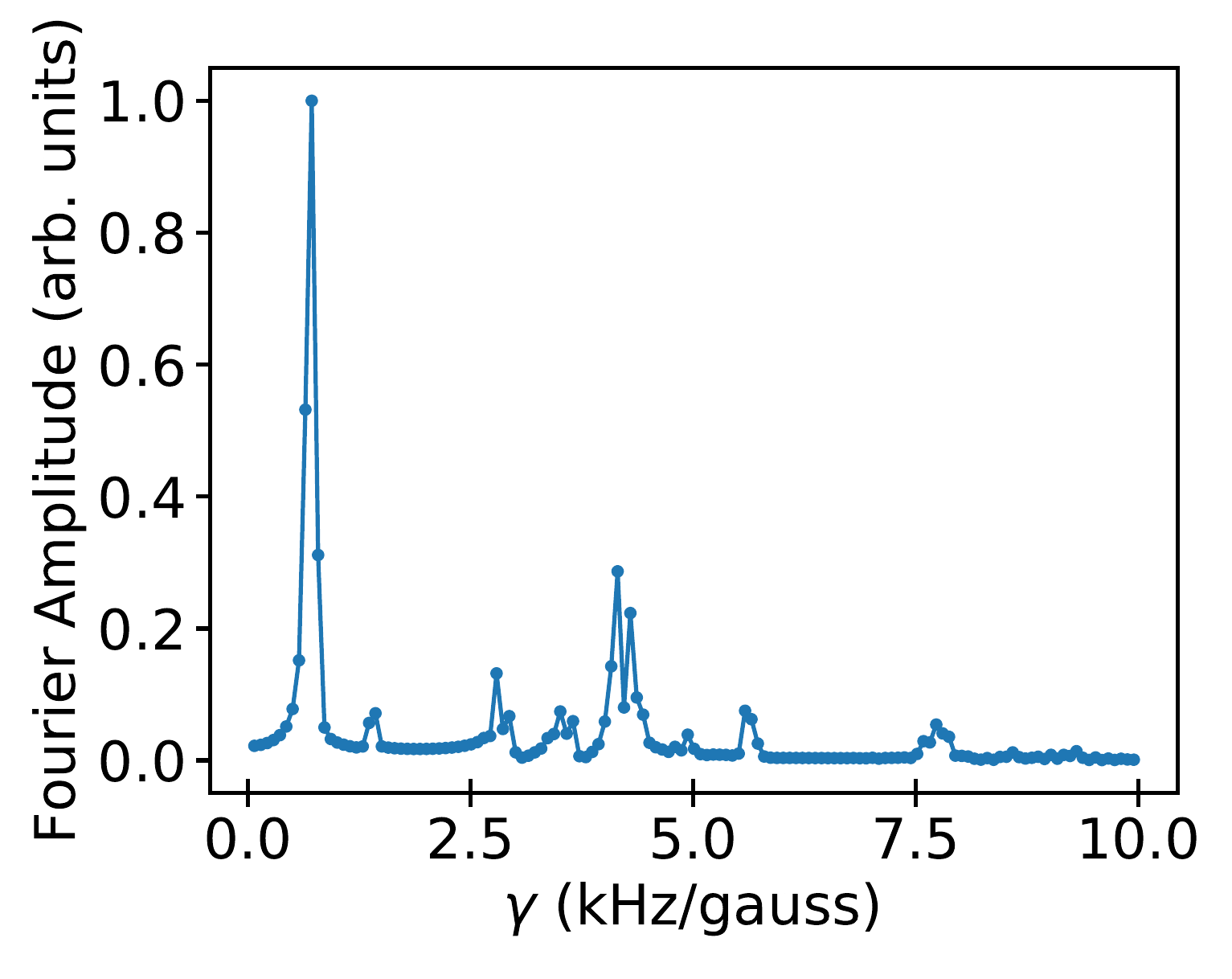}
	\end{minipage}
	\caption{Fourier transforms of the signals shown in Figure \ref{Fig_wigglePlots_differentMatrices} as functions of the generalized gyromagnetic ratio $\gamma$. The panel labels are described in the figure caption to Figure \ref{Fig_wigglePlots_differentMatrices} . 
	}
	\label{Fig_1DFFT_differentMatrices}
\end{figure}

We now examine the sensitivity of the calculated signals to various changes in the scattering transfer matrices. Figures \ref{Fig_wigglePlots_differentMatrices} and \ref{Fig_1DFFT_differentMatrices} 
demonstrate the impact of random variations of the scattering transfer matrix $\mathbf{\Sigma}$ on the oscillatory plots (for $B_1=\SI{440}{\gauss}$) and their spectra,  respectively. For simplicity, we keep the matrix elements of $\mathbf{\Sigma}$ independent of energy.

The first row of each figure (labelled \textbf{RP}, Random Phases) reflects the impact of differing phases imparted to each hyperfine state after scattering. Specifically,  $\mathbf{\Sigma} = \bigoplus_{i=1}^9 e^{i\theta_i}$ is a diagonal unitary matrix whose nine phases $\theta_i$ are randomly chosen from a uniform distribution of width $2\pi$. Such a form of scattering would result from purely elastic scattering where the different hyperfine states probe the surface for different lengths of time (i.e. each state penetrates to a different depth or encounters a resonance with a different lifetime). Significant differences in the relative peak amplitudes can already be seen at this point, indicating that the calculated signal is sensitive to these phases.

The second row of each figure (labelled \textbf{RDA}, Random Diagonal Amplitudes) reflects the impact of differing state losses due to scattering. Specifically,  $\mathbf{\Sigma} = \bigoplus_{i=1}^9 A_i$ is a diagonal matrix whose diagonal elements are randomly chosen from a uniform distribution on the interval $\left[0,1\right)$. This form models the impact of different losses of each hyperfine state to different scattering directions, reactions with the surface, or adsorbtion to the surface. Again, significant changes are observed, indicating sensitivity to these features.

The third and fourth rows (respectively labelled \textbf{ROM}, Random Orthogonal Matrices, and \textbf{RUM}, Random Unitary Matrices) probe the impact of inelastic (projection $m_Im_J$-changing) scattering on the calculated signal. 
For the third row, $\mathbf{\Sigma}$ is an orthogonal matrix randomly drawn according to the Haar measure on $O(9)$, while $\mathbf{\Sigma}$ is a unitary matrix randomly drawn according to the Haar measure on $U(9)$ for the fourth row of each figure. Here, randomly drawing according to the Haar measure can be understood as analogous to drawing from the ``uniform distribution'' over the space of possible matrices \cite{Mezzadri2007}. The orthogonal matrices model inelastic scattering where no relative phase changes occur, while the unitary matrices model inelastic scattering where relative phase changes do occur. In both cases, there is no loss of total population during scattering. Clearly, the calculated signals are also sensitive to inelastic scattering events, both with and without relative phase changes. 
Finally, we can see that the number of peaks in the signal between 430 and 450 gauss varies as a function of the scattering matrix (compare the RDA and \replaced[id=JTC_2]{RUM}{ROM} plots in Figure \ref{Fig_wigglePlots_differentMatrices}, for example)\added[id=JTC_2]{.}

\section{Comparison with a Semi-Classical Method}
\label{Sec_method_comparison}


The present approach is fully quantum mechanical, while, in their Supplemental Material, 
Godsi et al. \cite{Godsi2017} have described  a semi-classical method for calculating $P_{detection}$ that they used to model the propagation of \oH~in their molecular hyperfine interferometer (see  Ref.~\cite{Litvin2019} for the case of spin 1/2 particles). This semi-classical method treats the internal degrees of freedom of the molecules quantum mechanically and the centre-of-mass motion classically. As a result,  the momentum changes induced by the magnetic field are ignored and every internal state is described as propagating at the initial velocity $v_0$ of the molecule. The internal degrees of freedom are treated 
by applying the time evolution operator for the time period $t_i \equiv \frac{L_i}{v_0}$ spent in each magnetic field of length $L_i$. That is, the propagation is calculated in the molecular reference frame with a time-dependent Hamiltonian. Here, we compare the results of the semi-classical and fully-quantum approaches for \oH.


We compare the two methods under conditions close to the original application of the semi-classical method  to flux-detection measurements \cite{Godsi2017}. We work with   a field profile as shown in Figure \ref{Fig_oH2experimentAbstraction}, but with the second arm assumed to be of zero length and $B_2=0$. The field $B_1$ is varied.
For the sake of comparison, we also set the state selector and detector fields in the transfer matrix method  to \SI{100000}{\gauss} so that the basis changes performed by the transfer matrix method out of and into these regions match well the Clebsch-Gordon transformation from $\ket{m_Im_J}$ to $\ket{Fm}$ and its inverse, as used by the semi-classical method. Note that the off-diagonal elements of the full discontinuity matrix $\mathbf{K}$ (\ref{Eqn_fullDiscontMat}) are still only ${\sim} 10^{-5}$ at \SI{100000}{\gauss}, such that their neglect still does not invalidate our fully quantum formalism at these large field strengths. We also retain the rotation from the first branch to the second branch and set the scattering matrix to $\mathbb{1}_9$. To maximize sensitivity of the comparison, we use only a single velocity when calculating $P_{detection}$ in both methods. All other parameters, including the state selector and detector relative state probabilities $\eta_{m_Im_J}$ and $\kappa_{m_Im_J}$, are as per \replaced[id=JTC_2]{Appendix \ref{App_compParam_semiClassical}}{Appendix \ref{App_compParam}}. This allows for a test that includes all incoming and outgoing states and their relative phases at experimentally relevant conditions.

\begin{figure}[H]
	\centering
	\includegraphics[width=0.5\textwidth]{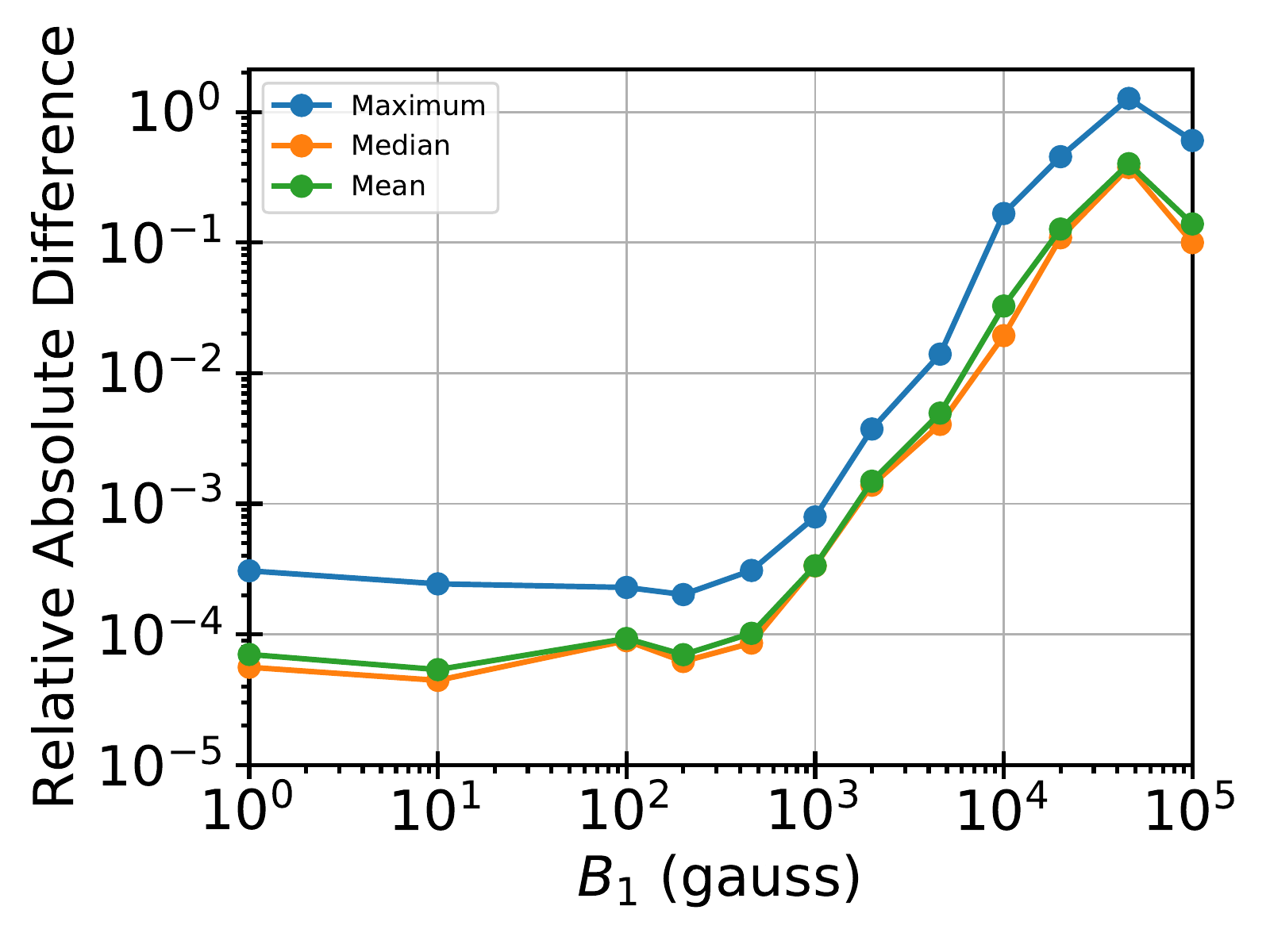}
	\caption{Maximum, median, and mean relative absolute difference between the calculated signals obtained by the semi-classical method discussed in the Supplementary Material of Godsi et al.~\cite{Godsi2017} and the present method, for various values of the controlling magnetic field. See Section \ref{Sec_method_comparison} for a description of the semi-classical method and the field profiles used. The relative absolute difference between the calculated signals is calculated point-by-point as a function of the magnetic field. The mean, median, and maximum values are then calculated over the magnetic field interval spanned by the calculated signal. The calculated signal is sampled at a rate of 1500 points per \SI{10}{\gauss};  the magnetic field varies from $B_1$ to $B_1+\SI{10}{\gauss}$ for each calculated signal; a single velocity was included in the calculations; the scattering transfer matrix $\mathbf{\Sigma} = \mathbb{1}_9$ and is constant for all energies. All other parameters are listed in Appendix \ref{App_compParam_semiClassical}.
	}
	\label{Fig_comparison_RAD}
\end{figure}

The signals $P_{detection}\left(B_1\right)$ are calculated from  $B_1$ to $B_1+\SI{10}{\gauss}$ for various values of $B_1$. We include 1500 datapoints in these ten-gauss intervals. The calculated signals are compared between the two methods by calculating their relative absolute difference at identical conditions. This produces a relative absolute difference at each of the 1500 magnetic field points. We then calculate the maximum, mean, and median relative absolute difference over the ten-gauss interval. Figure \ref{Fig_comparison_RAD} shows how these maximum, mean, and median values vary as a function of $B_1$. At low fields, there is no significant dependence on the magnetic field and the relative absolute difference is below the expected experimental error. This lack of dependence on $B_1$ is possibly due to some residual numerical error present in the implementation of one or both methods, which masks any underlying field-dependence. At approximately \SI{460}{\gauss}, however, the error begins to increase with the magnetic field until it saturates at approximately \SI{46000}{\gauss} at a relative absolute difference of approximately one. This increase in error as a function of magnetic field points to a systematic difference between the two methods. 


\begin{figure}[H]
	\centering
	
	\begin{minipage}{.49\linewidth}
		\centering
		\includegraphics[width=\textwidth]{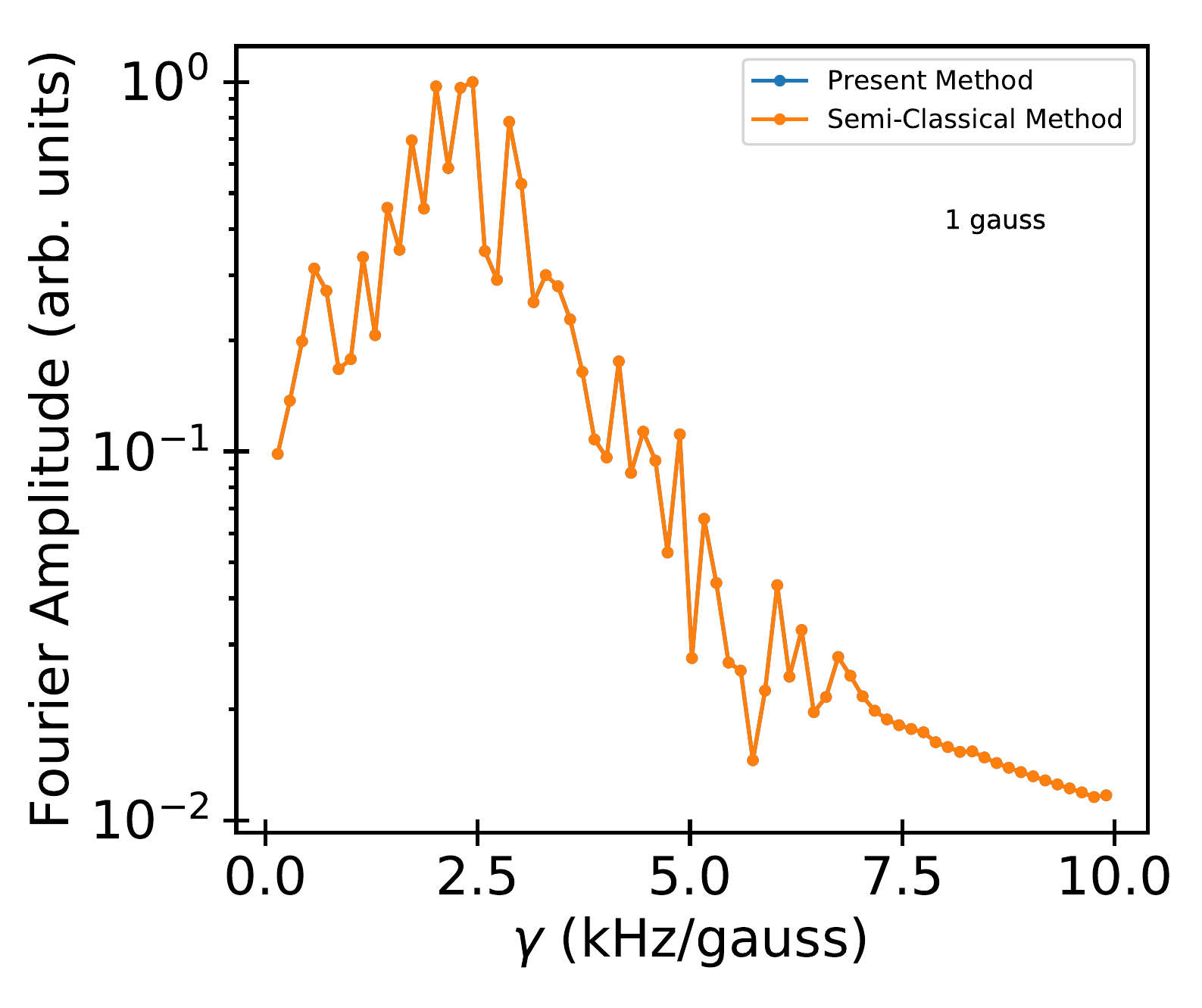}
	\end{minipage}
	\begin{minipage}{.49\linewidth}
		\centering
		\includegraphics[width=\textwidth]{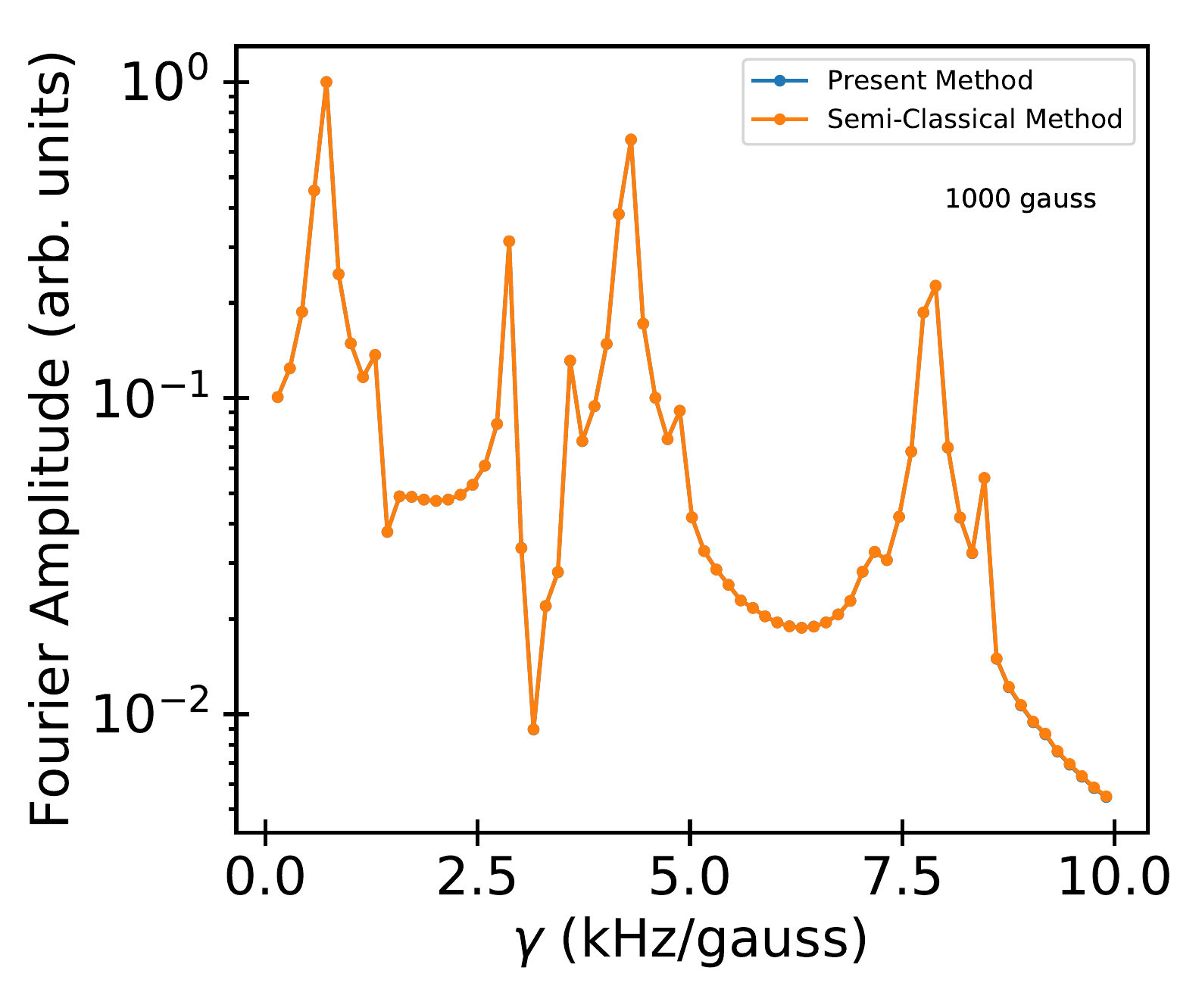}
	\end{minipage}
		
	
	\begin{minipage}{.49\linewidth}
		\centering
		\includegraphics[width=\textwidth]{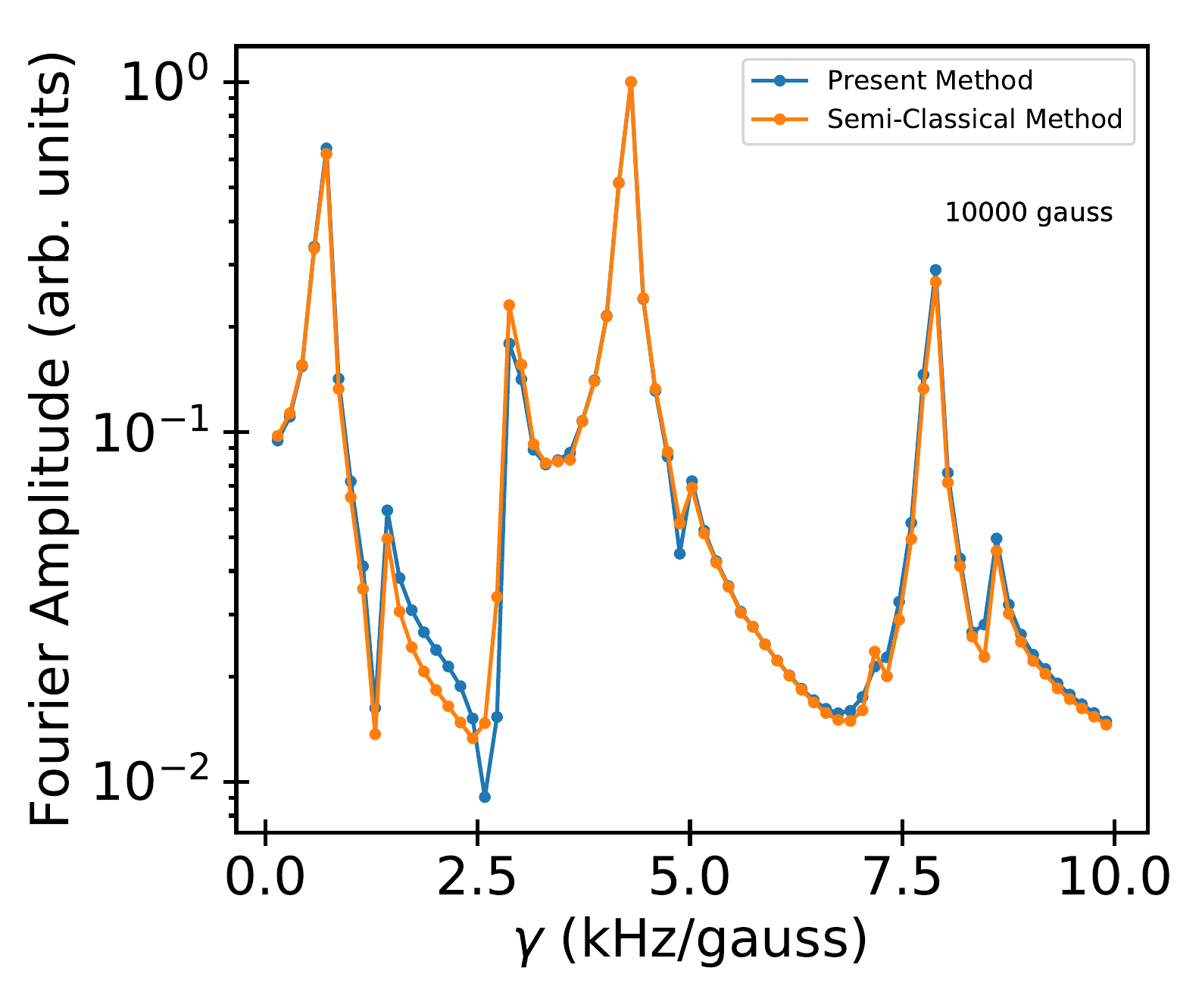}
	\end{minipage}
	\begin{minipage}{.49\linewidth}
		\centering
		\includegraphics[width=\textwidth]{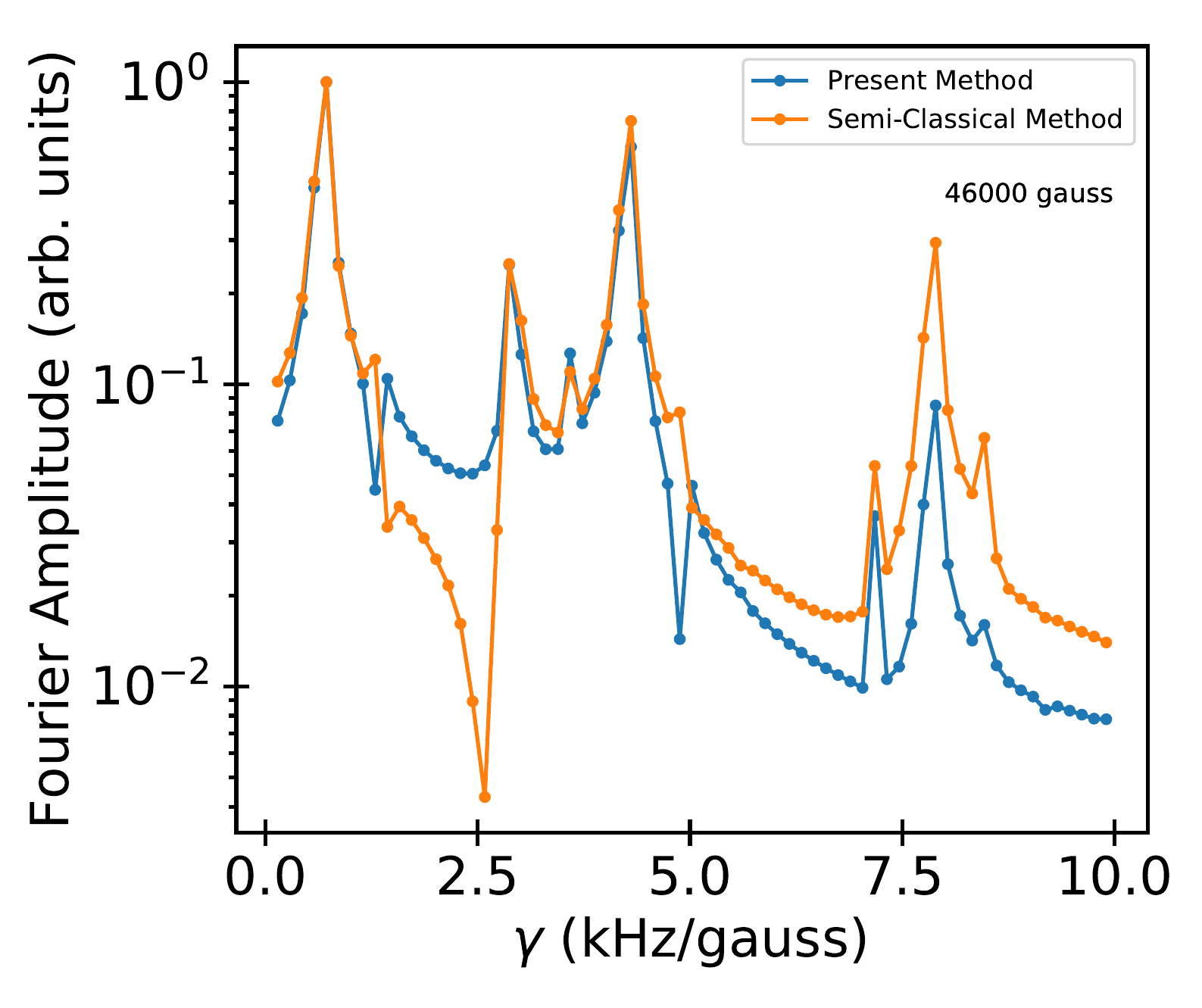}
	\end{minipage}
	\caption{Fourier amplitudes of the calculated signals at various magnetic field values computed with the semi-classical method discussed in the Supplementary Material of Godsi et al.~\cite{Godsi2017} (orange) and the present method (blue) as functions of the generalized gyromagnetic ratio $\gamma$. The calculation conditions are identical to those of Figure \ref{Fig_comparison_RAD}.
	}
	\label{Fig_comparison_FFT_plots}
\end{figure}

To illustrate the difference between the two approaches in more detail, we plot the Fourier transforms of the calculated signals at various magnetic field strengths in Figure \ref{Fig_comparison_FFT_plots}. At field strengths below \SI{1000}{\gauss}, little difference is observed. At higher field strengths, the feature locations agree, while the Fourier amplitudes differ. The feature locations are determined by the eigenvalues of the Hamiltonian, identical in both methods, while the amplitudes are a function of the relative phases and amplitudes of the wavefunction components. These amplitudes and phases are expected to differ between the two methods at sufficiently high fields because of the approximations made in the semi-classical method.

In particular, the semi-classical method accounts for \textit{most} of the relative phase and amplitude changes induced by the controlling magnetic fields. It does this by time-evolving the internal state vector for times that correspond to the time $t_i$ spent in each magnetic field by a molecule moving at its unchanged initial velocity. However, the semi-classical method ignores the small changes in the molecular velocity caused by the magnetic fields. These changes to the velocity modify the time spent in each magnetic field for each individual component of the internal state vector. Thus, $t_i$ should depend on the internal state $\ket{R}$. It is not immediately clear how to 
include these state-dependent velocity changes into the semi-classical method, however.

At low fields, these velocity changes and the dependence of $t_i$ on $\ket{R}$ are negligible and the fully quantum calculations agree with the semi-classical results  
to at least 0.1\%  for fields below \SI{1000}{\gauss}. However, this agreement can only be expected to occur for surfaces that do not change between the surface-impact events of the spatially-separated wavepacket components (discussed in Section \ref{Sec_exptDescription}). The maximum temporal separation between these impact events, caused by the velocity changes,  varies from a few to several hundreds of picoseconds.
Many surfaces do change on this timescale, as has been measured in several helium-3 spin echo experiments 
\cite{Jardine2009a,Jardine2009b,Jardine2009a,Jardine2009b, Jardine2009a,Jardine2009b,Hedgeland2016,Godsi2015,Jardine2009a,Jardine2009a,Jardine2009b}. In other words, the semi-classical method cannot be used to probe the dynamics of surfaces, while the method presented in this manuscript opens the possibility to account for the surface dynamics with molecular scattering experiments. 


%

\section{Conclusion}
\label{Sec_Conclusion}
In this paper, we have developed a theoretical framework for simulating a surface-sensitive molecular hyperfine interferometer. The approach treats the interferometer as an effective one-dimensional system, accounting for the real experimental geometry by rotating the quantization axis of the hyperfine states at the scattering point. 
The time evolution of the molecular states is described fully coherently and accounts for the mixing of the hyperfine states and momentum changes induced by the magnetic fields in the experiment. 
The present approach is fully quantum mechanical and includes a full description of the internal-state-dependent spatial superpositions imposed on the molecular wavepackets by the controlling magnetic fields.
This opens the possibility for a description of molecular scattering experiments that aim to probe surface dynamics on the picosecond to hundreds of picosecond time scale. 
To build the framework, we have derived and implemented a transfer matrix formalism that accounts for the internal (hyperfine) degrees of freedom of molecules and that allows for efficient computation of the experimental signal.

In the present work, 
the molecule-surface interaction is accounted for by a scattering transfer matrix (a transformed version of the scattering matrix) that is suitable for the description of experiments where the surface changes either much more slowly or much more quickly than the molecule-surface or wavepacket-surface interaction times (i.e., the molecule-surface scattering event does not involve energy transfer between the surface and the molecule). The extension to arbitrary surface dynamics (currently under investigation) requires a time-dependent scattering transfer matrix that reflects the underlying time-dependence of the molecule-surface interaction potential. Such a formalism would  naturally incorporate energy transfers between the surface and the molecule during the scattering event. 
 We have demonstrated, using the specific case of \oH, how the different features of the time-independent scattering transfer matrix, such as the phases of the diagonal elements, impact the experimental signal. In addition, we have shown that the experimental signal is sensitive to off-diagonal scattering matrix elements describing collisions that change the  projection quantum numbers of the molecular hyperfine states without energy transfer between the molecule and the surface.
 

The present approach also sets the stage for solving the inverse scattering problem in molecular hyperfine interferometry by means of machine learning approaches, such as Bayesian optimization \cite{Vargas2017,BML}.
For example, one can use the results of the transfer-matrix computations presented here to train Gaussian process models of the predicted experimental signal \cite{BML}. The difference between the experimental observations and the results of the transfer-matrix computations can then be minimized by varying the scattering matrix elements, as described in our previous work \cite{Vargas2017}. 
The results of Bayesian optimization will determine the properties of the scattering matrix elements compatible with a given experimental measurement. These scattering matrix properties can then be used to gain physical insight into molecule-surface interactions and surface properties. They can also be used to test approximations used in \emph{ab initio} calculations.


The formalism presented here is general to all closed-shell molecules and is flexible to describe various experimental setups. It can be used to explore various experimental protocols and evaluate their effectiveness at determining various molecule-surface interactions and surface properties. Thus, this paper provides the theoretical framework necessary to interpret a wide range of molecular hyperfine interferometry experiments, which are poised to apply molecular beam techniques to provide new information about molecule-surface interactions, surface morphologies, and surface dynamics.

\clearpage
\newpage

\section*{Acknowledgements}
This work is supported by NSERC of Canada and the Horizon 2020 Research and Innovation Programme grant 772228. We acknowledge useful discussions with \added[id=JTC_2]{Helen Chadwick,} Geert Jan-Kroes, Ilya Litvin, and Tsofar Maniv \added[id=JTC_2]{and thank Oded Godsi for allowing us to use the source code of his implementation of the semi-classical method}.

\appendix

\section{Schr\"odinger Equation for Eigenstate Coefficients}
\label{App_schroedingEqn} 

Using Eqn.~\ref{Eqn_eigstateExpanDef}, the time-independent Schro\"odinger equation is
\begin{align}
\hat{H}\ket{E\tilde R} &= E\ket{E\tilde R} \\
\hat{K}\ket{E\tilde R} &= \left(E-\RamHam\right)\ket{E\tilde R} \nonumber \\
\sum_{R}\int\mathrm{d}x~ \Phi_{R}^{E\tilde R}(x) \hat{K}\ket{xR} &= \sum_{R}\int\mathrm{d}x~ \Phi_{R}^{E\tilde R}(x) \left(E-\RamHam\right)\ket{xR} \nonumber \\
\sum_{R}\int\mathrm{d}x~ \Phi_{R}^{E\tilde R}(x) \bra{x_0R_0}\hat{K}\ket{xR} &= \sum_{R}\int\mathrm{d}x~ \Phi_{R}^{E\tilde R}(x) \bra{x_0R_0}\left(E-\RamHam\right)\ket{xR} \label{Eqn_schroedStart} 
\end{align}
where $\hat{H}$ is the total Hamiltonian (\ref{Eqn_totalHami}) in the current region, $\hat{K}\equiv\frac{{\hat p}^2}{2m}$, we use $\RamHam$ as a shorthand for $\RamHam(\vec{B}_\mathrm{loc})$, and the last line was multiplied by $\bra{x_0R_0}$.

The different terms can be evaluated as
\begin{align}
\bra{x_0R_0}\hat{K}\ket{xR} &= \delta_{R_0R}\bra{x_0}\int\mathrm{d}k~ \frac{\hbar^2k^2}{2m}\ket{k}\braket{k|x} \nonumber \\
&=  \int\frac{\mathrm{d}k}{2\pi}~ \label{Eqn_KOperatorElement} \delta_{R_0R}\frac{\hbar^2k^2}{2m}e^{ik\left(x_0-x\right)}, \\
\bra{x_0R_0}E \ket{xR} &= \delta_{R_0R}\delta(x-x_0)E, \\
\bra{x_0R_0}\RamHam \ket{xR} &= \delta(x-x_0)\RamHamEl_{R_0R},
\end{align}
where $\ket{k}$ is a momentum state with wavenumber $k$, and $m$ is the mass of the molecule. The additional factor of $\left(2\pi\right)^{-1}$ in Eqn.~(\ref{Eqn_KOperatorElement}) comes from $\braket{x|k}\equiv\left(2\pi\right)^{-\frac{1}{2}}e^{ikx}$. After inserting these three equations into Eqn.~(\ref{Eqn_schroedStart}) and evaluating most of the sums, we obtain
\begin{align}
\int\mathrm{d}x~ \int\frac{\mathrm{d}k}{2\pi}~ \frac{\hbar^2k^2}{2m}e^{ik\left(x_0-x\right)} \Phi_{R_0}^{E\tilde R}(x) &= \Phi_{R_0}^{E\tilde R}(x_0)E-\sum_{R}\RamHamEl_{R_0R}\Phi_{R}^{E\tilde R}(x_0) 
\end{align}
Noting that $k^2e^{ikx_0}=-\frac{\partial^2}{\partial x_0^2}e^{ikx_0}$ and $\int\frac{\mathrm{d}k}{2\pi}~ e^{ik\left(x_0-x\right)} = \delta(x_0-x)$, we obtain Eqn.~(\ref{Eqn_coeffSchroedEqn}) after the relabelling $x_0 \to x$.

\section{Coefficient Relations Across a Discontinuity}
\label{App_discontCoeff}

Since $\Phi_{R}^{E\tilde R}(x)\in C^1(x)$ for a specific value of $R$ and given Eqn.~(\ref{Eqn_coeffBasisTransform}), we get the defining equations for the continuity of  the wavefunction as
\begin{align}
\lim_{x\to 0^{-}} \Phi_{R^{+}}^{E\tilde R}(x) &= \lim_{x\to 0^{+}} \Phi_{R^{+}}^{E\tilde R}(x) &&\nonumber \\
\lim_{x\to 0^{-}} \sum_{R^{-}} \Phi_{R^{-}}^{E\tilde R}(x)S_{R^{-}R^{+}}^* &= \lim_{x\to 0^{+}} \Phi_{R^{+}}^{E\tilde R}(x) && \nonumber \\
\lim_{x\to 0^{-}} \sum_{R^{-}} S_{R^{-}R^{+}}^* \left(A_{R^{-}}e^{ik_{R^{-}}x}+B_{R^{-}}e^{-ik_{R^{-}}x}\right) &= \lim_{x\to 0^{+}} A_{R^{+}}e^{ik_{R^{+}}x}+B_{R^{+}}e^{-ik_{R^{+}}x} && \text{[Eqn.~(\ref{Eqn_coeffKspaceExpansion})]}\nonumber \\
A_{R^{+}}+B_{R^{+}} &= \sum_{R^{-}} S_{R^{-}R^{+}}^* \left(A_{R^{-}}+B_{R^{-}}\right), \label{AppEqn_coeffContinuity}
\end{align}
where $S_{R^{-}R^{+}}^* \equiv \braket{R^{+}|R^{-}}$,  $k_{R^{\pm}} \equiv \frac{\sqrt{2m \left(E-E_{\text{R}^{\pm}}\right) }}{\hbar}$, and $E_{\text{R}^{\pm}}\equiv \braket{R^{\pm}|\RamHam\boldsymbol{\left(\right.}\vec B(0^{\pm})\boldsymbol{\left.\right)}|R^{\pm}}$. There are $N_R$ such equations, one for each value of $R^{+}$. 

Correspondingly, the defining equations for the continuity of the first derivative of the coefficients are 
\begin{align}
\lim_{x\to 0^{-}} \frac{\partial}{\partial x} \Phi_{R^{+}}^{E\tilde R}(x) &= \lim_{x\to 0^{+}} \frac{\partial}{\partial x}\Phi_{R^{+}}^{E\tilde R}(x) &&\nonumber \\
\lim_{x\to 0^{-}} \sum_{R^{-}} \frac{\partial}{\partial x}\Phi_{R^{-}}^{E\tilde R}(x)S_{R^{-}R^{+}}^* &= \lim_{x\to 0^{+}} \frac{\partial}{\partial x} \Phi_{R^{+}}^{E\tilde R}(x) && \nonumber \\
\lim_{x\to 0^{-}} \sum_{R^{-}} S_{R^{-}R^{+}}^* \frac{\partial}{\partial x}\left(A_{R^{-}}e^{ik_{R^{-}}x}+B_{R^{-}}e^{-ik_{R^{-}}x}\right) &= \lim_{x\to 0^{+}} \frac{\partial}{\partial x} \left( A_{R^{+}}e^{ik_{R^{+}}x}+B_{R^{+}}e^{-ik_{R^{+}}x}\right) && \text{[Eqn.~(\ref{Eqn_coeffKspaceExpansion})]}\nonumber \\
\lim_{x\to 0^{-}} \sum_{R^{-}} S_{R^{-}R^{+}}^* ik_{R^{-}}\left(A_{R^{-}}e^{ik_{R^{-}}x}-B_{R^{-}}e^{-ik_{R^{-}}x}\right) &= \lim_{x\to 0^{+}} ik_{R^{+}} \left( A_{R^{+}}e^{ik_{R^{+}}x}-B_{R^{+}}e^{-ik_{R^{+}}x}\right) && \nonumber \\
A_{R^{+}}-B_{R^{+}} &= \sum_{R^{-}} S_{R^{-}R^{+}}^* \frac{k_{R^{-}}}{k_{R^{+}}} \left(A_{R^{-}}-B_{R^{-}}\right), \label{AppEqn_coeffDerivContinuity}
\end{align}
Solving Eqns.~(\ref{AppEqn_coeffContinuity}) and (\ref{AppEqn_coeffDerivContinuity}) for the coefficients $A_{R^{+}}$ and $B_{R^{+}}$, we obtain Eqns.~(\ref{Eqn_Acoeff}) and (\ref{Eqn_Bcoeff}).


\section{Computational parameters used for the application to \textit{ortho}-hydrogen}
\label{App_compParam} 

We take the mean velocity $v_0 = \SI{1436.14}{\meter \per \second}$ and the velocity spread to be \SI{4}{\%~FWHM}. When performing the integral of Eqn.~(\ref{Eqn_P_DetectFinal}), we take a $k$-space grid spacing $\Delta k = \SI{1E4}{\per \cm}$ and integrate from $-7\sigma_k$ to $+7\sigma_k$, where $\sigma_k$ is the Gaussian width in momentum space as defined in Section \ref{Sec_magLensImpact}. For the magnetic field profile and the angles between the two branches of the apparatus, see Figure \ref{Fig_oH2experimentAbstraction}. The relative probabilities used for the state selector probabilities $P_{R_0}$ and the detector coefficients  $c_{R_D}$ are given in Table \ref{Tab_detectorFunction}.

Where applicable, the parameters above were chosen to match those in the supplementary information of Godsi et al.~\cite{Godsi2017}\replaced[id=JTC_2]{, apart for the relative probabilities in Table \ref{Tab_detectorFunction}. The relative probabilities in Table \ref{Tab_detectorFunction} were obtained from improved semi-classical calculations of the molecular propagation through the magnetic lens \cite{Godsi2017,Kruger2018}.}{.} 

\begin{table}[H]
	\caption{Relative probabilities of the state selector $\eta_{m_Im_J}$ and the detector $\kappa_{m_Im_J}$. The state selector probabilities $P_{R_0}$ are calculated as $P_{R_0}=P_{m_Im_J}\equiv\eta_{m_Im_J}/\sum_{m_Im_J}\eta_{m_Im_J}$ and the detector coefficients $c_{R_D}$ are calculated as $c_{R_D}=c_{m_Im_J}\equiv\kappa_{m_Im_J}/\sum_{m_Im_J}\kappa_{m_Im_J}$.   }

	\begin{center}
	\resizebox{0.9\textwidth}{!} {
		\begin{tabular}{    l  c  c  c  c  c  c c  c  c }
			\toprule
			 $m_I$ & 1 & 1 & 1 & 0 & 0 & 0 & -1 & -1 & -1  \\ 
			 $m_J$& 1 & 0 & -1 & 1 & 0 & -1 & 1 & 0 & -1  \\ 
			\colrule
			$\eta_{m_Im_J}$& 0.0095&  0.0138&  0.0187&  0.0416&  0.0436&  0.0606&  0.3997&  0.9015&  1.0  \\ 
			$\kappa_{m_Im_J}$&  0.0611&  0.08&    0.1027&  0.3834&  0.5705&  0.8425&  1.0&     0.9422&  0.7209  \\ 
			\botrule
		\end{tabular}
	}
	\end{center}
	\label{Tab_detectorFunction}
\end{table}
%

\section{\added[id=JTC_2]{Computational parameters used for the comparison with the semi-classical method}}
\label{App_compParam_semiClassical} 

\added[id=JTC_2]{We take the mean velocity $v_0 = \SI{1436.14}{\meter \per \second}$. The relative probabilities used for the state selector probabilities $P_{R_0}$ and the detector coefficients  $c_{R_D}$ are given in Table \ref{Tab_detectorFunction_comparison}. Where applicable, the parameters were chosen to match those in the supplementary information of Godsi et al.~\cite{Godsi2017}.}

\begin{table}[H]
	\caption{Relative probabilities of the state selector $\eta_{m_Im_J}$ and the detector $\kappa_{m_Im_J}$, as used in the comparison to the semi-classical method of Godsi et al.~\cite{Godsi2017}. The state selector probabilities $P_{R_0}$ are calculated as $P_{R_0}=P_{m_Im_J}\equiv\eta_{m_Im_J}/\sum_{m_Im_J}\eta_{m_Im_J}$ and the detector coefficients $c_{R_D}$ are calculated as $c_{R_D}=c_{m_Im_J}\equiv\kappa_{m_Im_J}/\sum_{m_Im_J}\kappa_{m_Im_J}$.   }
	
	\begin{center}
		\resizebox{0.9\textwidth}{!} {
			\begin{tabular}{    l  c  c  c  c  c  c c  c  c }
				\toprule
				$m_I$ & 1 & 1 & 1 & 0 & 0 & 0 & -1 & -1 & -1  \\ 
				$m_J$& 1 & 0 & -1 & 1 & 0 & -1 & 1 & 0 & -1  \\ 
				\colrule
				$\eta_{m_Im_J}$& 1.0000 & 0.9755 & 0.7901 & 0.1465 & 0.1111 & 0.0738 & 0.0343 & 0.0299 & 0.0258  \\ 
				$\kappa_{m_Im_J}$& 1.00 & 0.96 & 0.93 & 0.53 & 0.42 & 0.37 & 0.21 & 0.19 & 0.16  \\ 
				\botrule
			\end{tabular}
		}
	\end{center}
	\label{Tab_detectorFunction_comparison}
\end{table}


\bibliography{HHE_TheoryPaper}

\end{document}